\documentclass[aps,twocolumn,showpacs,superscriptaddress]{revtex4}
\usepackage{epsfig}
\usepackage{graphicx}
\usepackage{amsfonts}
\usepackage[figuresright]{rotating}
\usepackage{amssymb}
\usepackage{amsmath}
\usepackage{subfigure}

\def\avg#1{\langle#1\rangle}

\def\be{\begin{equation}}
\def\ee{\end{equation}}
\def\bea{\begin{eqnarray}}
\def\eea{\end{eqnarray}}

\def\nn{\nonumber}
\def\pp{\parallel}

\begin{document}

\title{Quantum anomalous Hall states in the $p$-orbital honeycomb optical
lattices}
\author{Machi Zhang}
\affiliation{Department of Physics, University of California, San Diego,
CA 92093}
\affiliation{Department of Physics, Tsinghua University, Beijing, China 100084}
\author{Hsiang-hsuan Hung}
\affiliation{Department of Physics, University of California, San Diego,
CA 92093}
\author{Chuanwei Zhang}
\affiliation{Department of Physics and Astronomy, Washington State
University, Pullman,WA 99164}
\author{Congjun Wu}
\affiliation{Department of Physics, University of California, San Diego,
CA 92093}

\begin{abstract}
We study the quantum anomalous Hall states in the $p$-orbital bands of
the honeycomb optical lattices loaded with the single component fermions.
Such an effect has not been realized in both condensed matter and
cold atom systems yet.
By applying the available experimental technique developed by Gemelke
\textit{et al.} to rotate each lattice site around its own center
\cite{gemelke2010,gemelke2007,gemelke2009}, the band structures become
topologically non-trivial.
At a certain rotation angular velocity $\Omega$, a flat band structure
appears with localized eigenstates carrying chiral current moments.
With imposing the soft confining potential, the density profile
exhibits a wedding-cake shaped distribution with insulating plateaus at
commensurate fillings.
Moreover, the inhomogeneous confining potential induces dissipationless
circulation currents whose magnitudes and chiralities vary with the
distance from the trap center.
In the insulating regions the Hall conductances are quantized,
and in the metallic regions the directions and magnitudes of
chiral currents cannot be described by the usual
local-density-approximation.
The quantum anomalous Hall effects are robust at temperature scales
small compared to band gaps, which increases the feasibility of
experimental realizations.
\end{abstract}
\maketitle

\section{Introduction}
\label{sect:intro}
The anomalous Hall effect appears in ferromagnets in the absence of
external magnetic fields which was discovered soon after the Hall effect.
The mechanism of the anomalous Hall effect has been debated for a long time,
including theories of the anomalous velocity from the interband matrix elements
\cite{karplus1954}, the screw scattering \cite{smit1958},
and the side jump \cite{berger1970}.
Recently, a new perspective has been developed from topological band
properties, the Berry curvature of the Bloch wave eigenstates
\cite{jungwirth2002,nagaosa2006,nagaosa2010,niuqian2009,niuqian2010,
tomizawa2009}, which has been very successful.
The Berry curvatures serve as an effective
magnetic field in the crystal momentum space, leading to an anomalous
transversal velocity of electrons when electric fields are
applied \cite{niuqian2009}.
The anomalous transversal velocity of electrons gives the
intrinsic contribution to the observed anomalous Hall conductivity in
ferromagnetic semiconductors.

The integer quantum Hall effect (QHE) is the quantized version of the
Hall effect, in which Hall conductances are precisely quantized at
integer values.
This effect arises in the two dimensional electron gases in magnetic fields
with integer fillings of Landau levels.
The quantization of the Hall conductance is protected by the non-trivial
band structure topology characterized by the Thouless-Kohmoto-Nightingale-den
Nijs (TKNN) number, or the Chern number \cite{thouless1982,kohmoto1985}.

In order to achieve a non-zero Chern number pattern, time-reversal 
symmetry needs
to be broken, but Landau levels are not necessary.
Integer QHEs can appear as a result of the parity anomaly of the 2D
Dirac fermions \cite{jackiw1984,fradkin1986,haldane1988}.
Haldane constructed a tight binding model in the
honeycomb lattice with Bloch wave band structures, and showed 
that it exhibits quantum Hall states with $\nu =\pm 1$ \cite{haldane1988}.
This effect is termed {\textquotedblleft quantum anomalous Hall effect\textquotedblright} (QAHE) because the net magnetic flux is zero in
each unit cell and there are no Landau levels.
The Haldane model has been taken as a prototype model for QAHEs.

The Hall effect has been generalized into electron systems with spin
degrees of freedom as {``spin Hall effect''}, in which transverse spin
currents instead of charge currents are induced by electric fields \cite
{dyakonov1971,hirsch1989,sinova2003,murakami2003,kato2004, wunderlich2005}.
Different from the Hall effect, the spin Hall effect maintains time-reversal
symmetry. 
Topological insulators are the quantum version of the spin Hall systems, 
which exist in both 2D and 3D systems. 
Their band structures are characterized by the $Z_2$-topological index
\cite{bernevig2006a,qi2008a,kane2005,sheng2006,moore2007,roy2009,fu2007,
fu2007a,zhang2009}.
These states have robust gapless helical edge modes with odd number
of edge channels in 2D systems \cite{kane2005,wu2006,xu2006}, and odd number
of surface Dirac cones in 3D systems \cite{fu2007,fu2007a,zhang2009}.
Topological insulators have been experimentally observed in 2D quantum
wells through transport measurements \cite{konig2007}, and also in 3D
systems of Bi$_x$Sb$_{1-x}$, Bi$_2$Te$_3$,
Bi$_2$Se$_3$, and Sb$_2$Te$_3$ through the angle-resolved photoemission
spectroscopy \cite{hsieh2008,hsieh2009,xia2009,chen2009} and the absence of
backscattering from scanning tunnelling microscopy spectroscopy
\cite{roushan2009,alpichshev2010,zhang_chen_xue2009}.

Among all these Hall effects mentioned above, only the QAHE
has not been experimentally observed yet. Several proposals have
been suggested to realize this novel Hall effect in semiconductor systems
with topological band structures by breaking time-reversal symmetry,
such as ferromagnetic ordering \cite {qi2006a,liu2008,onoda2003a,yu2010}. 
Because no external magnetic fields are
involved, QAHE states are expected to realize the
dissipationless charge transport with much less stringent conditions than
those of the quantum Hall effect.
This is essential for future device applications.

On the other hand, the development of cold atom physics has provided a new
opportunity for the study of QHEs and QAHEs.
Several methods to realize these effects have been proposed
including globally rotating traps and optical lattices, or introducing
effective gauge potential generated by laser beams \cite{ho2000,scarola2007,
umucalilar2008,zhu2006,zhangcw2010,shao2008,stanescu2010,liu_liu2010}. In
particular, the Haldane-like models were proposed in Refs 
\cite{shao2008,stanescu2010,liu_liu2010}. Furthermore, the realization of the
quantum spin Hall systems has also been proposed \cite{goldman2010}. All
these proposals involve experimental techniques to be developed.

In a previous paper \cite{wu2008}, one of us has proposed to realize
the QAHE in the $p$-orbital bands in the honeycomb
optical lattices through orbital angular momentum polarizations. This can be
achieved by rotating each optical site around its own center, but there is
no overall lattice rotation. The net effect of this type of rotation is the
``orbital Zeeman effect'', which breaks the degeneracy of the onsite $p_x\pm
i p_y$ orbitals. This gives rise to non-trivial topological band structures,
and provides a natural way to realize the Haldane model. Increasing the
rotation angular velocity induces the topological phase transition by
changing the band structure Chern numbers. In the regime of large rotation
angular velocities, the band structures reduce into two copies of Haldane's
model for each of the $p_x\pm i p_y$ orbitals, respectively.

The main advantage of this proposal is that all the experimental techniques
involved are available. The honeycomb optical lattice was constructed long
time ago \cite{grynberg1993}. Recently, the superfluid-Mott insulator phase
transitions of bosons have been observed in the honeycomb lattice by
Sengstock's group \cite{sengstock2010}. The rotation technique has been
developed by Gemelke, Sarajlic, and Chu \cite{gemelke2010}. They have
applied it to rotate the triangular lattice filled with bosons to study the
fractional quantum Hall physics \cite{gemelke2009,gemelke2007}.
For the purpose of studying QAHE, we only need to apply this technique
to the honeycomb lattice and load it with fermions.

This proposal brings a natural connection between the QAHE
and orbital physics in optical lattices. Orbital is a degree of
freedom independent of charge and spin, which was originally investigated in
solid state systems. It plays an important role in superconductivity,
magnetism, and transport properties in transition metal oxides. The key
features of orbital physics are orbital degeneracy and spatial anisotropy.
Optical lattices bring new features to orbital physics which are not easily
accessible in solid state orbital systems. First, optical lattices are rigid
and free from Jahn-Teller distortions, thus orbital degeneracy
is robust. Second, the metastable bosons pumped into high orbital bands
exhibit novel superfluidity beyond Feynman's ``no-node" theory
\cite{liu2006,wu2006,stojanovic2008,wu2009,isacsson2005,kuklov2006},
which does not appear in $^4$He and the previous study of cold bosons.
Excitingly, this unconventional type of BECs have been 
experimentally observed recently \cite{mueller2007,wirth2010}.
Third, $p$-orbitals have a stronger spatial
anisotropy than that of $d$ and $f$-orbitals, while correlation effects in
$p$-orbital solid state systems (e.g. semiconductors) are not that strong.
In contrast, interaction strength in optical lattices is tunable.
We can integrate strong correlation effects with strong spatial anisotropy
more closely than ever in $p$-orbital optical lattice systems
\cite{wu2007,wu2008a,wu2008b,zhang_hung_wu2010}. 
Recently, we also extend the research of orbital physics with cold atoms 
into unconventional Cooper pairings, which includes the $f$-wave 
Cooper pairing \cite{lee2009} in the honeycomb lattice, and the 
``frustrated Cooper pairing'' in the triangular lattice \cite{hung2009}.

This paper is as an expanded version of the previous publication of Ref.
\cite{wu2008} on QAHE in the $p$-orbital band in optical lattices.
We will also present new results including the chiral flat band
structures which occur at an intermediate rotating angular velocity.
The effects of the confining
potential are investigated in detail, including the distributions of
densities and anomalous Hall currents.
The quantized anomalous conductances appear in the band insulating
regime at commensurate fillings.
The magnitudes and chiralities of anomalous Hall currents
in the metallic regions cannot be described by the usual
local-density-approximations.

The rest of the paper is organized as follows. In Sec. \ref{sect:ham}, we
give an introduction to the experimental setup and the orbital Zeeman
coupling. In Sec. \ref{sect:haldane}, a heuristic picture is given to arrive
at the Haldane model at large rotation angular velocities. In Sec. \ref
{sect:band}, the band structures including Berry curvatures and flat bands
are studied. In Sec. \ref{sect:trap}, the spatial distributions of the
particle density and anomalous Hall currents in the inhomogeneous harmonic
trap is studied. Finite temperature effects are also studied.
In Sec. \ref{sect:dect}, a brief discussion on the
detection of the anomalous Hall current is presented. Conclusions are made
in Sec. \ref{sect:conclusions}.


\section{The tight-binding Hamiltonian with the on-site rotation}
\label{sect:ham}

\begin{figure}[tbp]
\centering\epsfig{file=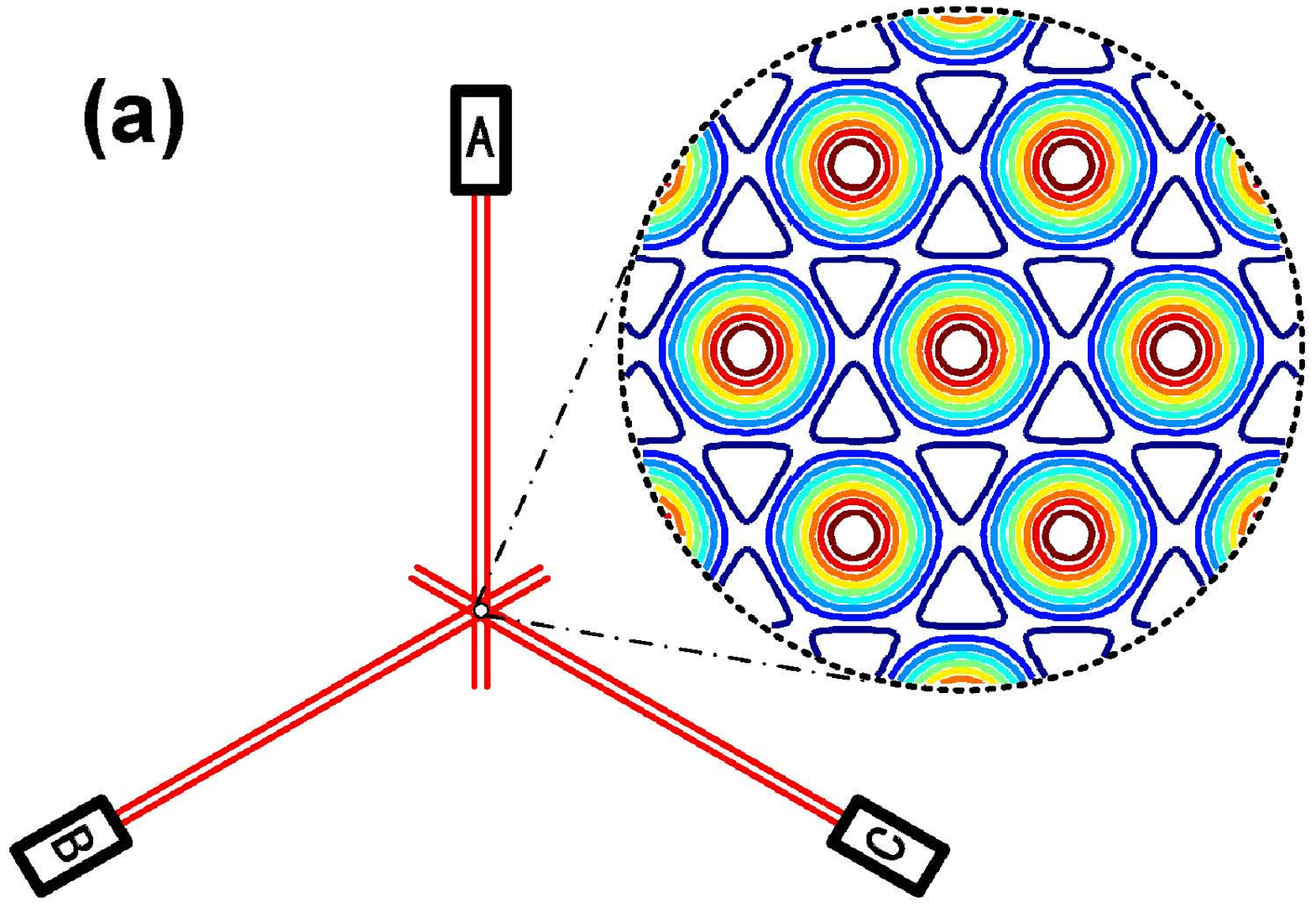,clip=1,width=0.65\linewidth,angle=0}
\centering\epsfig{file=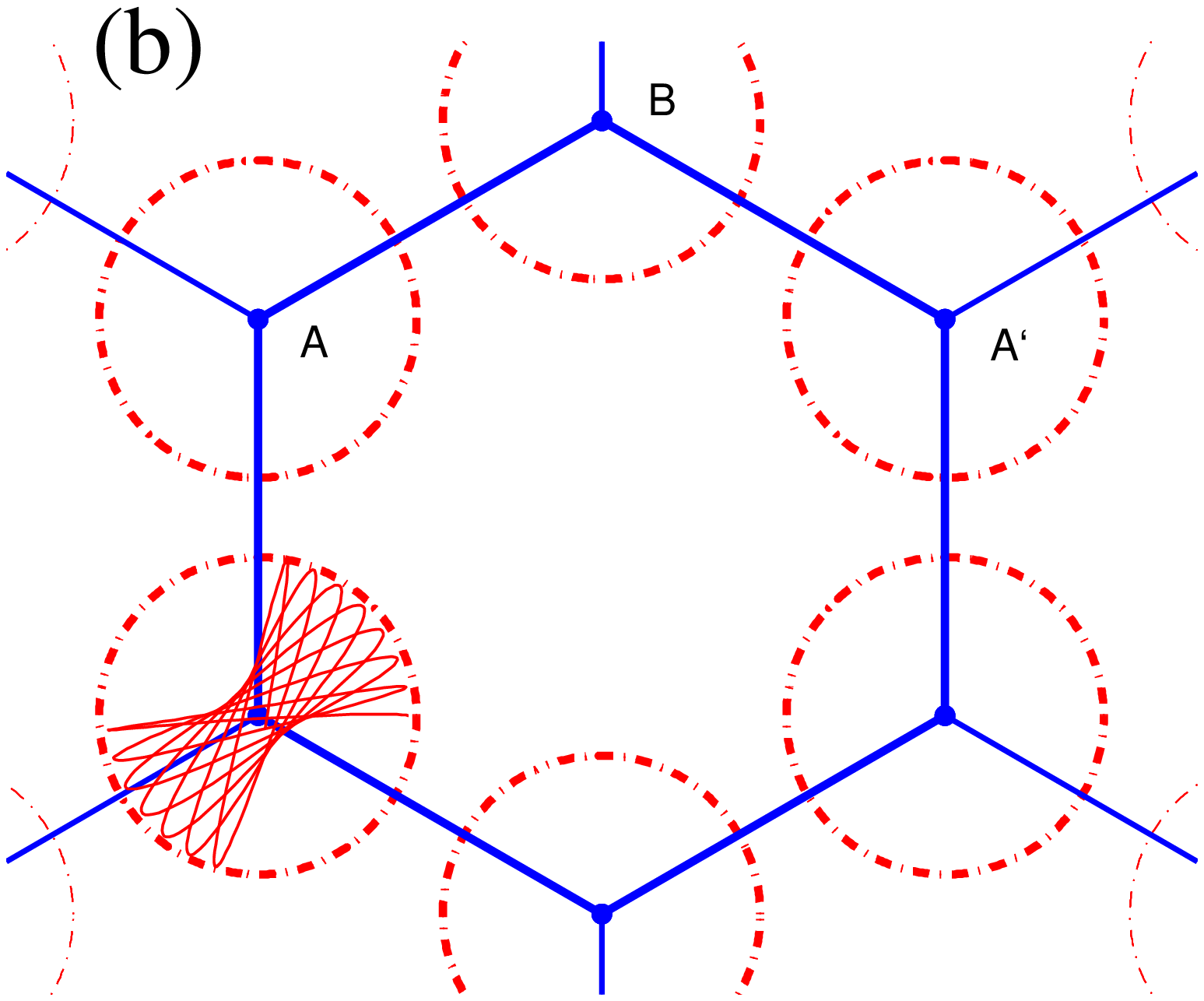,clip=1,width=0.65\linewidth,angle=0}
\caption{(Color online)
~a) The sketch of the honeycomb optical lattice.
The three laser beams cross each other at 120$^{\circ }$ in the $xy$-plane.
Phase modulators are placed in the paths of the two beams. b) The
illustration of the on-site rotation by Gemelke \textit{et al.} \protect\cite
{gemelke2007,gemelke2009,gemelke2010}. The entire lattice takes the motions
of a fast oscillation with the frequency $\protect\omega _{\text{RF}}$ and a
slow precession with the frequency $\Omega $, as schematically plotted with
the red solid lines in one lattice site. After taking the time average of
the fast oscillation, atoms feel that each site is rotating around its own
center with the precession frequency $\Omega $, which is plotted with the
red dot dash line around each site. }
\label{fig:expsketch}
\end{figure}

In this section we describe the experimental setup by Gemelke \textit{et al.}
to realize the on-site rotation of optical lattices \cite{gemelke2010,
gemelke2007,gemelke2009}, and then construct the effective tight-binding model
for such a system.

\subsection{The experiment setup by Gemelke \textit{et al.}}
The honeycomb optical lattice was experimentally realized quite some time ago
\cite{grynberg1993,zhu2007,kllee2009,sengstock2010}.
It is constructed by three phase coherent coplanar laser beams with
polarization along the $z$-axis, intersecting each other with $120^{\circ }$
in the $xy$-plane.
The schematic diagram of the experiment setup is shown in
Fig. \ref{fig:expsketch}.
The potential minima of the interference pattern form the honeycomb lattice if
the laser frequency is blue-detuned from the atom resonance frequency.
The advantage of this technique is that the phase shift in the laser beams only
leads to shift entire lattices without destroying the lattice geometry.

The on-site rotation technique by Gemelke \textit{et al.} was originally
applied to the triangular lattice \cite{gemelke2009,gemelke2010}. It would
be straightforward to apply the same method to the honeycomb lattice. Two
electro-optic modulators are placed in two of the laser beams, and the phase
modulated potential is \cite{gemelke2007}
\begin{eqnarray}
V(\vec{r}) &=&V_{0}[\cos (\vec{k}_{1}\cdot \vec{r}+\phi ^{+})+\cos (\vec{k}
_{2}\cdot \vec{r}+\phi ^{-})  \notag \\
&+&\cos (\vec{k}_{3}\cdot \vec{r}-\phi ^{+}-\phi ^{-})],
\label{eq:opt}
\end{eqnarray}
where $\vec{k}_{i}=\frac{1}{2}\epsilon _{ijk}(\vec{q}_{j}-\vec{q}_{k})$;
$\vec{q}_{i}(i=1,2,3)$ are the wavevectors of the three coplanar laser beams
satisfying $|q_1|=|q_2|=|q_3|=q$.
In the following of this paper, we use the
definition of recoil energy $E_r=\hbar^2 q^2/(2M)$ where $M$ is the atom mass.
Please note that this definition of $E_r$ is 3 times smaller than that used
in Ref. \cite{wu2008b} which is defined as $\hbar^2 k^2/(2M)$.
In Eq. \ref{eq:opt},
$\phi^{\pm }=\eta \sin (\Omega t \pm {\frac{2\pi }{3}})\sin (\omega _{\text{RF}
}t)$ where $\Omega $ is the slow precession frequency; 
$\eta $ is a phase modulating constant
which determines the amplitude of the oscillation;
$\omega _{\text{RF}}$ is the fast rotation frequency at radio frequency.
Atoms do not follow the fast oscillation and only feel a time average 
of the potential as
\bea
V(\vec{r},t)=V_{0}\sum_{i=1}^{3}\big[A_{i}(t)\cos (\vec{k}_{i}\cdot \vec{r})
\big],
\label{eq:time_dep_lat}
\eea
where $A_{i}(t)=J_{0}[\eta \sin (\Omega t +\frac{2\pi}{3})]$;
$J_{0}$ is the zeroth order Bessel function; $\eta $ is a small parameter
\cite{gemelke2007}.

Eq. \ref{eq:time_dep_lat} still maintains the same lattice translational
symmetry.
Around each potential minimum $\vec r_0$ in the original lattice without
rotation, the potential can be expanded to the second order, yielding
a slightly anisotropic harmonic potential:
\begin{eqnarray}
V(\vec{r}-\vec r_0,t)&\approx& {V_{0}\over 4}\Big\{ \frac{8\pi
^{2}}{3}\left( 1-\frac{\eta ^{2}}{8}
\right) | \vec{r}-\vec{r}_{0}| ^{2}
+\frac{\eta ^{2}\pi ^{2}}{6}\nn \\
&\times& \left\vert \vec{r}-\vec{r}
_{0}\right\vert^2 \cos \big[ \frac{\pi}{3}
+2(\Omega t+\varphi_{\vec r-\vec r_0}) \big] \Big\}, \ \ \
\end{eqnarray}
which rotates with a slow frequency of $\Omega $.
Here $\varphi_{\vec r-\vec r_0} $ is the
polar angle of $\vec{r}-\vec{r}_{0}$.
The slight deformation of the optical potential processes
around each site center, which can be regarded
as an on-site rotation.

\subsection{The tight-binding Hamiltonian}
Now we construct the effective tight-binding model to describe the above
system.
First of all, each lattice site is rotating around its own center,
and there is no overall rotation of the entire lattice.
In other words, the system still has the lattice translational symmetry.
There should be no vector potential for inter-site hopping associated
with the Coriolis force.
Within each site, the rotation angular velocity couples to the onsite
orbital angular momentum through the orbital Zeeman coupling.
Such a coupling also exists in solid state systems in the presence of
external magnetic fields.
However, the typical energy scales of the Zeeman couplings, including both
spin and orbital channels, are at most at meV which are tiny compared to 
band widths.
They usually do not change the band topology.
The advantage of the experiments by Gemelke \textit{et al.}
\cite{gemelke2007,gemelke2009,gemelke2010} is that the orbital Zeeman energy
scale can easily reach the order of kHz, which is comparable to band widths.

The orbital Zeeman term from the onsite rotation becomes important in
orbital bands with angular momenta higher than $s$.
For the $p_{x,y}$-orbital bands, one of us \cite{wu2008} introduced
the following coupling as
\begin{eqnarray}
H_{L}=-\Omega \sum_{\vec r} L_z(\vec r) = i\hbar\Omega \sum_{\vec{r}} 
\big\{p^{\dag}_{x,\vec{r}} p_{y,\vec{r}} -p^{\dag}_{y,\vec{r}} p_{x,\vec{r}} \big\}.
\label{eq:rotation}
\end{eqnarray}
It breaks the degeneracy between $p_x \pm i p_y$ states, and 
induces topologically non-trivial band structures 
as presented in later sections.

The remaining part of the tight-binding Hamiltonian is as usual.
In Ref \cite{wu2007,wu2008b}, one of us studied the $p_{x,y}$-orbital
bands in the honeycomb optical lattice filled with spinless fermions,
which is the counterpart of graphene described by the $p_z$-orbital but
exhibits fundamentally different properties.
The tight-binding Hamiltonian reads
\begin{eqnarray}
H_0&=&t_{\parallel} \sum_{ \vec{r} \in A} \sum^3_{i=1} \big\{ p^{\dag}_{i,
\vec{r}} p_{i,\vec{r}+\hat{e}_i} +h.c. \big\} -\mu\sum_{\vec{r}\in A \oplus B} 
n (\vec{r}), ~\label{eq:singlep}
\end{eqnarray}
where $\hat{e}_{1,2}=\pm \frac{\sqrt{3}}{2} \hat{x}+\frac{1}{2} \hat{y}$ and
$\hat{e}_{3}=-\hat{y}$; $A$ and $B$ are indices of two different
sublattices; $t_\parallel$ is the $\sigma$-bonding describing the
longitudinal banding of $p$-orbitals along the bond direction; $\mu$ is the
chemical potential; $n(\vec{r})=p^{\dag}_{x,\vec{r}}p_{x,\vec{r}
}+p^{\dag}_{y,\vec{r}}p_{y,\vec{r}}$ is the filling number at site $\vec{r}$.
The operators $\hat{p}_{i,\vec{r}}$ are defined as the projection
of $p$-orbital along the vector  $\hat{e}_i$ as
$\vec{p}_{\vec{r}}= p_{x,\vec{r}} \hat{x}+ p_{y,\vec{r}} \hat{y}$.
Rigorously speaking, $t_\parallel$ should be time-dependent which
depends on the oscillation amplitude $\eta$.
Here we neglect this time-dependence by assuming $\eta$ is small.
$t_\parallel$ is positive as a result of the odd parity of the $p$
-orbitals. The $\pi$-bonding $t_\perp$ is much weaker than the $\sigma$-bonding.
For example, $t_\perp/t_\parallel$ can be easily suppressed around $1\%$
\cite{wu2008} within realistic experimental parameters of $V_0/E_r=15$,
thus the $t_\perp$ is not considered in most of this paper
unless in Sect. \ref{subsect:tpi}.
The band Hamiltonian to be investigated
below is the combination between Eq. \ref{eq:rotation} and Eq. \ref
{eq:singlep} as
\begin{eqnarray}
H=H_0+H_L.  \label{eq:Hcoor}
\end{eqnarray}

\section{The appearance of the Haldane model at large rotation angular
velocities}
\label{sect:haldane}
\begin{figure}[tbp]
\centering\epsfig{file=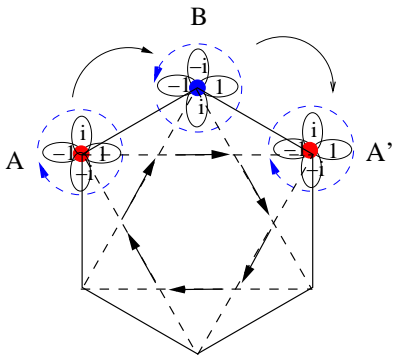,clip=1,width=0.6\linewidth,angle=0}
\caption{(Color online)
The pattern of the induced complex-valued NNN hopping
at $\Omega \gg t_\parallel$, which is generated by the virtual hopping
between orbitals with opposite chiralities. From Wu \protect\cite{wu2008}. }
\label{fig:haldane}
\end{figure}

Haldane proposed a tight-bonding model for the QAHE effect whose 
Bloch wave band structure is topologically non-trivial \cite{haldane1988}.
The Hamiltonian of the Haldane model is defined in the honeycomb lattice,
which reads
\begin{equation}
H=-t\sum_{\langle ij \rangle}\Big\{a^{\dag}_i a_j +h.c.\Big\}+
\sum_{\langle\langle ij\rangle\rangle} \Big\{t^\prime_{ij} a^{\dag}_i a_j+h.c.\Big\},
\label{eq:haldane}
\end{equation}
where $\langle ij \rangle$ represents the nearest-neighbor (NN)
hopping; $\langle\langle ij \rangle\rangle$ represents
the next-nearest-neighbor (NNN) hopping.
The NNN hopping
$t^\prime=|t^\prime| e^{\pm i\phi}$ is complex-valued whose argument
takes $\pm\phi$ if the hopping from $i$ to $j$ is anticlockwise 
(clockwise) with respect to the plaquette center.
Eq. \ref{eq:haldane} breaks time-reversal symmetry.
The band spectra exhibit two gapped Dirac cones in the Brillouin zone
(BZ), whose mass values have opposite signs.
The band structure has the non-vanishing Chern numbers $\pm 1$,
which leads to the QAHE with $\nu=\pm 1$.
In the system with the open boundary conduction, unidirectional
edge currents appear surrounding the system, i.e.,
the edge currents are chiral.

Before the detailed study on the band structures of our $p$-orbital
Hamiltonian Eq. \ref{eq:Hcoor}, 
we present an intuitive picture that the on-site rotation induces 
complex-valued NNN hopping terms in the limit of
$\Omega/t_\parallel \gg 1$ as in the Haldane model.
As a result, the Gemelke type rotation provides a possibility to
realize the QAHE state in the cold atom experiments.
In the presence of rotation, the on-site eigen-orbitals become $p_x\pm i p_y$
with an energy splitting of $2\Omega$. When $\Omega\gg t_\parallel$, each
level of $p_x\pm i p_y$ broadens into a band without overlapping each other.
We consider the case that $\Omega>0$, such that the low energy sector of the
Hilbert space consists of the $p_x + ip_y$ orbital state. The leading order
term of the effective Hamiltonian in this sector is just the
NN hopping with the hopping integral of $\frac{1}{2}t_\parallel$.
Moreover, the second order perturbation process generates the NNN hopping
with complex-valued integral as explained below.

Let us consider the two-step virtual hopping process illustrated in Fig. \ref
{fig:haldane}. In the first step, the atom starting from the low energy
sector of the $p_x + i p_y$ orbital in the A-site hops into the high energy
sector of the $p_x-i p_y$ orbital in the nearest neighbor B-site. 
The phases along the AB-bond is $30^\circ$ from the A-site and $150^\circ$ 
from the
B-site, thus there is a phase mismatch of $120^\circ$. The corresponding
hopping integral is complex-valued with
$\frac{1}{2}t_\parallel e^{i\frac{2}{3}\pi}$.
Similarly, during the second step the atom hops back into the $
p_x+ip_y$ orbital in the NNN A-site with the complex hopping integral
$\frac{1}{2}t_\parallel e^{i\frac{2}{3}\pi}$.
The hopping process is $(A+)\to (B-) \to (A^{\prime }+)$ where $\pm$
represents the chirality of $p_x\pm i p_y$ orbitals.
The corresponding amplitude is calculated as follows:
\begin{eqnarray}
t_{\text{NNN}} &=& \frac{\avg{A^\prime+|H_0|B-}\avg{B-|H_0|A+}}{-2\Omega}\nn \\
&=& -{\frac{t_{{\small
\parallel}}^2}{8\Omega}}e^{i4\pi/3}. \ \ \   \label{eq:nnnhopping}
\end{eqnarray}
All the NNN hoppings have the same phase value following the
arrows, which is exactly the same as in the Haldane model.
The above analysis applies to the high energy sector as well. Thus we have two
copies of the Haldane model, each for the $p_x\pm i p_y$ bands, respectively.


\section{Band structures in the homogeneous system}
\label{sect:band} 
In this section, we present the band spectra in the homogeneous system 
with the periodical boundary conditions (PBC). 
The general structure is studied in Sec. \ref{subsect:general}, and the
interesting flat band structure is presented in Sec. \ref{subsect:flat}.

\subsection{The general band structures}
\label{subsect:general}
\begin{figure}[tbp]
\centering
\epsfig{file=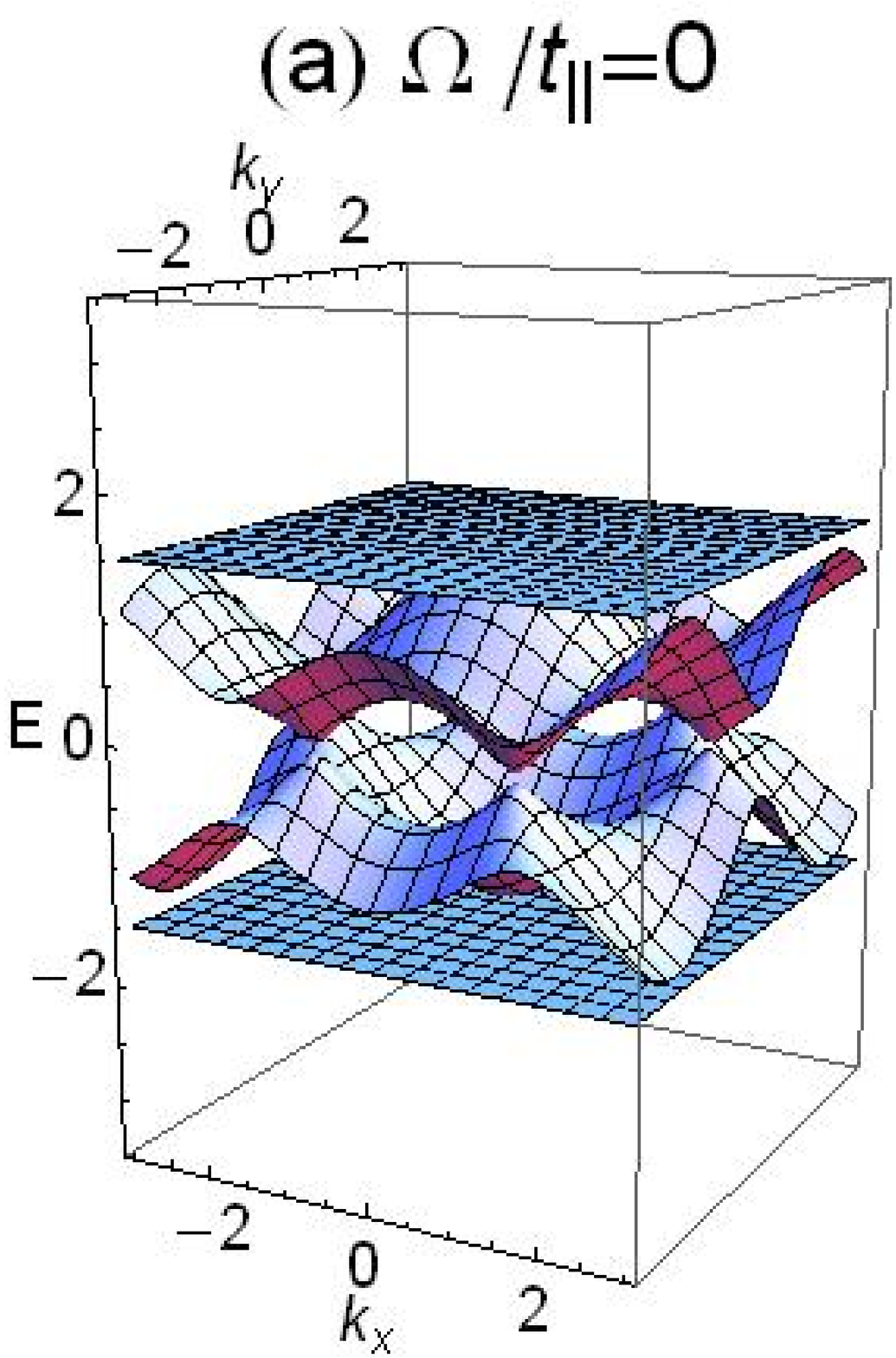,clip=1,width=0.4\linewidth, height=0.55\linewidth, angle=0}
\epsfig{file=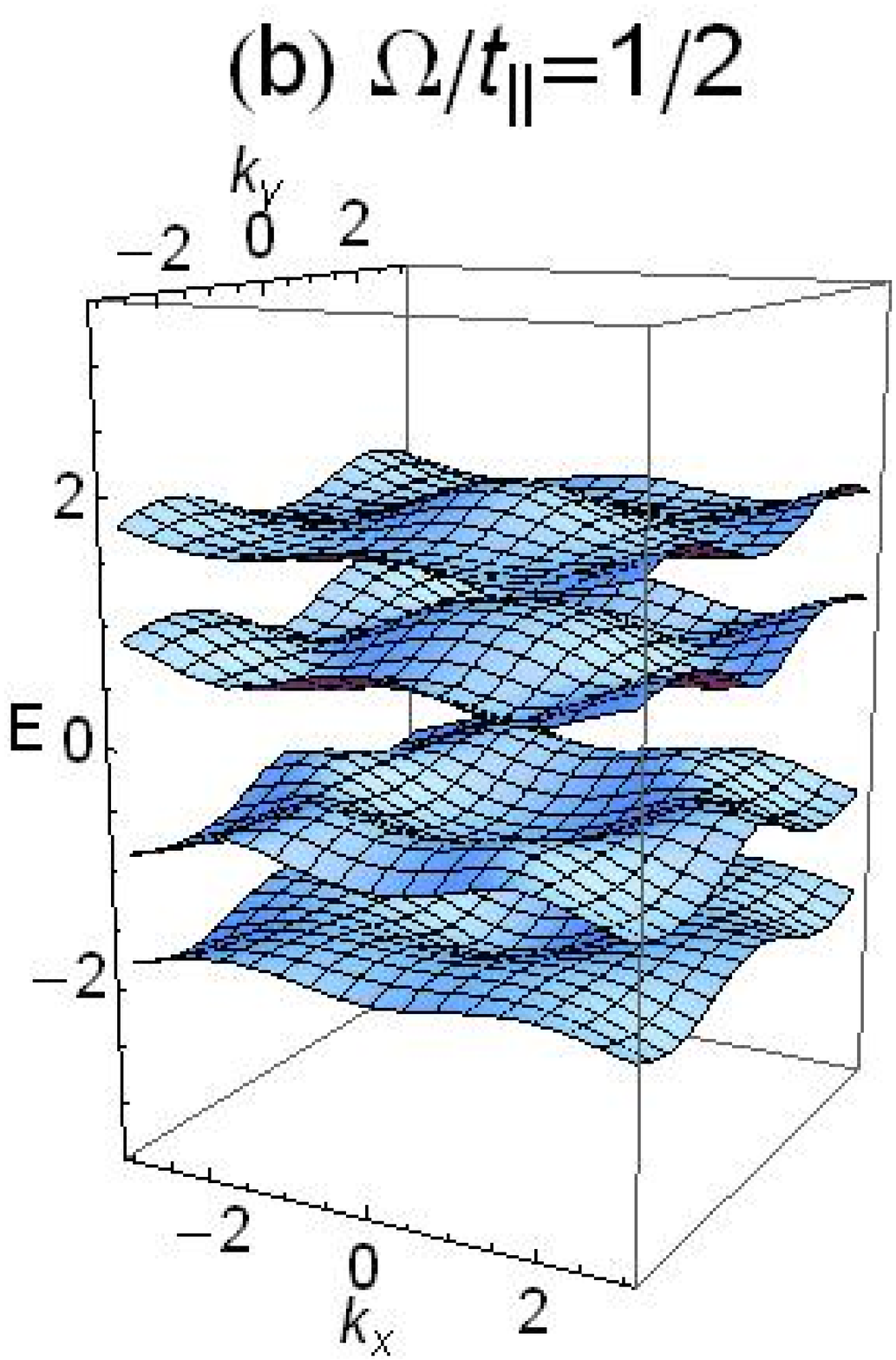,clip=1,
width=0.4\linewidth, height=0.55\linewidth, angle=0}
\epsfig{file=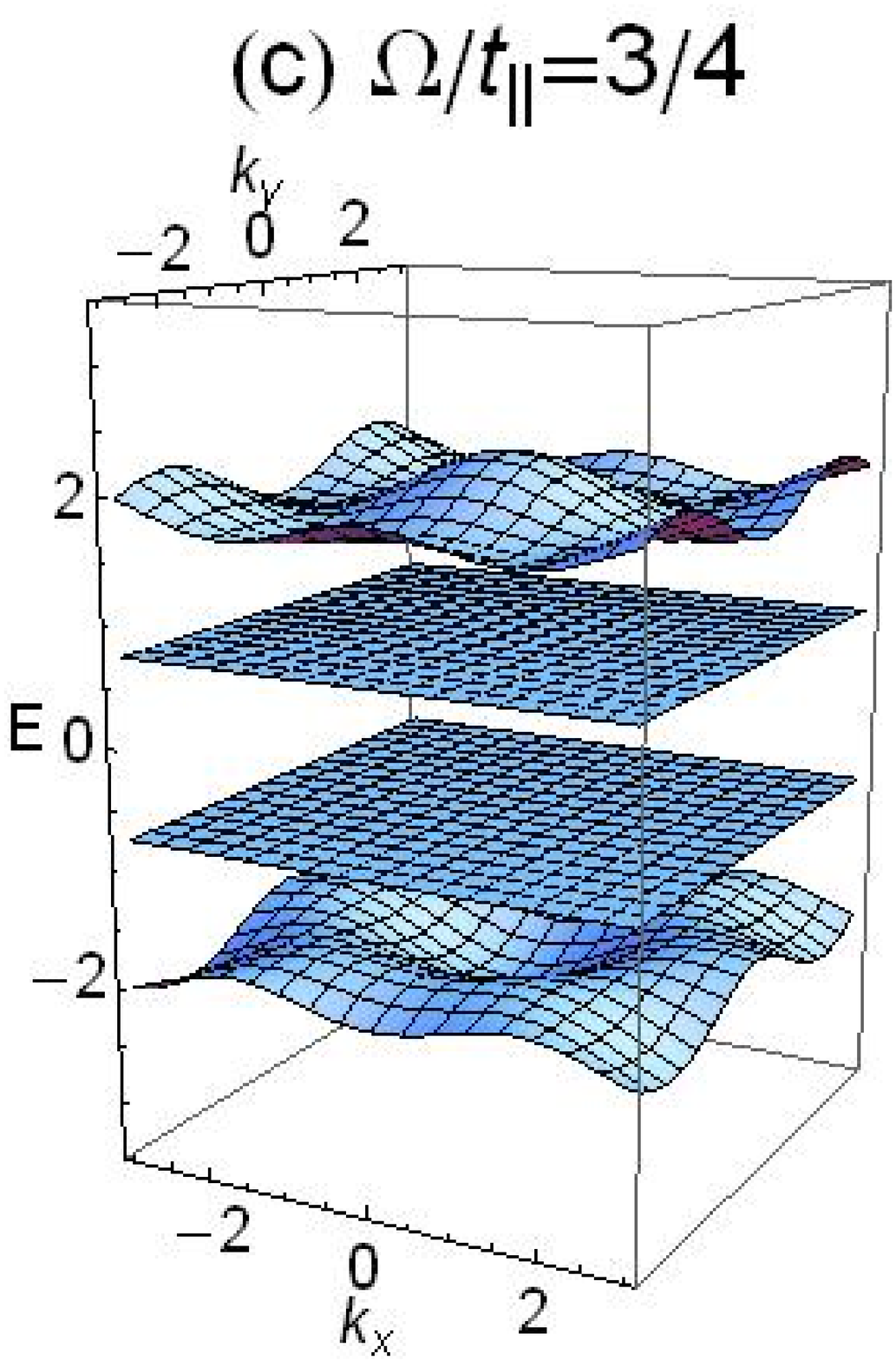,clip=1,
width=0.4\linewidth, height=0.55\linewidth, angle=0}
\epsfig{file=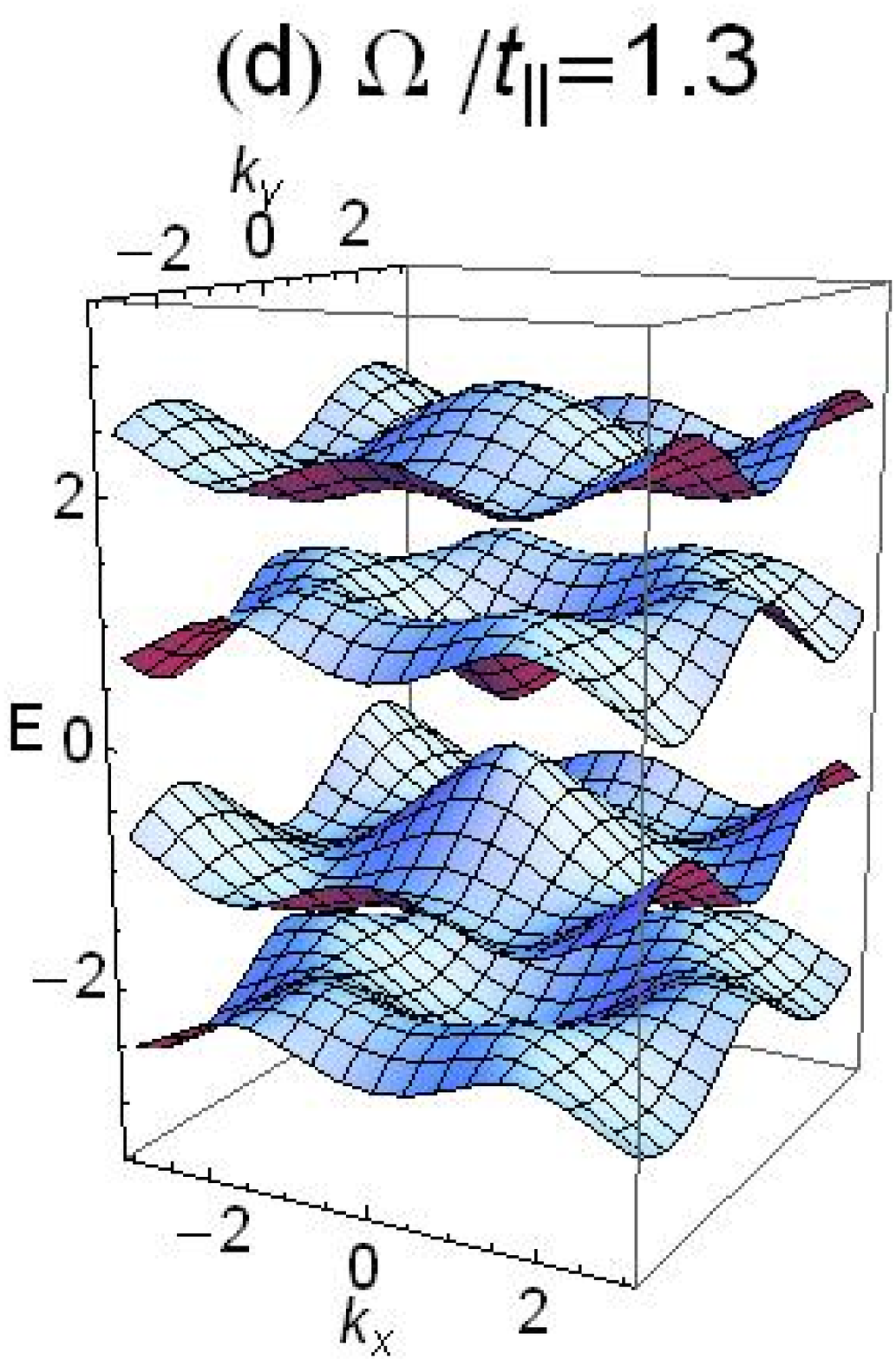,clip=1,
width=0.4\linewidth, height=0.55\linewidth, angle=0} 
\epsfig{file=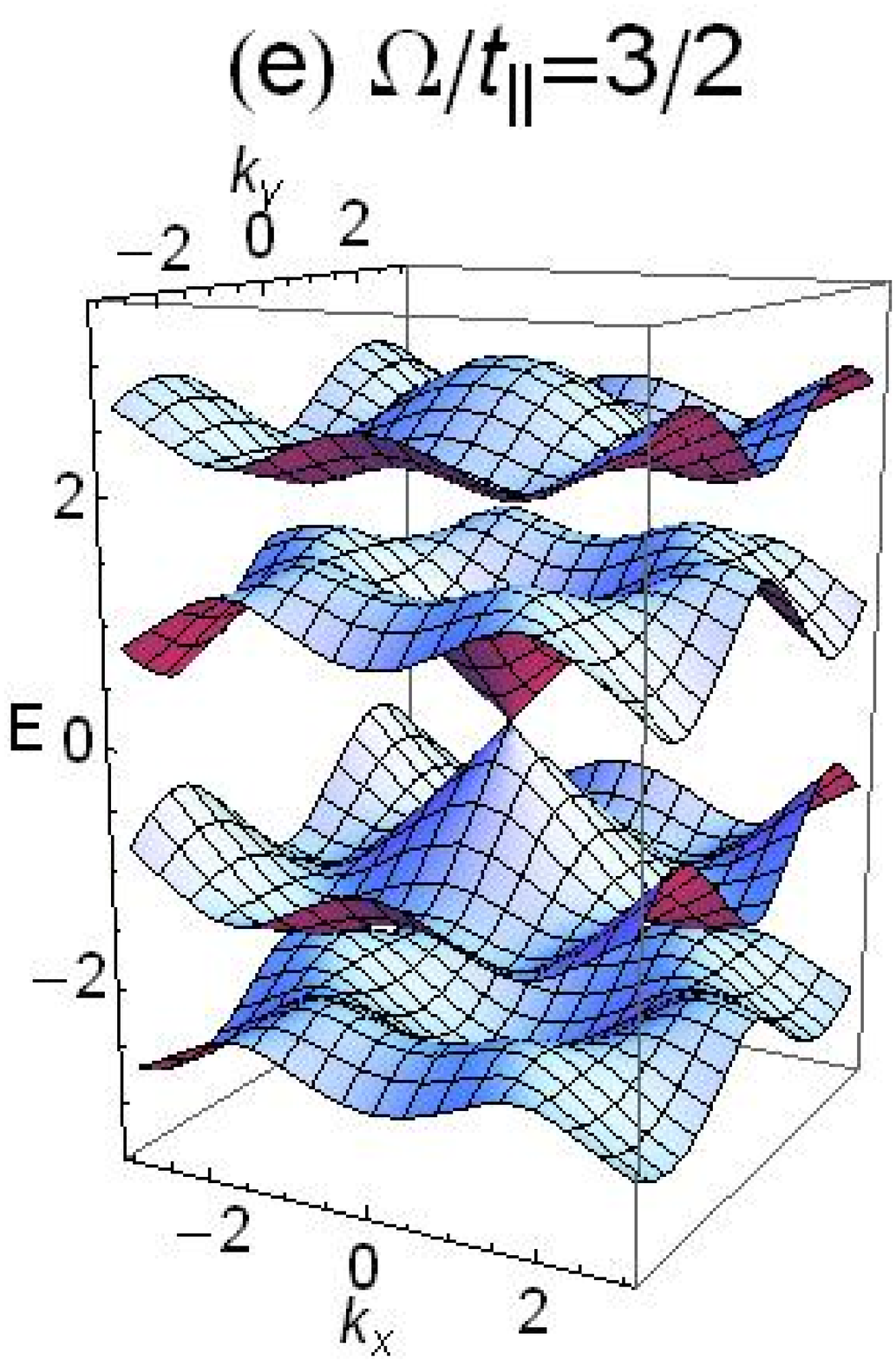,clip=1,
width=0.4\linewidth, height=0.55\linewidth, angle=0}
\epsfig{file=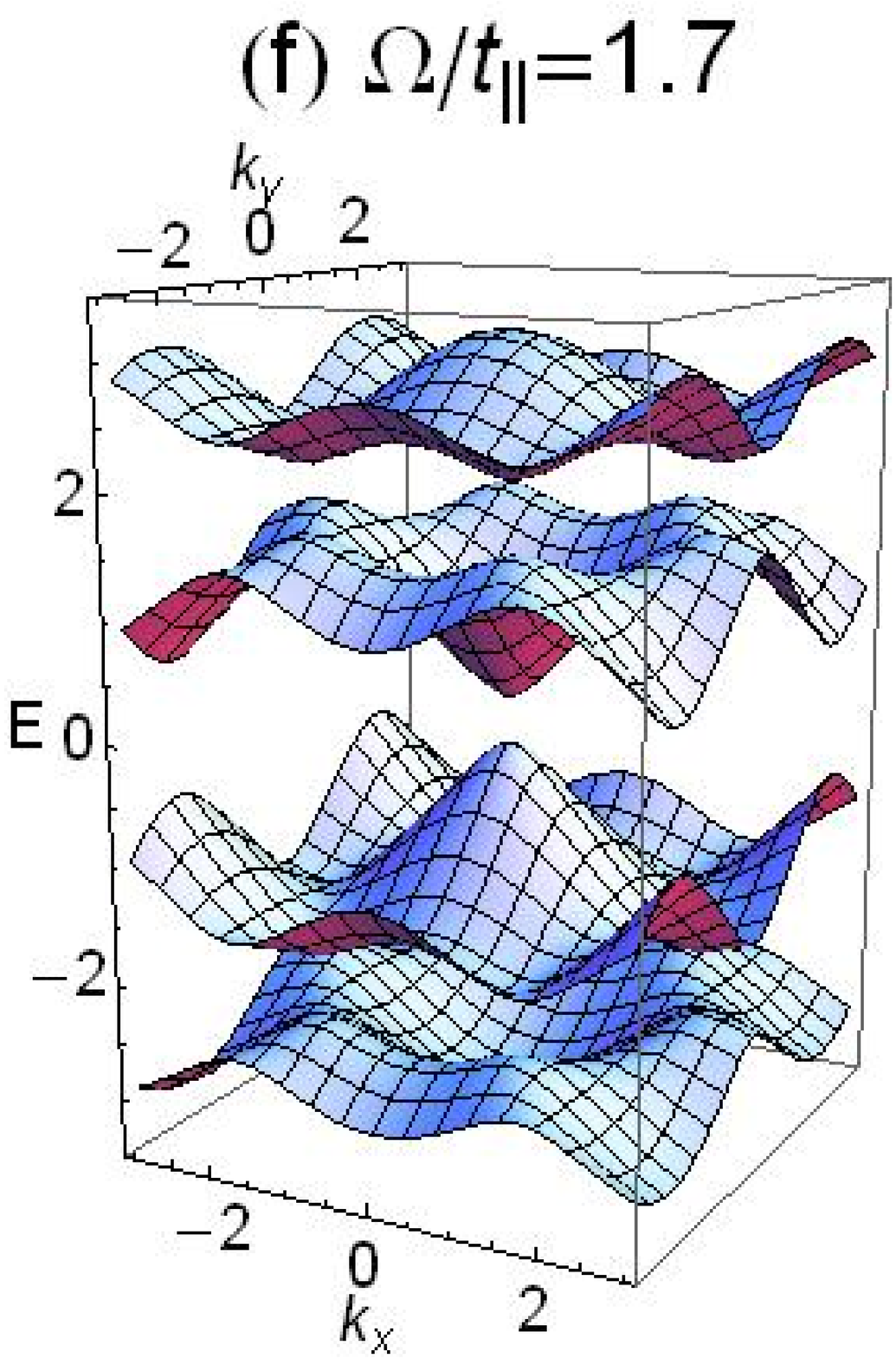,clip=1,
width=0.4\linewidth, height=0.55\linewidth, angle=0}
\caption{(Color online)~~The band structures of the Hamiltonian Eq.
[\protect\ref{eq:Hmatrix}].
With increasing rotation velocity from (a) $%
\Omega/t_\parallel =0$ to (f) $\Omega/t_\parallel =1.7 $.
The flat bands appear at $\Omega/t_\parallel=\frac{3}{4}$.
Energies are in unit of hopping strength $t_\parallel$ and
wavevectors are in the unit of  $1/a$, which
is the reciprocal of the lattice constant.
The energies of these two flat bands are $E/t_\parallel =\pm \frac{3}{4}$. }
\label{fig:spectra}
\end{figure}

We define the four-component spinor representation for the two-orbital
wavefunctions in the two sublattices as
\begin{eqnarray}
|\psi_{\vec k}\rangle =\left[p_{Ax}(\vec k),p_{Ay}(\vec k), p_{Bx}(\vec
k),p_{By}(\vec k)\right]^T.
\end{eqnarray}
After performing the Fourier transform, the Hamiltonian Eq. \ref{eq:Hcoor}
becomes
{\small
\begin{eqnarray}
H=\sum_{\vec{k}} \psi^{\dag}_{a}(\vec{k}) [H_{ab}(\vec{k})-\delta_{ab} \mu]
\psi_{b}(\vec{k}),
\end{eqnarray}
where $H_{ab}(\vec{k})$ is written as 
\begin{eqnarray}  \label{eq:Hmatrix}
\left(
\begin{array}{cccc}
0 & i \Omega & \frac{3}{4} (e^{i \vec k \cdot \vec e_1} +e^{i \vec k \cdot
\vec e_2}) & \frac{\sqrt 3}{4} (e^{i \vec k \cdot \vec e_1} -e^{i \vec k
\cdot \vec e_2}) \\
-i \Omega & 0 & \frac{\sqrt 3}{4} (e^{i \vec k \cdot \vec e_1} -e^{i \vec k
\cdot \vec e_2}) & \frac{1}{4} (e^{i \vec k \cdot \vec e_1} +e^{i \vec k
\cdot \vec e_2}) +e^{i \vec k \cdot \vec e_3} \\
h.c. &  & 0 & i \Omega \\
&  & -i \Omega & 0,
\end{array}
\right). 
\end{eqnarray}
}

The band structure in the absence of rotation \textit{i.e.,} $\Omega=0$, has
been studied in Ref. \cite{wu2007,wu2008b}, which includes both flat bands
(the bottom and top bands) and two dispersive bands with Dirac cones as
depicted in Fig. \ref{fig:spectra} (a). The flat bands and dispersive bands
touch at the center of the first BZ, and two dispersive
bands touch at Dirac cones.
The location of the Dirac cones are at $\vec{k}=(\pm\frac{4\pi}
{3 \sqrt{3}a},0)$ \cite{wu2007,wu2008b}.
The band flatness means that the corresponding band eigenstates can be
constructed as localized states in real space.
Each hexagonal plaquette supports one localized eigenstate whose
orbital configuration on each site is along the tangent direction as
presented in Fig. 2a in Ref. \cite{wu2007}.
When the filling is inside the flat bands, interaction effects are
non-perturbative.
This results in the exact solutions of the Wigner crystallization for
spinless fermions \cite{wu2007} and the flat band ferromagnetism for
spinful fermions \cite{zhang_hung_wu2010}.
For the dispersive bands, although their spectra are the same as in
graphene, their eigen-wavefunctions are fundamentally different
exhibiting rich orbital structures as presented in Ref. \cite{wu2008b}.

When the on-site rotation is turned on, \textit{i.e.}, $\Omega>0$, band gaps
open.
The previous touching points between the first and second bands at $%
\Omega=0$ split. The lowest band is no longer flat, and the center of the
second band is pushed up as depicted in Fig. \ref{fig:spectra} (b). The
Dirac cones between the middle two dispersive bands also become gapped. In
this case, the topology of each band $n~(n=1\sim 4)$ is characterized by the
Chern number defined as
\begin{eqnarray}
C_n= \frac{1}{2\pi}\int d^2 k ~ F_{n,xy} (\vec k),
\end{eqnarray}
where the Berry curvature $F_{n,xy}$ is defined as
\begin{eqnarray}
F_{n,xy}(\vec k)=\partial_{k_x} A_{n,y} (\vec k) - \partial_{k_y} A_{n,x}
(\vec k),
\end{eqnarray}
and $A_{n,\mu}(\mu=x,y)$ is the Berry gauge potential
in momentum space defined as
\begin{eqnarray}
A_{n,\mu}=i\langle \psi_n(\vec k)| \partial_{k_\mu}| \psi_n (\vec k)\rangle.
\end{eqnarray}

The Chern number patterns at $\Omega>0$ have been calculated in Ref. \cite%
{wu2008}, and the distribution of the Berry curvatures $F_{n,xy}$ in the BZ
is depicted in Fig. 2 of Ref. \cite{wu2008}. Below a critical value of the
rotation angular velocity $\Omega_c/t_\parallel=\frac{3}{2}$, the Chern
number pattern reads
\bea
C_1=-C_4=1; ~~~ C_2=-C_3=0.
\eea
At $\Omega_c/t_\parallel=
\frac{3}{2}$, a single Dirac cone connecting the second and third bands
shows up at $\vec k=(0,0)$, which triggers a topological phase transition.
Beyond $\Omega_c$, this Dirac point becomes gapped, and the Chern number
pattern becomes
\bea
C_1=-C_2=1; ~~~ C_3=-C_4=-1.
\eea
In this case, the band
structure is qualitatively the same as the two copies of the Haldane model
as discussed in Sec. \ref{sect:haldane}.

\subsection{Flat bands at $\Omega/t_\parallel=\frac{3}{4}$}
\label{subsect:flat}
It is evident in Fig. \ref{fig:spectra} (c) that the second and third bands
become flat at $\Omega/t_\parallel=\frac{3}{4}$. In this part, we discuss
various properties of the flat bands, including the localized eigen-states,
the distribution of the Berry curvature, and the interaction effects.

\subsubsection{Localized eigen-states}
\begin{figure}[htb]
\centering\epsfig{file=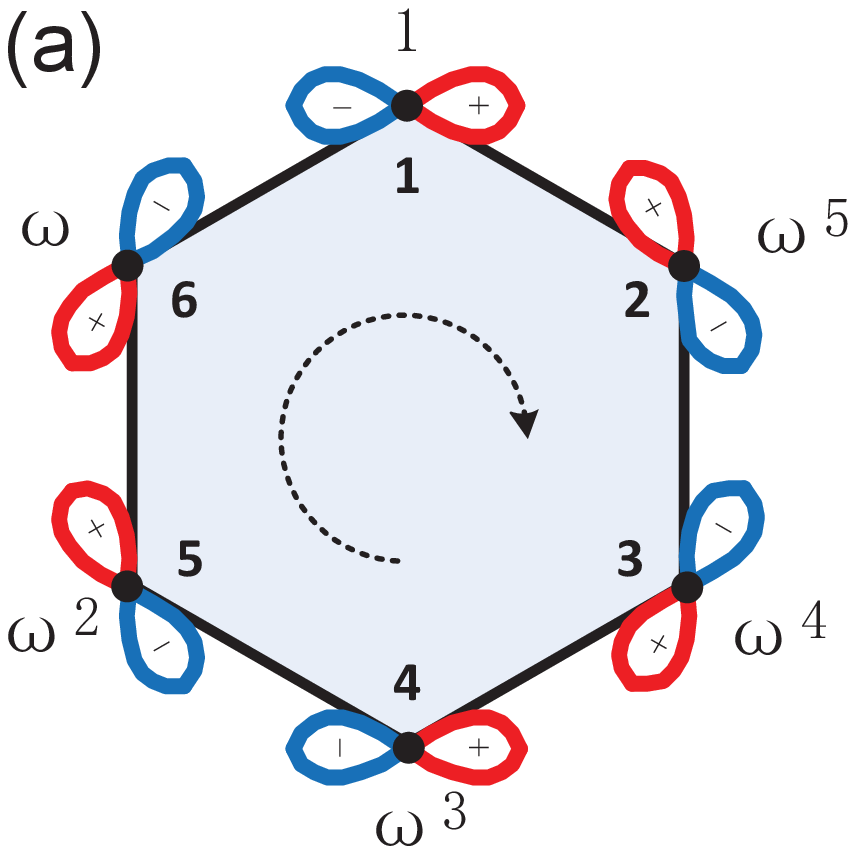,clip=1,width=0.205\textwidth,angle=0}
\centering\epsfig{file=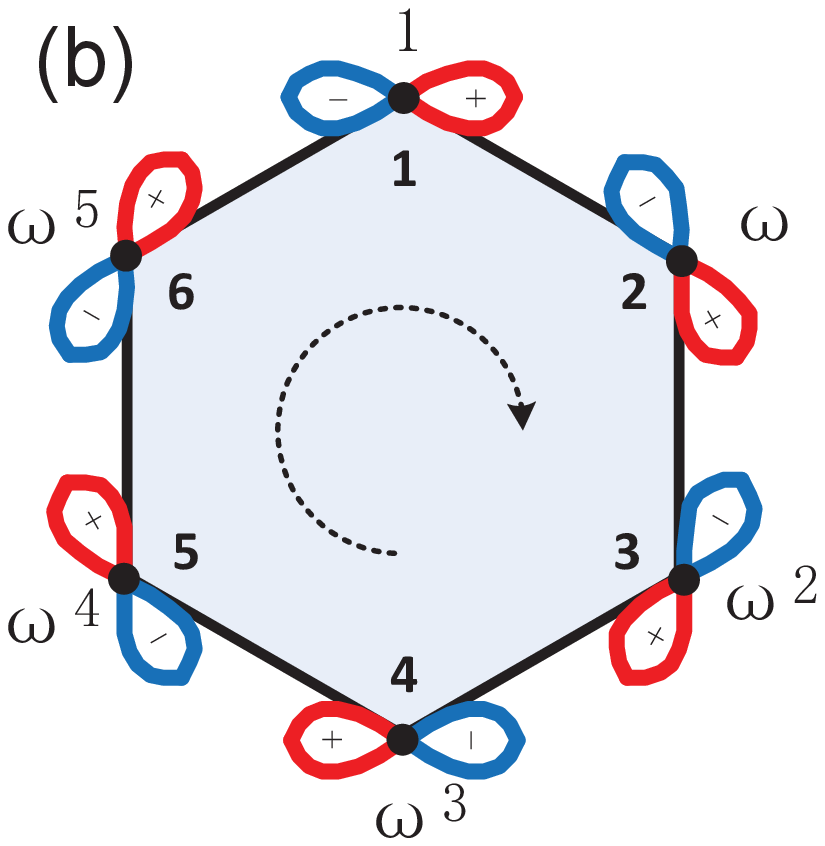,clip=1,width=0.2\textwidth,angle=0}
\caption{(Color online)~~Configurations of the localized eigenstates
of the two flat bands at $\Omega/t_\parallel=\frac{3}{4}$ for (a) $E/t_\parallel=-%
\frac{3}{4}$ and (b) $E/t_\parallel=\frac{3}{4}$, respectively. $\protect%
\omega=e^{i\frac{\protect\pi}{3}}$ is the relative phase factor between
neighboring sites. The plaquette currents
directions are clockwise for both (a) and (b) in the opposite direction of
the $\Omega$. }
\label{fig:localizedstate}
\end{figure}

The band flatness usually means that the eigen-states can be
reconstructed as localized states in real space. We assume that each
localized eigenstate exists within a single hexagon plaquette constructed as
follows
\begin{equation}
|\psi _{\vec{R}}\rangle =\sum_{j=1}^{6}(-)^{j-1}e^{i(j-1)\phi }\Big\{\cos
\theta _{j}|p_{j,x}\rangle -\sin \theta _{j}|p_{j,y}\rangle \Big\},\ \ \
\label{eq:wannier}
\end{equation}
where $\vec{R}$ is the coordinate of the plaquette center; $j$ is the site
index within the same plaquette and $\theta _{j}=(j-1)\frac{\pi }{3}$; 
$e^{i\phi }$ is the phase factor to be determined satisfying the
periodical boundary condition $e^{i6\phi }=1$; the factor of $(-)^{j-1}$ is a
sign convention because of the odd parity of the $p$-orbitals.
The $p$-orbital configuration on each site $j$ is along the tangent direction.
Substituting Eq. \ref{eq:wannier} into the band Hamiltonian, we arrive at
the condition for Eq. \ref{eq:wannier} to be the eigenstate as
\bea
\Omega =-{\frac{\sqrt{3}}{2}}\sin {\phi }, \ \ \, E=-{\frac{3}{2}}\cos {\phi },
\eea
where $\phi =0,\pm \frac{\pi }{3},\pm \frac{2\pi }{3},\pi $. 
For the cases of $\phi =0$ and $\pi $, they are the situations studied 
before in Ref. \cite{wu2007} without the on-site rotation.
The other four cases are with the on-site rotation. 
Without loss of any generality, we take $\phi =-\frac{\pi }{3}$
and $\phi =-\frac{2\pi }{3}$ such that $\Omega /t_{\parallel }=\frac{3}{4}>0$
and $E/t_{\parallel }=\pm \frac{3}{4}$. 
The schematic diagram of these two typical localized state is shown 
in Fig. \ref{fig:localizedstate} (a) and (b), respectively.

The main difference between these two groups of localized states at $%
\Omega=0 $ and $\Omega/t_\parallel=\frac{3}{4}$ is that there exists a
current around each plaquette for the latter case. 
The current operator along each bond is defined as
\begin{eqnarray}
\hat J_{\vec r, \vec r^+\hat e_i}=i \frac{t_\parallel}{\hbar} 
\Big\{ \hat p^\dagger_{\vec r} \cdot \hat e_i) (\hat p_{\vec r+\hat e_i} \cdot
\hat e_i) -h.c. \Big\}.
\end{eqnarray}
For the localized plaquette eigen-states of both bands with 
$E=\pm \frac{3}{4}t_\pp$, the currents have the same value and 
chirality as
\begin{equation}  
\label{eq:currentresult}
J=-\sqrt{3} \Omega.
\end{equation}
Eq. \ref{eq:currentresult} indicates that the current direction is opposite
to the rotation and the magnitude is proportional to the angular velocity 
$\Omega$.

At $\Omega/t_\parallel=\frac{3}{4}$, we solve the eigenvectors for the two
flat bands in momentum space as
\begin{widetext}
\bea
|\psi_{\vec k,\mp}\rangle=\frac{1}{\sqrt{N_\mp(\vec k)}}
\left (\begin{array}{c}
\frac{1}{2} e^{-i k\cdot e_1\mp i\chi_1} + \frac{1}{2} e^{-i k\cdot e_2\mp
i\chi_2} - e^{-i k\cdot e_3\mp i\chi_3} \\
\frac{\sqrt 3}{2} e^{-i k\cdot e_1\mp i \chi_1} - \frac{\sqrt 3}{2} e^{-i
k\cdot e_2\mp i\chi_2} \\
\pm (\frac{1}{2} e^{i k\cdot e_1\mp i\chi_1} + \frac{1}{2} e^{i k\cdot
e_2\mp i\chi_2} - e^{i k\cdot e_3\mp i\chi_3})  \\
\mp (\frac{\sqrt 3}{2} e^{i k\cdot e_1\mp i\chi_1} - \frac{\sqrt 3}{2} e^{i
k\cdot e_2\mp i\chi_2})
\end{array}
\right),
\label{eq:flat-bloch}
\eea
\end{widetext}
where $\psi_\mp(\vec k)$ represent eigenvectors for the bands with $
E/t_\parallel=\mp \frac{3}{4}$, respectively;
$\chi_1={\frac{\pi }{6}}, \chi_2={\frac{5\pi }{6}}$, and $\chi_3={%
\frac{3\pi }{2}}$.
The normalization factors $
N_\mp(\vec k)$ read
\begin{eqnarray}
N_\mp(\vec k)=2\Big[3-\sum_i \cos (\vec{k}\cdot\vec{b}_i\mp \frac{2}{3}\pi)%
\Big].
\end{eqnarray}
These flat band Bloch wave states can be represented as
the linear superpositions of the localized eigenstates in Eq. \ref%
{eq:wannier} as :
\begin{eqnarray}
|\psi_{\vec k,\mp}\rangle= {\frac{1}{\sqrt{N_\mp(\vec k)}}} \sum_{\vec R}
e^{i \vec k \cdot \vec R} |\psi_R \rangle
\end{eqnarray}
where $|\psi_R\rangle$ is defined in Eq. \ref{eq:wannier}.

\subsubsection{Brief discussions on interaction effects}

The band flatness means that interaction effects are always important
compared to the vanishing kinetic energy scale.
In our previous studies \cite{wu2007,wu2008b,zhang_hung_wu2010}, we have 
examined the non-perturbative effects in the flat bands (the lowest and 
highest bands) in the same system at $\Omega=0$. 
In Ref. \cite{wu2007,wu2008b}, we have shown that the flat
bands result in the exact solution of Wigner crystal configuration for
spinless fermions in the lowest band. 
At $\langle n\rangle=\frac{1}{6}$, which corresponds to that $\frac{1}{3}$ 
of the flat band plaquette states are occupied, the occupied plaquettes form a 
triangular lattice structure without touching each other. 
As filling increases, exact solutions are no longer available.
Self-consistent mean-field theory calculation shows a serials of insulating
states with different orbital orderings at commensurate fillings \cite{wu2008b}.
Similarly, in Ref. \cite{zhang_hung_wu2010}, we found the exact flat-band
ferromagnetism for spinful fermions in the flat bands.

For the flat bands occurring at $\Omega /t_{\parallel }=\frac{3}{4}$, the
physics will be similar to the previous studies at $\Omega =0$. 
However, the flat bands here are in the middle. 
When $\mu $ lies in the flat bands, there are always background particles 
or holes filling in the dispersive bands. 
The solutions of the Wigner crystal and flat-band
ferromagnetism is only valid if the interaction energy scale is smaller than
the band gaps between the flat bands and the dispersive bands. For example,
when the filling is inside the second band, the effect from the background
filling cannot be neglected, if the interaction energy scale is stronger
than the band gap.

\subsubsection{Berry curvatures v.s. local eigen-states}
\begin{figure}[tbp]
\centering\epsfig{file=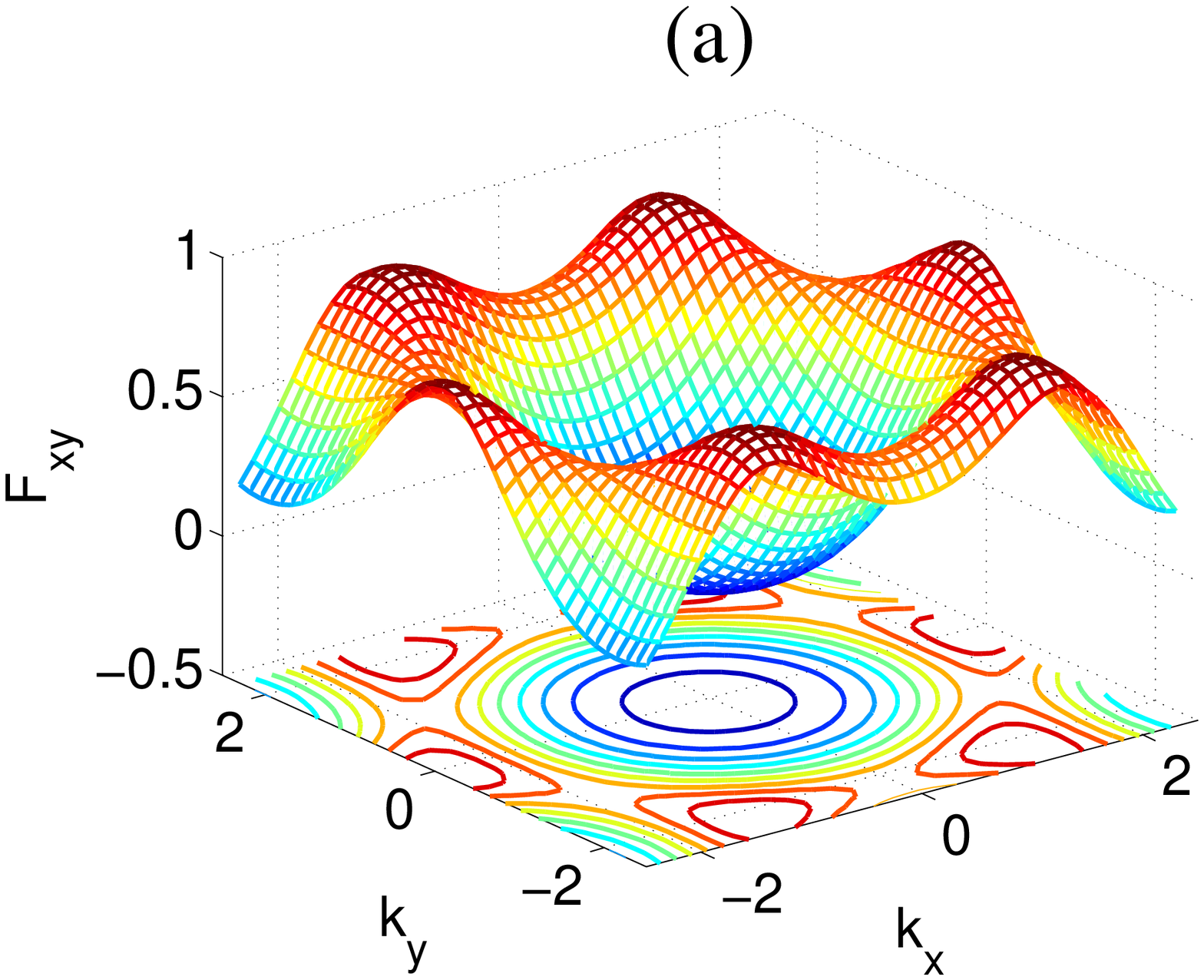,clip=1,width=0.49\linewidth, angle=0}
\centering\epsfig{file=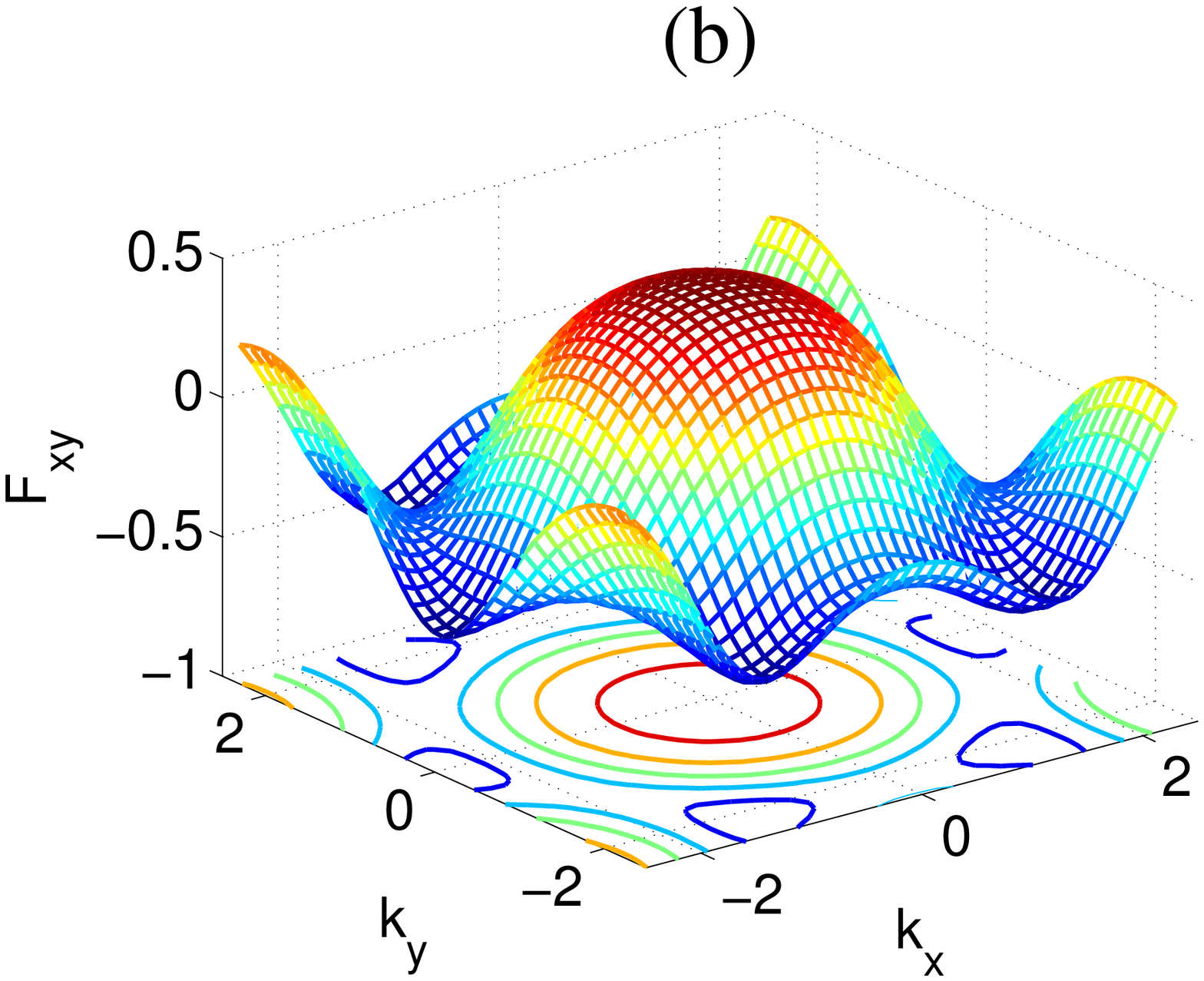,clip=1,width=0.49\linewidth,  angle=0}
\caption{(Color online)~~ The Berry curvature $F_{xy}$(in unit of $a^2$)
distribution at $\Omega/t_\parallel=\frac{3}{4}$ for the 1st (a) and 2nd (b)
bands. The Chern numbers for the 1st and 2nd bands are 1 and 0,
respectively. The unit of wave vector is $1/a$. }
\label{fig:chern}
\end{figure}

The current carried by the localized eigenstates of the flat bands 
depicted in Fig. \ref{fig:localizedstate} is chiral.
It looks very similar to the classic picture of cyclotron orbit of
electrons in the external magnetic fields.
We would expect that in a system with the open boundary condition, the
fully filled flat band would result in edge currents and contributes to the
quantized anomalous Hall conductance.
However, we need to be very careful with this, which turns out to be incorrect.
We have performed a preliminary diagonalization for a finite
size system with the open boundary condition.
The number of degeneracy for the flat bands equals
to the number of plaquettes plus 1.
We conjecture that this extra state should not belong to
a particular plaquette but rather distribute along the edge,
which carries a current in the opposite direction and 
cancels the contribution from other plaquette states.
Further examinations on this problem will be deferred to a later
publication.

We calculate the Berry curvature distributions at $\Omega/t_\parallel=\frac{3%
}{4}$ for the 1st and 2nd bands as presented in Fig. \ref{fig:chern} (a) and
(b), respectively.
Those of the 3rd (4th) band are just with an opposite sign compared to
the 2nd (1st) band due to the particle-hole symmetry of the band
Hamiltonian.
The 1st and 4th bands are topologically non-trivial with the
Chern number $\pm 1$. However, the 2nd and 3rd bands, which are flat, are
topologically trivial with the zero Chern number. In fact, these two bands
should not contribute to quantum anomalous Hall conductance when they are
fully filled. This is confirmed from the anomalous Hall current calculation
in the inhomogeneous trap as presented in Sec. \ref{sect:trap}.

\subsection{Effects of the $\pi$-bonding to band structures}
\label{subsect:tpi}

\begin{figure}[tbp]
\centering
{\epsfig{file=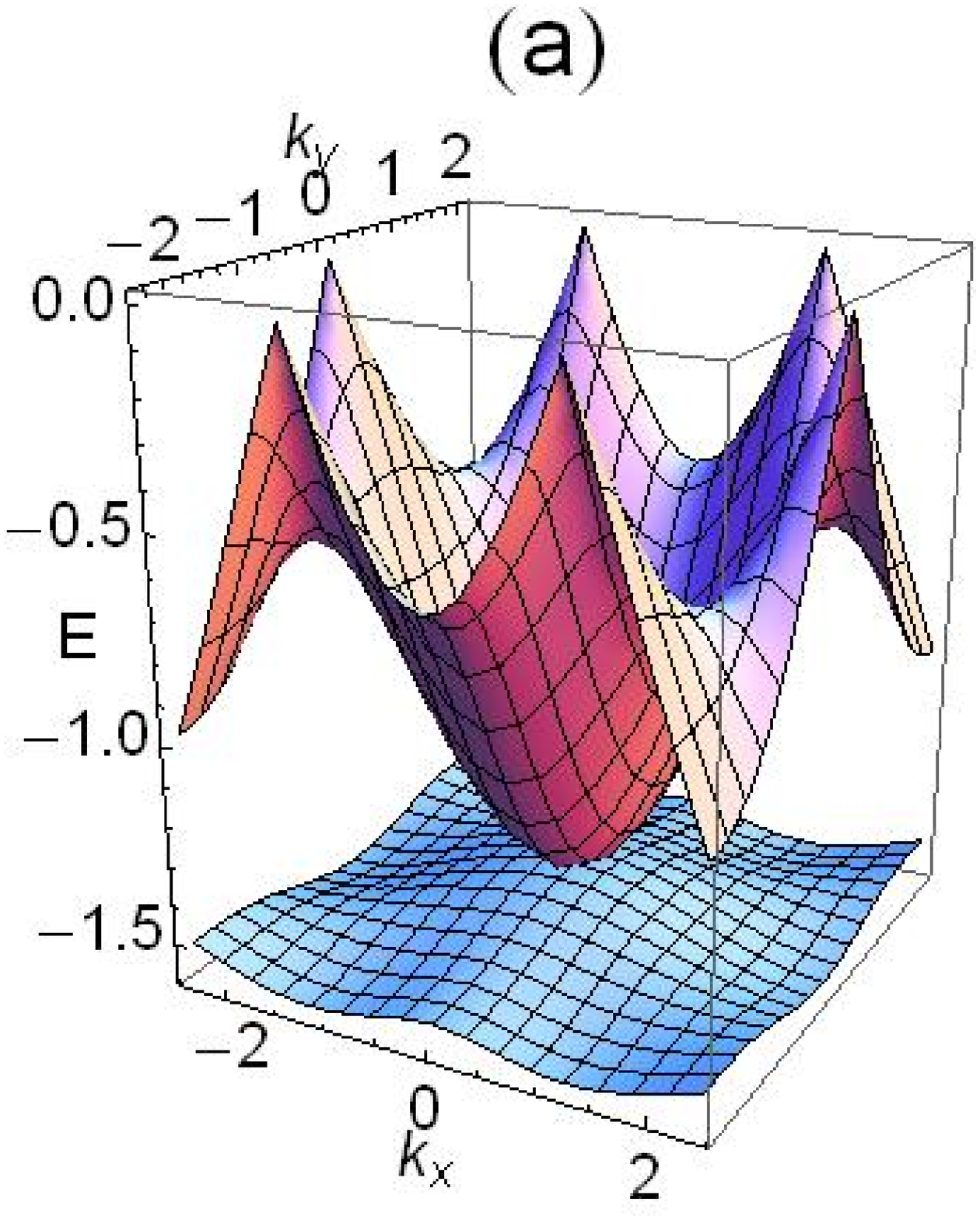,clip=1,
width=0.4\linewidth,  angle=0}}
\centering
{\epsfig{file=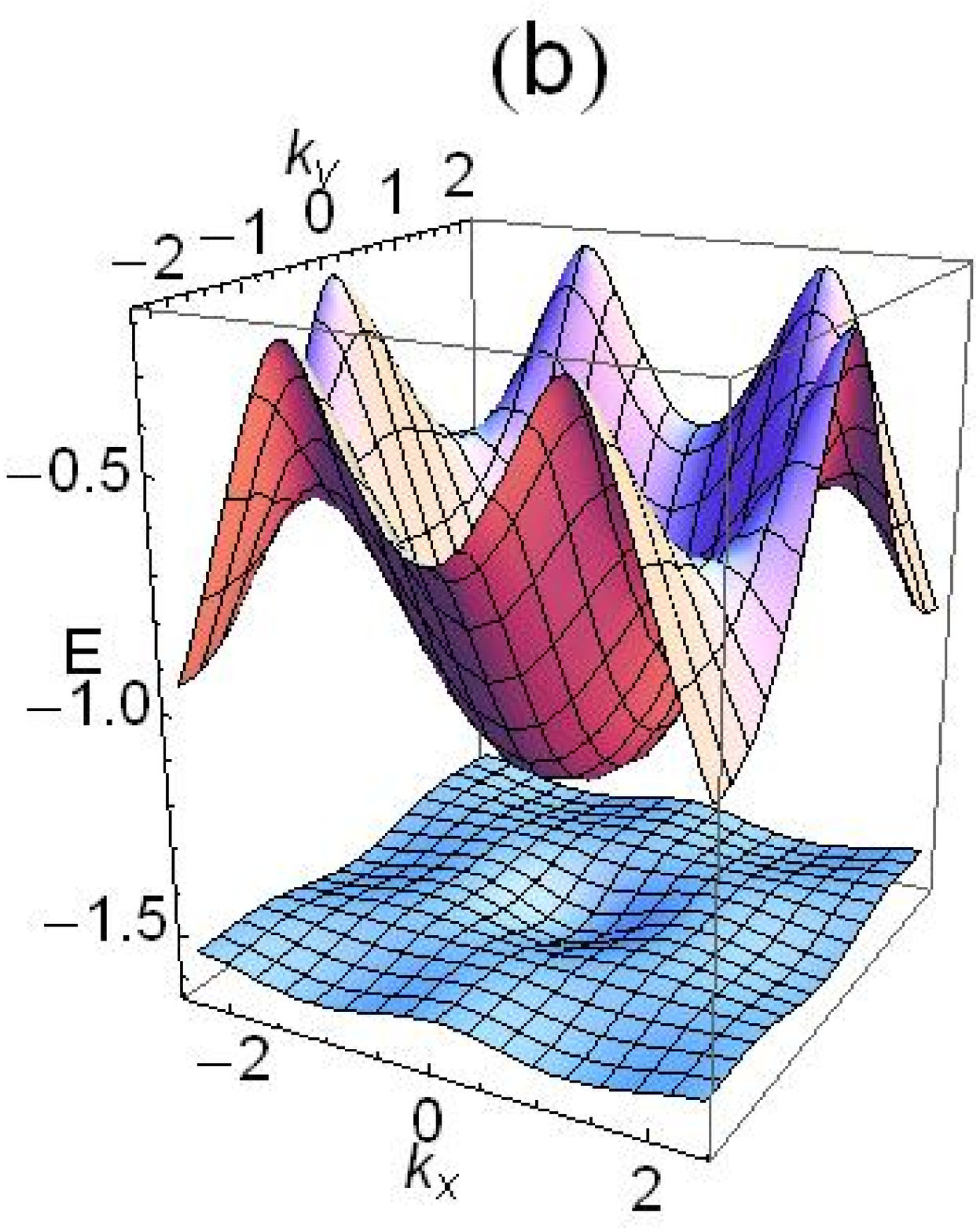,clip=1,
width=0.4\linewidth,  angle=0}}
\centering
\caption{(Color online)~~
The band structures with both $\pi$-bonding $(t_\perp/t_\pp=0.05)$
and on-site rotation $\Omega$.
Only the lower two bands are presented, and the spectra of
the other two bands are symmetric with respect to zero energy.
(a) At $\Omega/t_\pp=0$, the two bands remain touching at
the center of the BZ. The lowest band develops dispersion
at the order of $t_\pp$.
(b) At $\Omega/t_\pp=0.2$, the gap opens between the lower two bands.
The lowest band is topologically non-trivial and nearly flat.
}
\label{fig:spectra_tpi}
\end{figure}

So far, we have neglected $\pi$-bonding $t_\perp$ which can be
easily suppressed around $1\%$ of $t_\pp$ at intermediate optical
potential strength \cite{wu2008b}.
Here we explicitly present its effects to band structures for the
case of a realtively weak lattice potentials by taking $t_\perp/t_\pp=0.05$.
The projections of $p_{x,y}$-orbitals perpendicular to the
$\hat e_{1,2,3}$ directions are defined as
$p^\prime_{1,2}=-\frac{1}{2} p_x \pm \frac{\sqrt 3}{2} p_y,
p^\prime_3=p_x$, respectively.
The $\pi$-bonding Hamiltonian can be written as
\bea
H_\pi&=&-t_\perp\sum_{\vec r \in A, i=1\sim 3}
\Big \{ p^{\prime \dagger}_{\vec r,i} p^\prime_{\vec r + a \hat e_i,i}
+h.c. \Big \},
\label{eq:hampi}
\eea
where the hopping integral of the $\pi$-bonding has
the opposite sign to that of the $\sigma$-bonding.
In momentum space, Eq. \ref{eq:hampi} transforms into
\bea
H_\pi&=&-t_\perp \sum_k \psi^\dagger_\alpha(\vec k)
H_{\pi,\alpha\beta}(\vec k)
\psi_\beta(\vec k),
\eea
with the martix kernel $H_{\pi}(\vec k)$ as
{\small
\bea
\left(
\begin{array}{cccc}
0&0& \frac{1}{4} (e^{i \vec k \cdot \hat e_1} +e^{i \vec k \cdot \hat e_2})
+e^{i\vec k \cdot \hat e_3}
&\frac{3}{4} (-e^{i \vec k \cdot \vec e_1} +e^{i \vec k \cdot \hat e_2})
\\
0&0& \frac{ 3}{4} (-e^{i \vec k \cdot \vec e_1} +e^{i \vec k
\cdot \vec e_2})&
\frac{\sqrt 3 }{4} (e^{i \vec k \cdot \vec e_1} +e^{i \vec k \cdot \vec e_2})
\\
h.c.& & 0&0 \\
   & & 0&0
\end{array}
\right). \nn \\
\eea }

The effects of $\pi$-bonding $t_\perp$ are presented in Fig.
\ref{fig:spectra_tpi}.
The spectra remain symmetric with respect to zero energy, and thus
only the lower two bands are presented.
As presented in Ref. \cite{wu2008b}, at $\Omega=0$, the bottom
bands are no longer rigorously flat but develops a finite width
at the order of $t_\perp$.
The lower two bands remain touching at $\vec k=(0,0)$ with parabolic
spectra, and the bottom band has a negative curvature.
With increasing $\Omega$, as in the case of $t_\perp=0$, the band
gap at the order of $\Omega$ opens.
Furthermore, $\Omega$ lowers the energies of the bottom band
near the center of the BZ, which suppresses its dispersion.
As a result, we arrive at a nearly flat band with non-zero
Chern number.
The ratio between the width of the bottom band and the gap
between the lower two bands can reach the order of 5
as shown in Fig. \ref{fig:spectra_tpi} b.
Recently, we notice that the nearly flat bands with non-trivial
Chern number have been attracting attention, for its possible
realization of fraction quantum Hall states in the lattice
\cite{sun2010,neupert2010,tang2010}.

\section{Anomalous Hall currents in harmonic trap potentials}
\label{sect:trap}

\begin{figure}[tbp]
\centering\epsfig{file=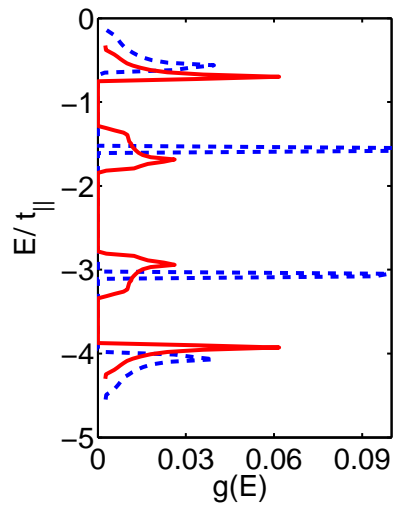,clip=1,width=0.6\linewidth,angle=0}
\caption{(Color online)~~The DOS of the $p_{x,y}$-orbital model Eq. \protect
\ref{eq:singlep} and Eq. \protect\ref{eq:rotation} in the homogeneous
system. The red solid lines are the DOS at $\Omega/t_\parallel =\frac{1}{2}$
whereas the blue dash lines are the DOS at $\Omega/t_\parallel =\frac{3}{4}$.
}
\label{fig:dos}
\end{figure}

In section \ref{sect:band}, the homogeneous $p$-orbital system with the PBC
has been studied, in which the wavevector $k$ is a good quantum number.
However, in reality the honeycomb lattice is inhomogeneous with a soft
harmonic confining trap. In this section, we shall consider the anomalous
Hall currents in such a realistic system.

The trapping potential adds a new term in the band Hamiltonian $H_0+H_L$ of
Eqs. \ref{eq:singlep} and \ref{eq:rotation} as
\begin{eqnarray}
H_T=\sum_{\vec r} V_T(r) n(\vec r).
\end{eqnarray}
The trapping potential $V_T(r)$ reads
\begin{eqnarray}
V_T(r) = \frac{1}{2} M \omega_T^2 r^2 =\frac{\beta t_\parallel}{2}\big(\frac{%
r}{a}\big)^2,
\end{eqnarray}
where $a$ is the lattice constant, $\beta=\frac{\hbar \omega_T}{t_\parallel}
(\frac{a}{l_0})^2$; $l_0=\sqrt{\frac{\hbar}{M \omega_T}}$ is the trapping
length scale. 
The typical value of the trapping frequency $\omega_T$ is in
the order of 10Hz, and that of the recoil energy $E_r$ is roughly several
kHz \cite{leggett2001}. 
In Ref. \cite{wu2008b}, we have calculated that $t_\parallel/E_r=0.24$ 
for $V_0/E_r=15$, thus $\hbar \omega/t_\pp$ is at the order of 0.1.
The typical trapping length scale is several lattice constants.
Taking into account all these factors, we choose a convenient value of $%
\beta=0.01$ for later calculations.

In the inhomogeneous system with trapping potential, the on-site rotation
induces the circulating currents along the azimuthal direction.
We will study the spatial distributions of the these anomalous Hall 
currents and particle density.
Because the band topology has a transition at
$\Omega_c/t_\parallel =\frac{3}{2}$, the results are presented at different
sets of parameters below, at, and above $\Omega_c$.
Our results are calculated by using the eigen-wavefunctions from the numerical
diagonalization of the free Hamiltonian in an open lattice with the trapping
potential.
We also use a modified local-density-approximation (LDA) to understand
the exact results.
The size of the lattice is with the radius of $r/a=40$.

\subsection{Low rotation angular velocity}
In this subsection the angular velocities are taken as $\Omega/t_\parallel=%
\frac{1}{2}$ and $\frac{3}{4}$ below $\Omega_c$.
The chemical potential is chosen as $\mu/t_\parallel=2.3$ to guarantee
that all bands are filled at the center of the trap.
The spatial distribution of the particle density exhibits a
four-layered wedding cake-like structure.
The density plateaus correspond to the band insulating regions, where
the local chemical potential, defined as $\mu_{loc} (r) =\mu-V_T(r)$,
lies inside band gaps.
Furthermore, the anomalous Hall currents flow along the tangent
direction, whose conductances are quantized in the insulating regions.

\subsubsection{The insulating plateaus of $\langle n(r)\rangle$}
\begin{figure}[tbp]
\centering
{\epsfig{file=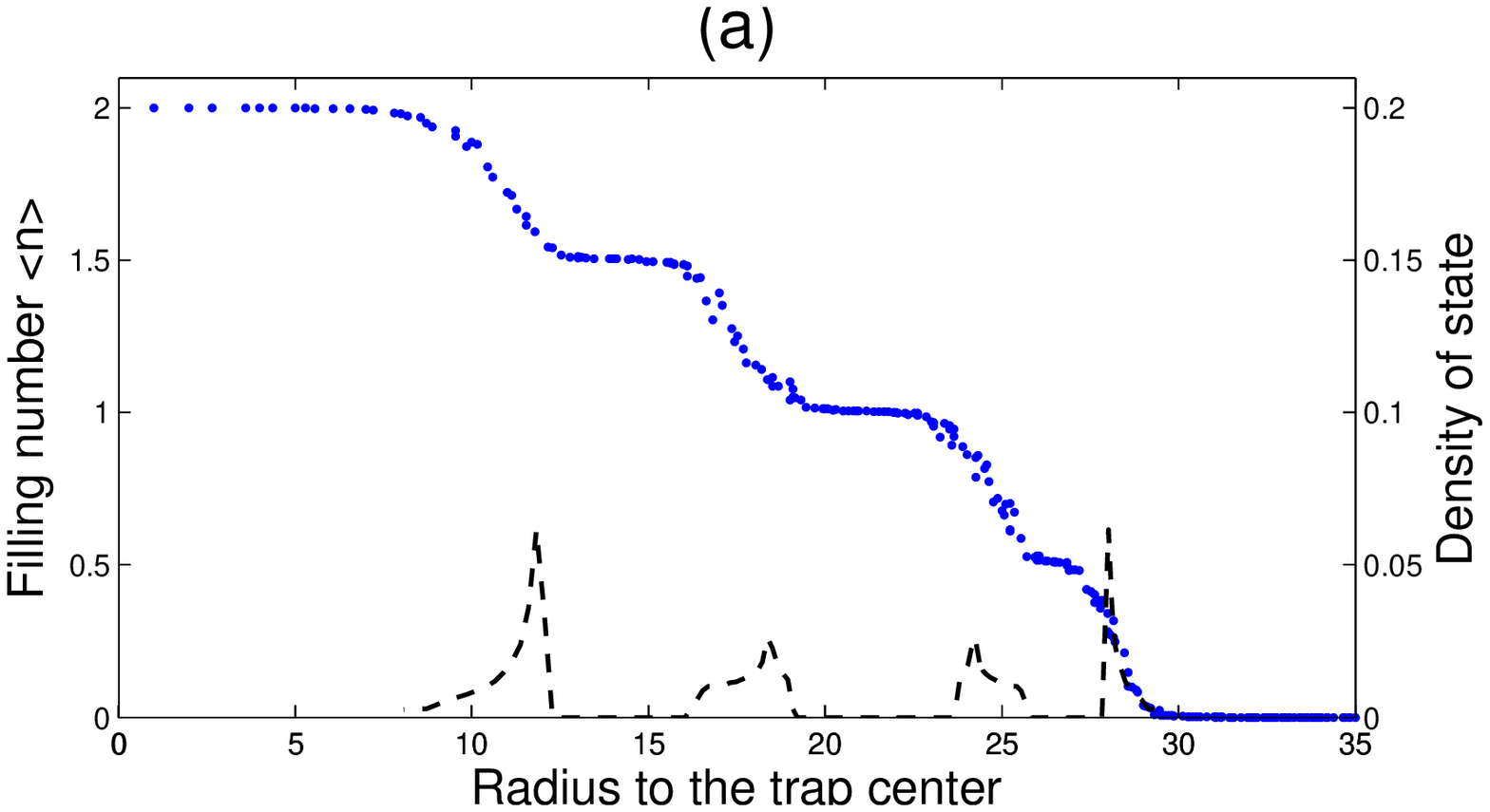,clip=1,width=0.8\linewidth,angle=0}}
\centering
{\epsfig{file=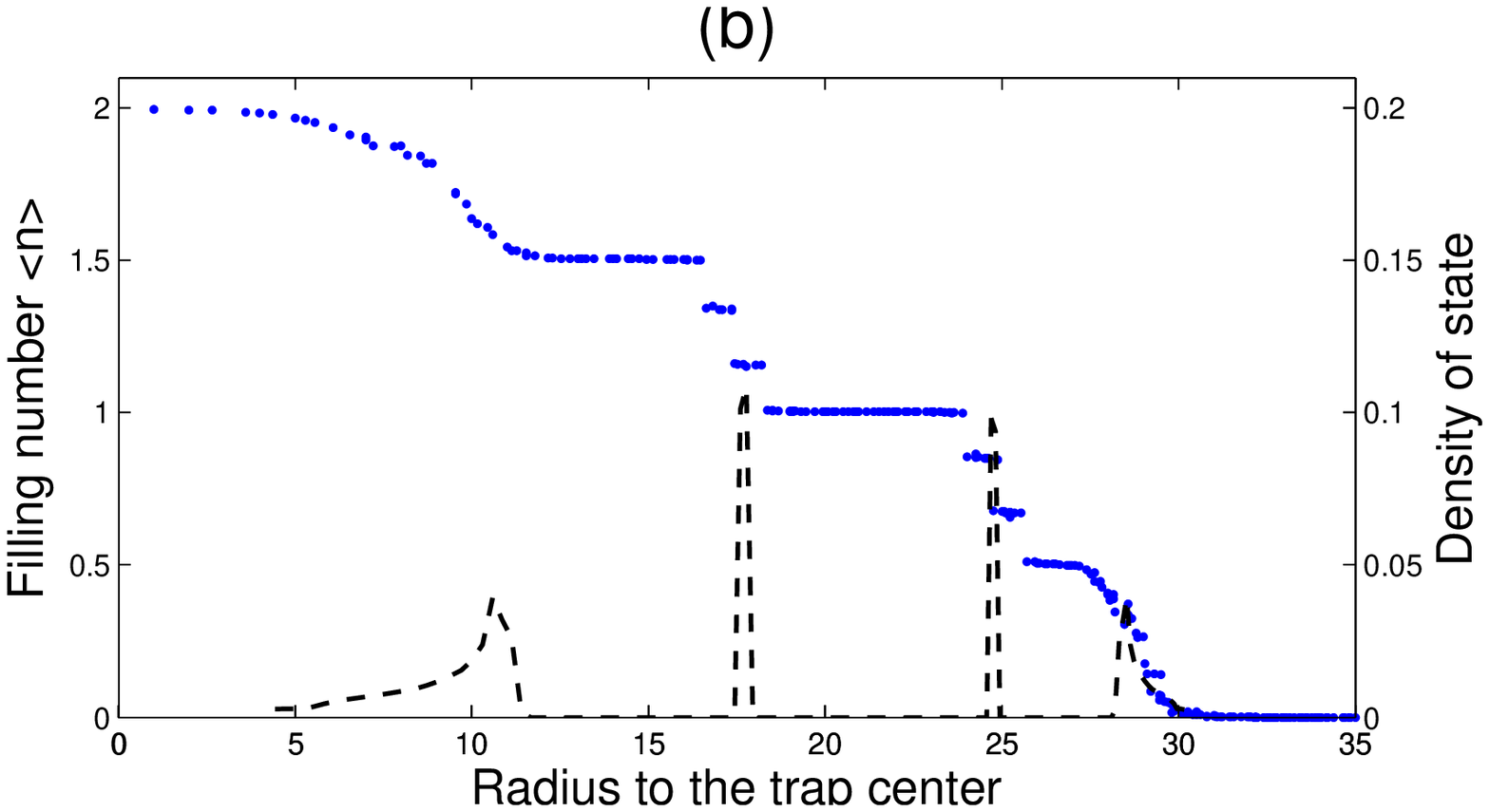,clip=1,width=0.8\linewidth,angle=0}}
\caption{(Color online)
The particle density distributions $\langle n(\vec r)\rangle$
\textit{v.s.} the radius $r$ at (a) $\Omega/t_\parallel=\frac{1}{2}$
and (b) $\frac{3}{4}$, which exhibit a four-layer wedding cake shape
(the blue dots).
The radius is in the unit of the lattice constant $a$. The DOS at the
local chemical potential $\protect\mu_{loc}(r)$ in the LDA approximation
is plotted with the black dash lines.
$\protect\mu_{loc}(r)/t_\parallel=2.3$ at the center of the trap and
$\protect\omega_T/t_\parallel=0.1$.}
\label{fig:plateau1}
\end{figure}

We first present the density of states (DOS) for the $p$-orbital bands in
the homogeneous system at $\Omega/t_\parallel=\frac{1}{2}$, and $\frac{3}{4}$
in Fig. \ref{fig:dos}. The DOS is defined as
\begin{equation}
g(E)=\int \frac{d^2 k}{(2\pi)^2} \delta(E(\vec k)-E).
\end{equation}
At $\Omega/t_\parallel=\frac{3}{4}$, the strong divergence of the DOS is
indicated at the 2nd and 3rd bands due to the appearance of the flat bands.
At the small value of $\Omega/t_\parallel=\frac{1}{2}$, the DOS of the 1st
and 4th bands are larger than the 2nd and 3rd bands, which is a reminiscence
of the band flatness at $\Omega=0$. Moreover, it is obvious that the band
gaps open at $\Omega_c>0$. For the chemical potential $\mu$ lying in the
band gaps, the system is in the band insulating states with the commensurate
values of the particle number per site $\langle n \rangle=\frac{1}{2}, 1,
\frac{3}{2}$, and $2$, respectively.

In the inhomogeneous trap, the real space distributions of the filling
number $\langle n(\vec r)\rangle$ are calculated by using the eigenstate
wavefunction obtained through diagonalizing the Hamiltonian, which are
depicted in Fig. \ref{fig:plateau1} for $\Omega/t_\parallel=\frac{1}{2}$ and
$\frac{3}{4}$. In both cases, plateaus appear at $\langle n\rangle=\frac{1}{2%
},1, \frac{3}{2}$. These plateaus can be understood within the LDA picture. 
Recall the band structure in Fig. \ref{fig:spectra} and the DOS in 
Fig. \ref{fig:dos}. 
When $\mu_{loc}(r)$ lies in the band gaps, the filling number stops 
increasing until $\mu_{loc}(r)$ reaches next band edge. 
The local DOS at site $\vec r$ at the energy $\mu_{loc}(r)$ is also 
plotted, which roughly proportional to $\partial_r n(r) $. 
It is clear that the locations of the plateaus of $\langle
n(r)\rangle$ are and band gaps are consistent.

Because the honeycomb lattice breaks the $SO(2)$ rotational symmetry down to
the 6-fold one, lattice sites with the same magnitude of $r$ may have 
different values of $\langle n(r)\rangle$.
They are slightly scattered in the metallic regions between different 
plateaus as depicted in Fig. \ref{fig:plateau1} (a) and (b). 
In the case of $\Omega/t_\parallel= \frac{3}{4}$, the distribution of 
$\langle n(r)\rangle$ exhibit devil's stair-like features 
\cite{bak1982,bak1986}. 
In the cliffs as filling in the middle two flat bands whose degeneracies 
are slightly lifted by the potential gradient. 
If with interactions, the flat band regime may further
exhibit plateaus of Mott-insulating states with orbital orderings, which
will be deferred to a later research.

\begin{figure}[tbp]
\centering
{\epsfig{file=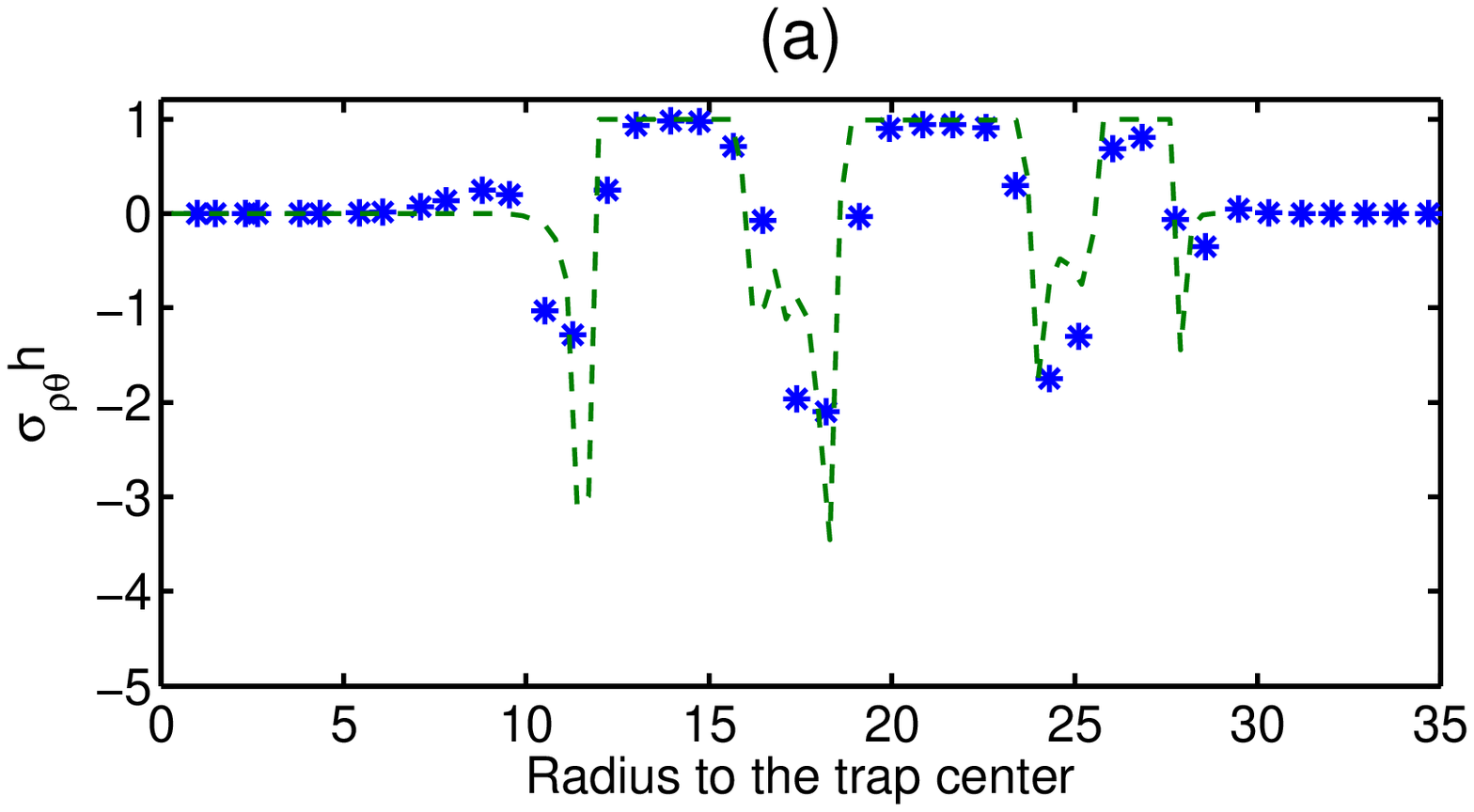,clip=1,
width=0.8\linewidth,angle=0}} \centering
{\epsfig{file=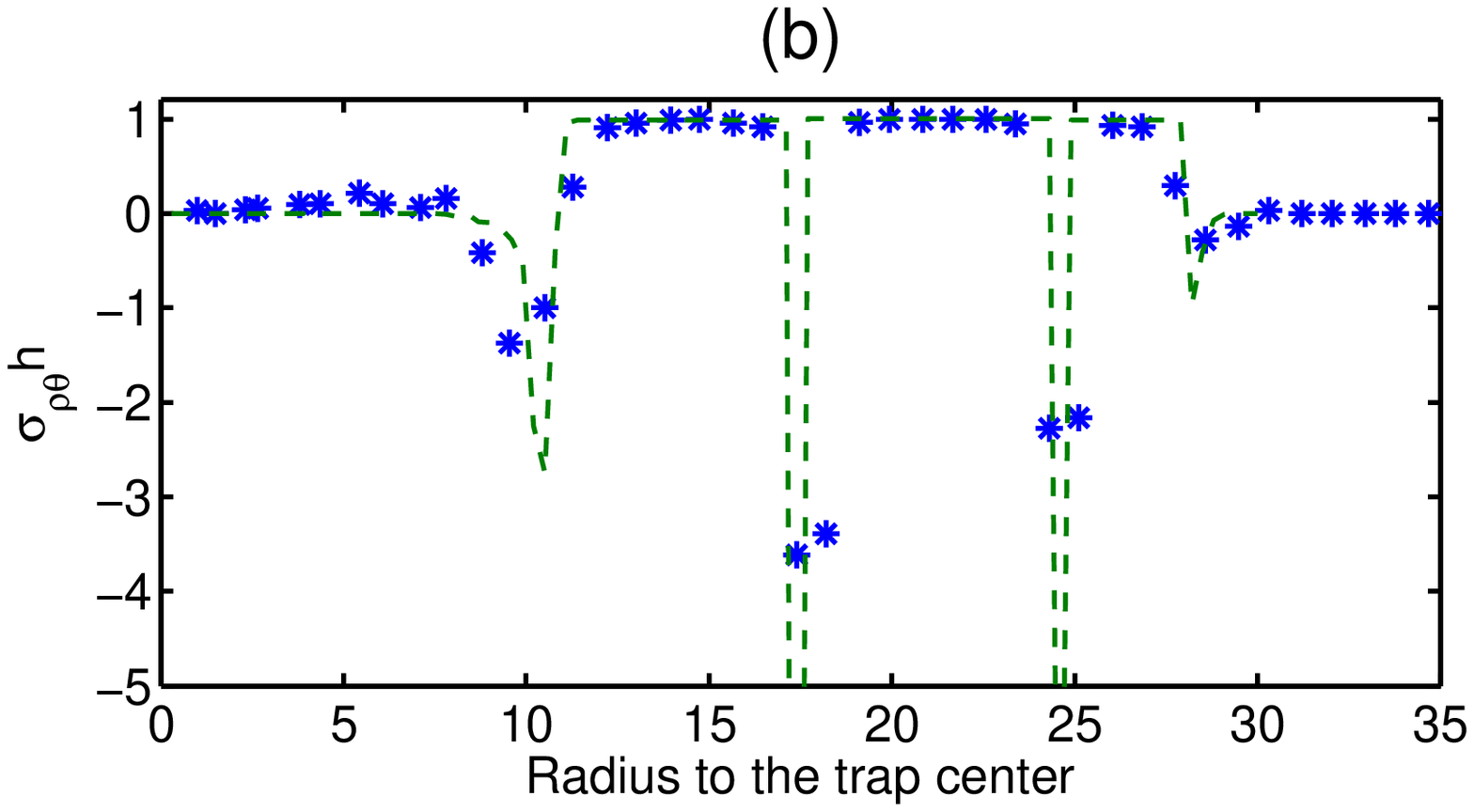,clip=1,width=0.8\linewidth,angle=0}}
\caption{(Color online)~~ The effective local Hall conductance $\protect%
\sigma_{\protect\rho\protect\theta}(r)$ v.s. $r$ defined in Eq. \protect\ref%
{eq:local_conductance} at $\Omega/t_\parallel=\frac{1}{2}$ (a) and $\frac{3}{%
4}$ (b). The radius is in the unit of the lattice constant $a$. The results
from diagonalizing the free Hamiltonian with the trapping potential is
marked with asterisks, and those from the modified LDA are plotted with
dashed lines. $\protect\sigma_{\protect\rho\protect\theta}$ is quantized in
the insulating plateaus with commensurate fillings of $\langle n(r)\rangle$.
}
\label{fig:sigmaonr1}
\end{figure}

\begin{figure}[tbp]
\centering\epsfig{file=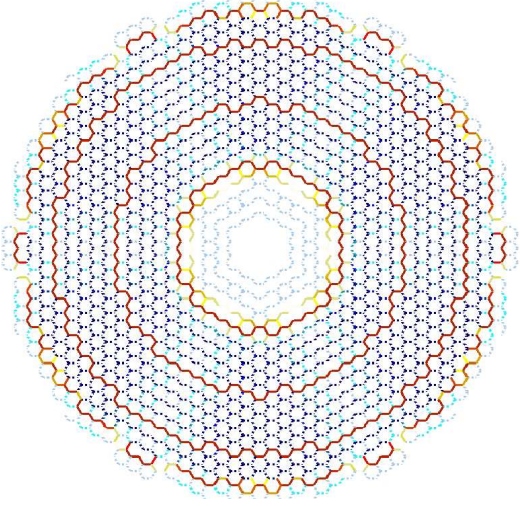,clip=1,width=0.8\linewidth,angle=0}
\caption{(Color online)~~ The pattern of the anomalous Hall current in the
honeycomb lattice with the confining trap at the rotation of
$\Omega/t_\parallel=0.75$. The blue dotted (red solid) lines represent the
counterclockwise (clockwise) anomalous Hall currents, respectively. The
color depth indicates the magnitude of the current. The reversed direction
of the anomalous Hall currents (red solid lines) in the metallic regions
between two neighboring plateaus can be explained as the anomalous
contribution from the gradient of $\langle n(r)\rangle$. }
\label{fig:fullcurrent}
\end{figure}

\subsubsection{QAHE currents in the insulating density
plateaus}
Due to the non-trivial topology of the band structure, anomalous Hall
currents circulate along the azimuthal direction due to the radial potential
gradient. The plateaus of the filling number $\langle n(r)\rangle$
correspond to the insulating quantum anomalous Hall regions with quantized
Hall conductance. Compared with the usual quantum Hall systems, the on-site
rotation breaks time-reversal symmetry and brings non-trivial topology to
the band structures without Landau levels.

The anomalous Hall current along each bond is calculated by using the
eigen-wavefunctions obtained from the diagonalization of the real space 
$p$-orbital Hamiltonian as
\begin{eqnarray}
J_{\vec r, \vec r^+\hat e_i}=i \frac{t_\parallel}{\hbar} \sum_n \langle |
(\hat p^\dagger_{\vec r} \cdot \hat e_i) (\hat p_{\vec r+\hat e_i} \cdot
\hat e_i) -h.c. |\rangle,
\end{eqnarray}
where $\langle | |\rangle$ represents the ground state at $T=0$ or
thermal average at finite temperatures.
Let us focus on those bonds orienting along the azimuthal direction, and
define the effective local Hall conductance as
\begin{eqnarray}
\sigma_{\rho\theta}^{\text{eff}} (r) = -{j_{\theta} \over \partial_r V_T},
\label{eq:local_conductance}
\end{eqnarray}
where $r$ is the radius of the middle point of the bond;
$j_{\theta}$ is the current density;
$ V_T$ is the trapping potential.
In our honeycomb lattice system, the current density is defined as the current
on each bond $J_{\vec r, \vec r^+\hat e_i}$ divides the distance between
neighboring parallel bonds.

In the homogeneous systems, the Hall conductance is represented as
\begin{equation}
\sigma _{\rho\theta} = \frac{1}{h}\frac{1}{2\pi} \sum_i \int\mathrm{d}^2 k
F_{i,xy}(\vec k) n_f (i,k),  \label{eq:conductance}
\end{equation}
where $n_f$ is the Fermi distribution function; $i$ is the band index.
When the chemical potential is inside band
gaps, $\sigma_{\rho\theta}$ is quantized as the sum of the Chern numbers of
the occupied bands \cite{niuqian2009,thouless1982,kohmoto1985}
\begin{equation}
\sigma _{\rho\theta} =\frac{1}{h} \sum_{i} C_i.  \label{eq:quan_conductance}
\end{equation}
For the cases of $\Omega/t_\parallel=\frac{1}{2}, \frac{3}{4}$, the Chern
number pattern is the same as $C_1=-C_4=1$ and $C_2 =-C_3=0$ \cite{wu2008}.
The quantized Hall conductances reads $0$, $1$, $1$ and $1$ as $\mu$ lies
from above the band top down to the three consecutive three band gaps.

The results of $\sigma_{\rho\theta}^{\text{eff}}(r)$
(defined in Eq. \ref{eq:local_conductance}) v.s $r$ are marked as
asterisks in Fig. \ref{fig:sigmaonr1} (a) and (b) for
$\Omega/t_\parallel=\frac{1}{2}$ and $\frac{3}{4}$,
respectively, which are obtained by diagonalizing the free but
inhomogeneous Hamiltonian.
The real space circulating current pattern at
$\Omega/t_\parallel=\frac{3}{4}$ is depicted in Fig. \ref{fig:fullcurrent}.
The quantized Hall conductances in the insulating plateau regions can be
understood within the LDA picture. At the center, the local chemical
potential $\mu_{loc}(r)$ lies above the band top, and the conductance is
therefore zero. As moving into the insulating density plateaus of $\langle
n(r)\rangle=\frac{3}{2},1$, and $\frac{1}{2}$, $\sigma^{\text{eff}%
}_{\rho\theta}$ is quantized at $1/h$.
Counterclockwise currents are plotted
as blue dash line under the harmonic trap potential.
When the radius $r>30$,
the Fermi level is lower than the band bottom, thus the current vanishes
again.


\subsubsection{Anomalous Hall currents in the metallic regions}

\begin{figure}[tbp]
\centering
{\epsfig{file=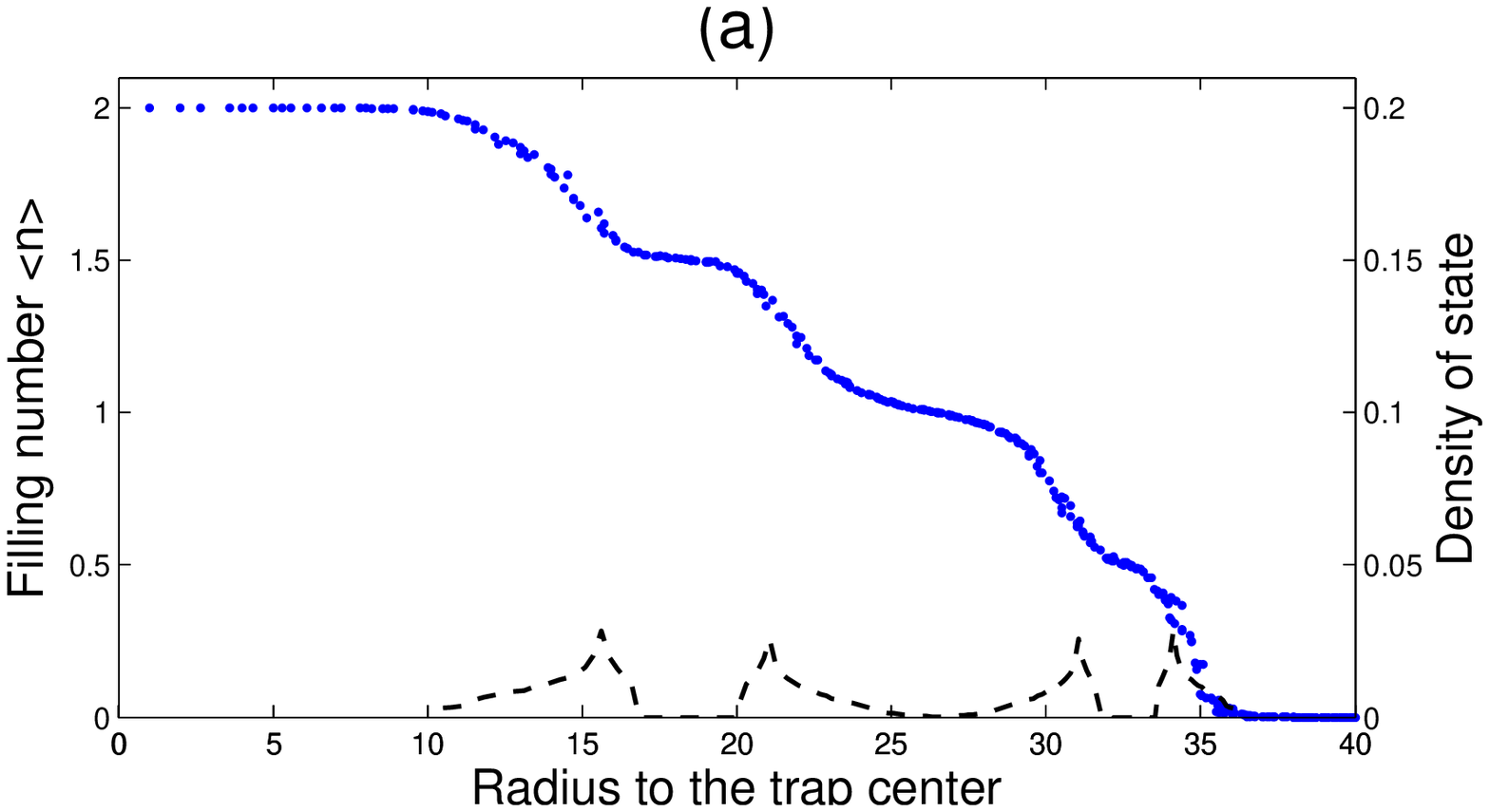,clip=1,width=0.8\linewidth,angle=0}} \centering%
{\epsfig{file=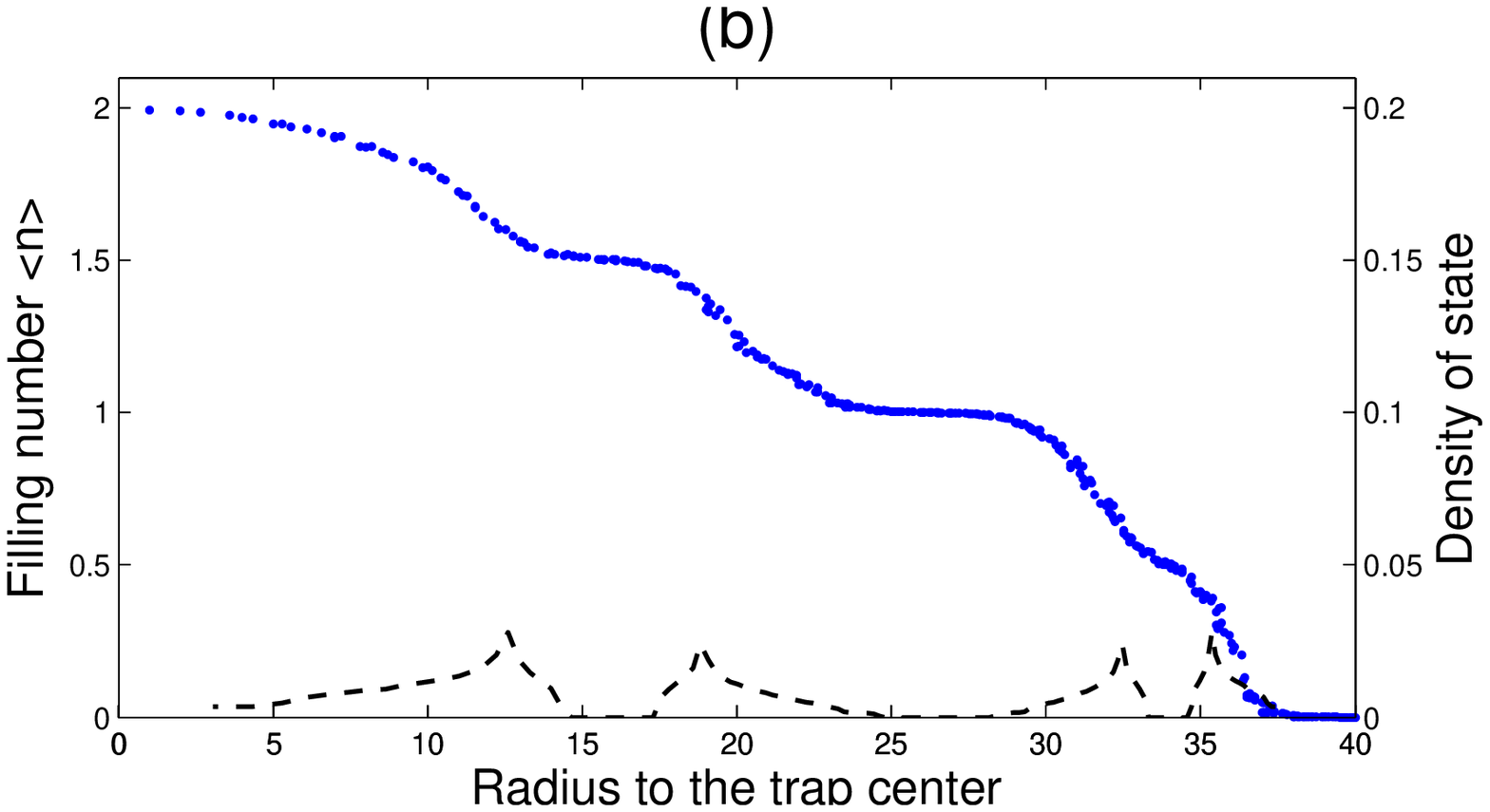,clip=1,width=0.8\linewidth,angle=0}}
\caption{(Color online)~~The particle density distributions $\langle n(\vec
r)\rangle$ \textit{v.s.} the radius $r$ at (a) $\Omega/t_\parallel=\frac{3}{2%
}$ and (b) $2$ (the blue dots). The radius is in the unit of the lattice
constant $a$. The DOS at the local chemical potential $\protect\mu_{loc}(r)$
in the LDA approximation is plotted with the black dash lines. $\protect\mu%
_{loc}(r)/t_\parallel=3.5$ at the center of the trap and $\protect\omega%
_T/t_\parallel=0.1$. }
\label{fig:plateau2}
\end{figure}

Between two adjacent insulating plateaus, the system is metallic with
incommensurate fillings of $\langle n(r)\rangle$, therefore the anomalous
Hall conductances are non-quantized. The Hall current response to the radial
potential gradient is non-local in these inhomogeneous metallic regions. The
effective Hall conductance $\sigma_{\rho\theta}$ defined in Eq. 
\ref{eq:local_conductance} 
\emph{cannot} be obtained from Eq \ref{eq:conductance}
of the homogeneous system by using LDA with a local chemical potential 
$\mu_{loc}(r)$. 
For example, the currents can reverse the direction to be
clockwise in the metallic regions in Fig. \ref{fig:sigmaonr1} (a) and (b).
However, because of the Chern number pattern of $C_2=C_3=0$ and $C_1=-C_4=1$,
the Hall conductance $\sigma_{\rho\theta}$ defined in Eq.
\ref{eq:conductance} is always positive within the LDA
at $\frac{1}{2}<\langle n\rangle<1$ which corresponds to 
that $\mu_{loc}(r)$ lies in the gap between the 1st and 2nd bands,
The naive LDA results would only give
rise to counterclockwise Hall currents in these two metallic regions.

Now we propose a modified LDA method to fit the above exact results from
diagonalization. We define the effective driving force as the derivative of
the spatial-dependent part of the ground state energy density as
\begin{eqnarray}
F(r)&=&\frac{1}{n(r)}\frac{\partial}{\partial_r} \Big\{(\mu_{loc}(r)
-E_B(\Omega)) n(r)\Big\}  \notag \\
&=& F_{\text{drift}} +F_{\text{diff}},
\end{eqnarray}
where $E_B(\Omega)$ is the band bottom energy; $F_{\text{drift}}$ and 
$F_{\text{diff}}$ are defined as
\begin{eqnarray}
F_{\text{drift}}&=& -\partial_r V_T(r),  \notag \\
F_{\text{diff}}&=& \frac{\mu_{loc}(r)-E_B(\Omega)}{n(r)}\partial_r n(r).
\end{eqnarray}
$F_{\text{drift}}$ comes from the gradient of the trapping potential, while
$F_{\text{diff}}$ is the chemical pressure from the particle density gradient.

Correspondingly, the anomalous Hall currents can be interpreted by two
contributions from the ``drift'' and ``diffusive'' Hall currents as
\begin{eqnarray}
J_\theta(r)&=& J_{\text{drift},\theta}(r)+J_{\text{diff},\theta}(r)  \notag
\\
&=& \sigma_{\rho\theta}(r) \big\{ F_{\text{drift}}(r) + F_{\text{diff}}(r)
\big\},
\end{eqnarray}
where $\sigma_{\rho\theta}(r)$ is obtained from Eq. \ref{eq:conductance} in
the LDA by using the local chemical potential $\mu_{loc}(r)$. Thus $\sigma^{%
\text{eff}}_{\rho\theta}(r)$ defined in Eq. \ref{eq:local_conductance} is
related to $\sigma_{\rho\theta}(r)$ through
\begin{eqnarray}
\sigma^{\text{eff}}_{\rho\theta}(r)=\sigma_{\rho\theta}(r) \Big\{1+ \frac{F_{\text{
diff}}(r)}{F_{\text{drift}}(r)}\Big\}.
\end{eqnarray}
The results of the effective Hall conductance $\sigma^{\text{eff}%
}_{\rho\theta}(r)$ using this modified LDA is presented in Fig.
\ref{fig:sigmaonr1} with dashed lines, which nicely agrees with the exact
results.

In the insulating plateaus, $F_{\text{diff}}=0$, thus the modified LDA
reduces back to the naive LDA. However, in the metallic regions, $\partial_r
n(r)$ has the opposite direction to $\partial_r V_T$. As a result, the
direction of $J_{\text{diff},\theta}$ is also opposite to that of $J_{\text{%
drift},\theta}$. The reversed direction of the Hall currents in the metallic
regions can be understood as the contribution of ``diffusive'' Hall current
dominates over that of the ``drift'' Hall current.

\subsection{Large rotation angular velocities ($\Omega/t_\parallel=\frac{3}{2},
2$)}
\begin{figure}[tbp]
\centering
{\epsfig{file=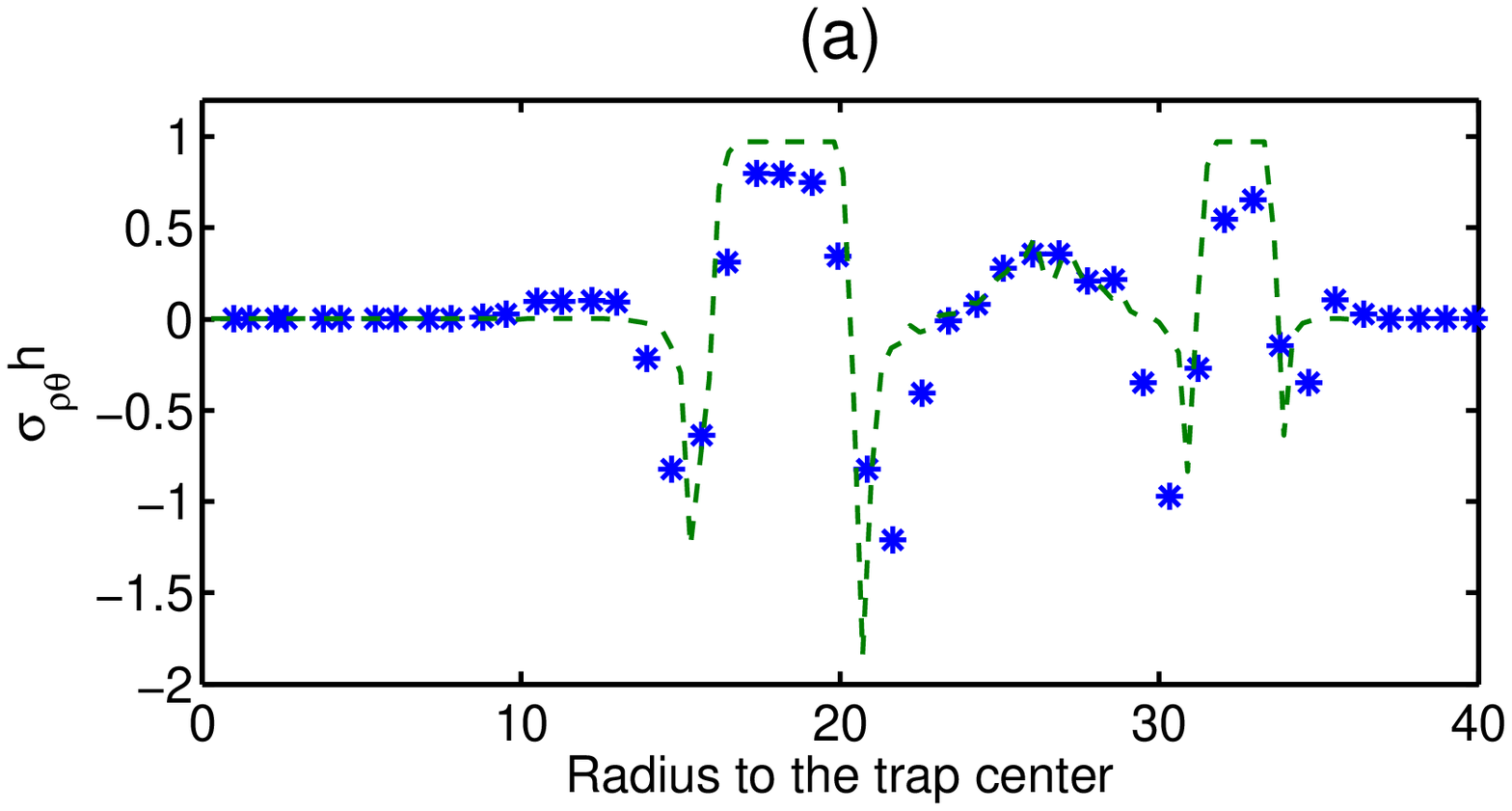,clip=1,width=0.8\linewidth,angle=0}} \centering
{\epsfig{file=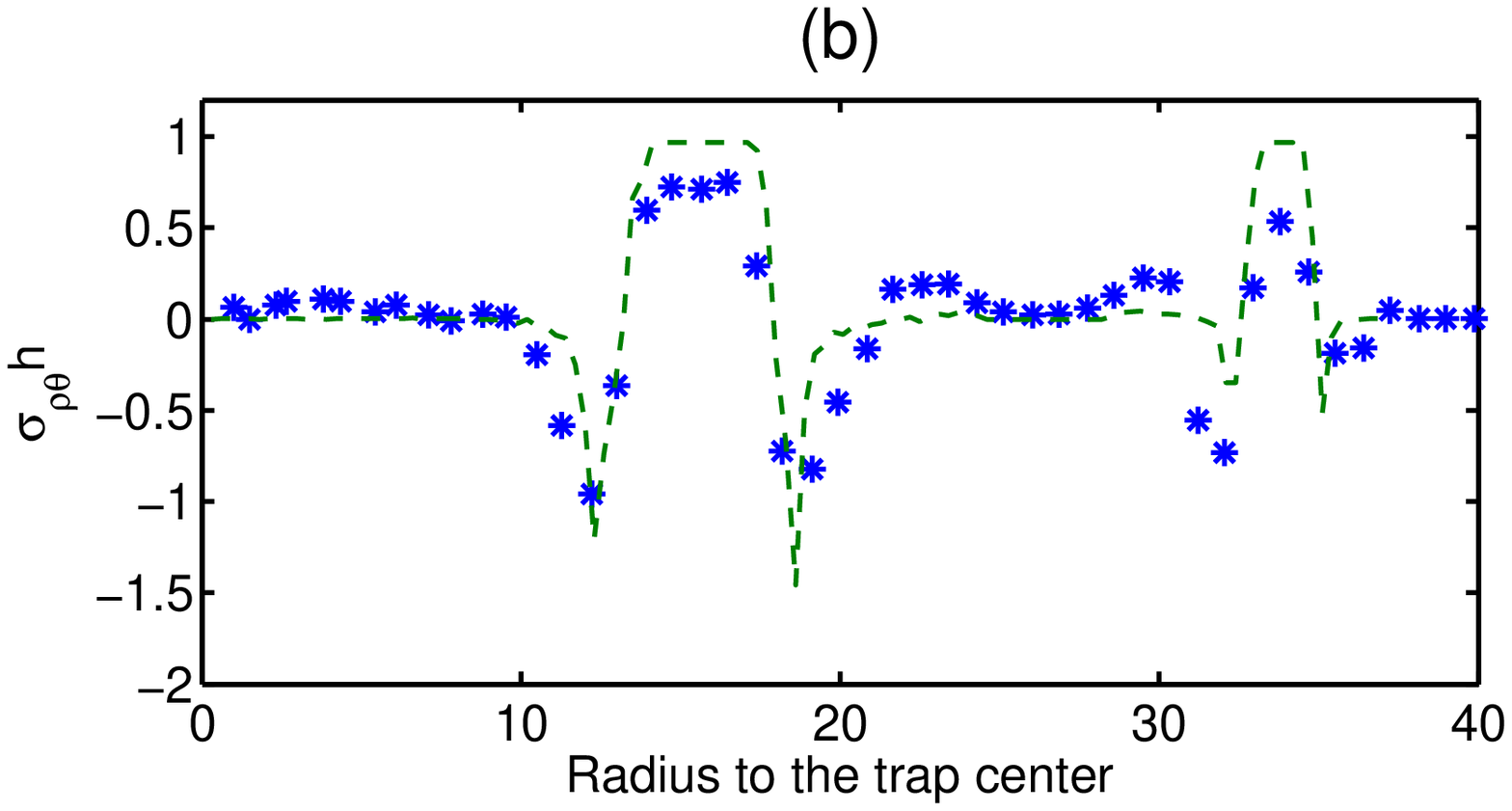,clip=1,width=0.8\linewidth,angle=0}}
\caption{(Color online)~~ The effective local Hall conductance $\protect%
\sigma_{\protect\rho\protect\theta}(r)$ v.s. $r$ defined in Eq. \protect\ref
{eq:local_conductance} at $\Omega/t_\parallel=\frac{3}{2}$ (a) and $2$ (b).
The radius is in the unit of the lattice constant $a$. The results from
diagonalizing the free Hamiltonian with the trapping potential is marked
with asterisks, and those from the modified LDA are plotted with dashed
lines. }
\label{fig:sigmaonr2}
\end{figure}

In this subsection, we consider the large rotation angular velocities at $
\Omega/t_\parallel=\frac{3}{2}$ and $2$ which are at and above $
\Omega_c/t_\parallel= \frac{3}{2}$, respectively. The band structure
topology above $\Omega_c$ changes to a different Chern number pattern of $
C_1=-C_2=C_3=-C_4=1$. The chemical potential $\mu/t_\parallel$ is chosen to
3.5, which guarantees that all bands are filled at $r=0$.

The distributions of the filling number $\langle n(\vec r)\rangle$ are
depicted in Fig. \ref{fig:plateau2} at (a) $\Omega/t_\parallel=\frac{3}{2}$
and (b) $\Omega/t_\parallel=2$, respectively. At $\Omega/t_\parallel=\frac{3%
}{2}$, the 2nd and 3rd bands touch each other at a Dirac cone located at the
center of the BZ. The DOS vanishes linearly, and thus the density profile
exhibits a soft slope instead of a flat plateau in Fig. \ref{fig:plateau2}
(a). In both Fig. \ref{fig:plateau2} (a) and (b), the density distributions
between the 1st and 2nd bands also exhibit soft slopes although there do
exist a band gap in the homogeneous system. This is because the potential
gradient increases as $r$ goes larger in the confining trap, which closes
the small gap between the 1st and 2nd bands at large values of $\Omega$.

The local anomalous Hall conductances defined in Eq. \ref%
{eq:local_conductance} are depicted in Fig. \ref{fig:sigmaonr2} at $%
\Omega/t_\parallel=\frac{3}{2}$ and $2$. 
In the insulating plateaus between the 3rd and 4th bands, $\sigma_{\rho\theta}$ 
is close to the quantized value
of $1/h$ for both rotation angular velocities. The small deviation comes
from the finite width of the insulating regions. 
In the region with soft slopes of the distributions of $\langle n(r)\rangle$ 
between the 1st and 2nd bands, $\sigma_{\rho\theta}$ is significantly
smaller than $1/h$ because this region is not rigorously insulating.
In the inhomogeneous metallic regions, the values of the local anomalous
Hall currents are non-quantized which are determined by the combined effects
from the gradients of the trapping potential and the density distribution.

\subsection{Temperature effects}

\begin{figure}[tbp]
\centering \epsfig{file=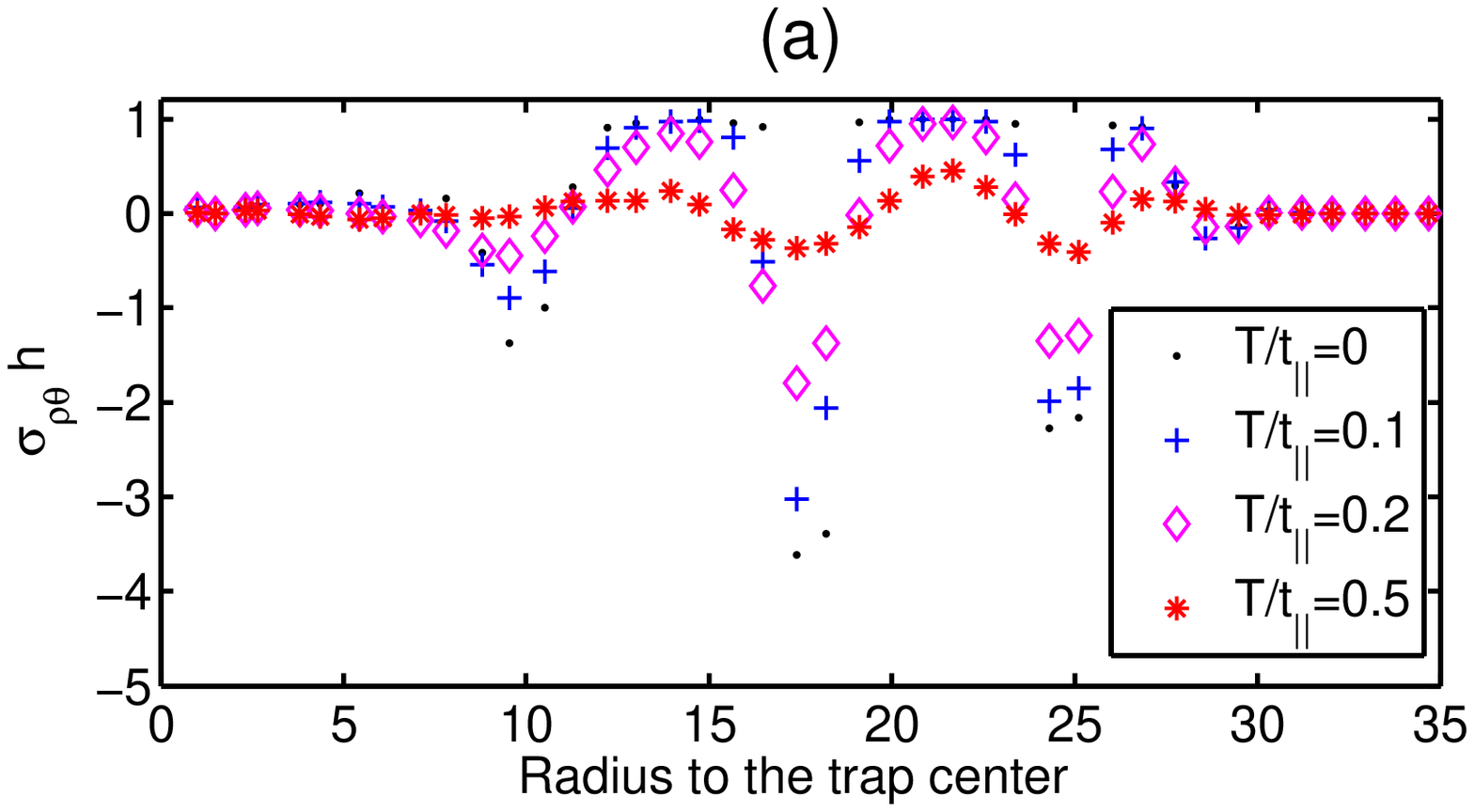,clip=1,width=0.8\linewidth,angle=0}
\centering \epsfig{file=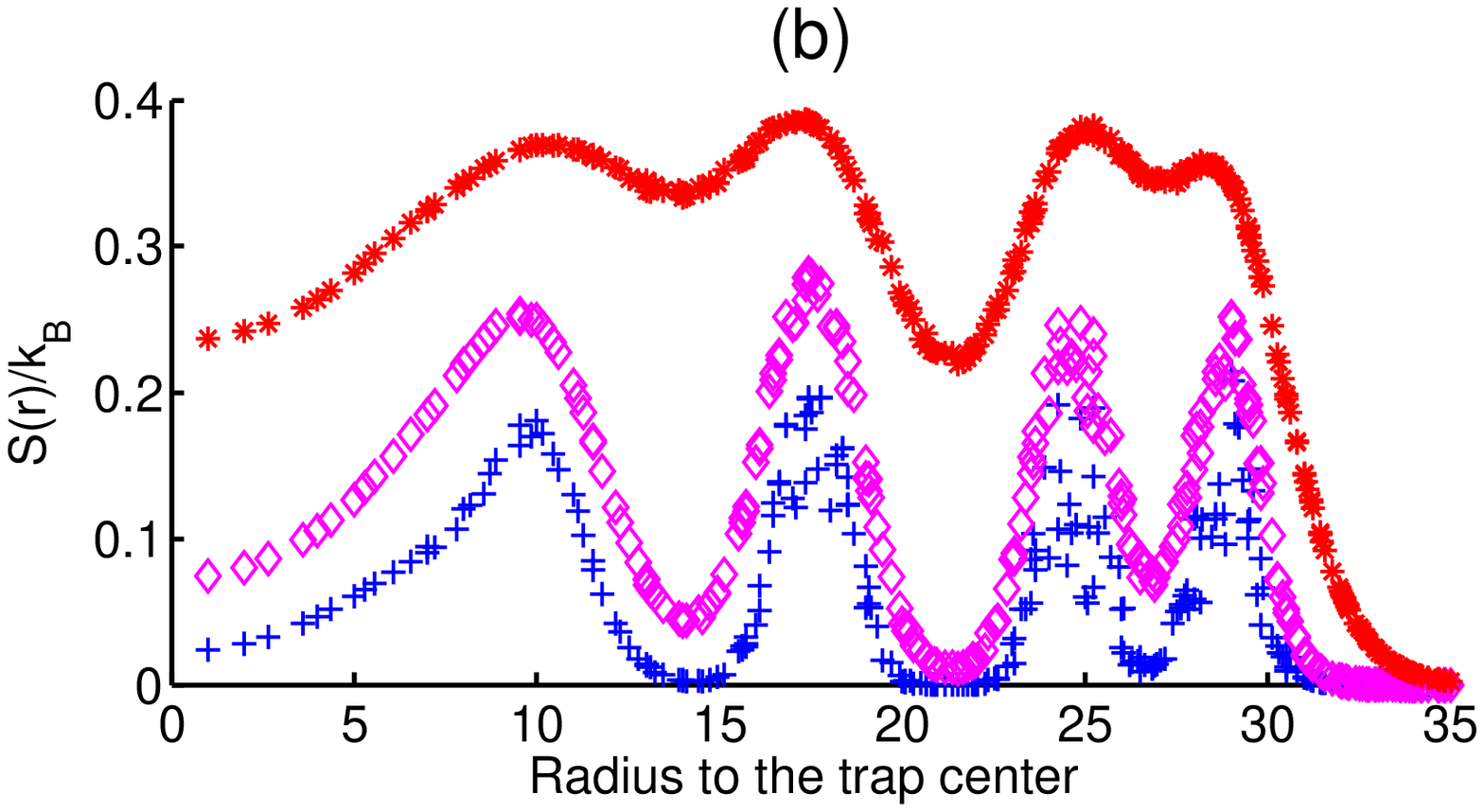,clip=1,width=0.8\linewidth,angle=0}
\caption{(Color online)~~ The radial distributions of (a) the anomalous
Hall conductance $\sigma_{\rho\theta}(\vec r, T)$ and (b) the entropy
$S(\vec r, T)$ with  $\Omega/t_\pp=\frac{3}{4}$.
The black dot, blue ``+'', magenta diamonds and red asterisks of the
data represent temperatures at $T=0$, $T/t_\pp=0.1$, $T/t_\pp=0.2$
and $T/t_\pp=0.5$, respectively.
 }
\label{fig:Temp}
\end{figure}

In this subsection, we briefly discuss the finite temperature effects
to the anomalous Hall conductance.
The QAHE is a topological property existing in band insulating regions,
thus it is robust against finite temperatures provided their scale
is small compared to the band gap.
On the other hand, we do expect that the anomalous Hall conductance
in the metallic regions will be significantly affected by finite
temperatures.

The radial distribution of the local anomalous Hall conductance
$\sigma_{\rho\theta}^{\text{eff}}(r)$ {\it v. s.} $r$ is plotted in
Fig. \ref{fig:Temp} (a) at different temperatures.
The rotation angular velocity is taken as $\Omega/t_\pp=\frac{3}{4}$.
In this case, the band gaps are at the same order of $t_\pp$
as shown in Fig. \ref{fig:dos}.
$\sigma_{\rho\theta}$ remains nearly quantized in the insulating
regions for $T/t_\pp=0.1\sim 0.2$.
According to the calculation of band structures in Ref. \cite{wu2008b},
$t_\pp\approx 0.24 E_r$ at $V_0/E_r=15$ and the typical
energy scale of $E_r$ is $0.1 \sim 0.2\mu K$,
thus the QAHE signature should survive at the order of $10$nK,
which is an experimentally accessible temperature scale.
In the metallic regions, naturally $\sigma_{\rho\theta}$'s are
more strongly affected by finite temperatures.
As $T$ further increases to the half value of $t_\pp$,
the quantized signatures of $\sigma_{\rho\theta}$ disappear.

We also present the entropy distributions in real space at various
temperatures as shown in Fig. \ref{fig:Temp} (b).
The local entropy is defined as
\bea
S(\vec r, T)&=&-k_B \sum_i |\psi_i (\vec r)|^2
\Big \{ n_i(T) \ln n_i (T) \nn \\
&+&(1-n_i(T)) \ln [1-n_i (T)] \Big \},
\eea
where $k_B$ is the Boltzmann constant; the subscript $i$ is the
index of energy levels; $\psi_i (\vec r)$ is the wavefunction at the
location $\vec r$; $n_i(T)$ is the Fermi distribution function.
At low temperatures (e.g. $T/t_\pp=0.1$ and $0.2$), $S(\vec r, T)$ 
concentrates in the gapless metallic regions, and remains
negligible in the band insulating regions.
We have the coexistence of the insulating QAHE regions
and the metallic regions to hold a significant amount of entropy.
The tolerance of the large residue entropy densities in the trap
greatly facilitates the experimental realization of the QAHE state.
As temperatures go high, say, $T/t_\pp=0.5$, the entropy
distribution becomes more uniform, and there are no clear
distinctions between insulating and metallic regions any more.
This agrees with the picture in Fig. \ref{fig:Temp} (a), in which
the plateaus of quantized anomalous Hall conductance disappear.

\section{Experimental detections}
\label{sect:dect}

In experiments, the plateaus of commensurate fillings of atoms in Fig. \ref%
{fig:plateau1} and Fig. \ref{fig:plateau2} can be observed by measuring the
in-trap density distribution or the compressibility of the lattice system,
as clearly demonstrated in recent experiments \cite%
{Esslinger2008,Kuhr2010,Bloch2008,Greiner2009}. However, the density
plateaus cannot distinguish the conventional band insulators and the quantum
anomalous Hall insulators.

In solid state systems, the Hall conductivity are obtained from transport
measurements, which are very difficult for the cold atom experiments.
Nevertheless, it has been proposed to detect through the response of the
atom density to an external magnetic field \cite{Zhai2008}, which can be
realized by further rotating the harmonic trap \cite{Haljan} or coupling
atoms with additional laser fields. In particular, the motion of atoms in
laser fields leads to an artificial magnetic field, which has been observed
in a recent experiment \cite{Lin2009}. In the presence of an artificial
magnetic field, the quantized anomalous Hall conductivity $%
\sigma_{xy}=\left( \frac{\partial n}{\partial B}\right)_{\mu }$, according
to the well-known Streda formula \cite{Streda} derived for the quantum Hall
effects in the solid state. Therefore the density of atoms changes linearly
with respect to the applied magnetic field when $\sigma _{xy}$ is quantized
in some regions of the harmonic trap.

Another possible method to detect the anomalous Hall current is as follows.
We assume all atoms are initially prepared in a hyperfine ground state $%
\left\vert 1\right\rangle $. To detect the anomalous Hall current shown in
Fig. 9, we apply a local two-photon Raman transition using two
co-propagating focused laser beams in a small area $S$ to transfer atoms in $%
S$ to another hyperfine state $\left\vert 2\right\rangle $. A subsequent
time-of-flight measurement of the velocity distribution of atoms in state $%
\left\vert 2\right\rangle $ gives the initial velocity distribution (thus
the current) of atoms in the state $\left\vert 1\right\rangle $ in the
optical lattice. The above density and current measurements provide the
experimental signature of the quantum anomalous Hall effects in the 
$p$-orbital honeycomb lattice.


\section{Conclusions and Outlook}

\label{sect:conclusions}

In summary, we have proposed the realization of the quantum anomalous Hall
states in the cold atom optical lattices based on the experimentally
available technique of the on-site rotation developed by Gemelke \textit{et
al.} This rotation generates the orbital Zeeman coupling whose energy scale
can reach the order of the band width. In the $p$-orbital bands of the
honeycomb lattice, the band structures become topologically non-trivial at
any nonzero rotation angular velocities. A topological transition occurs at $%
\Omega _{c}/t_{\parallel }=\frac{3}{2}$ with different band Chern number
patterns below and above $\Omega _{c}$. At $\Omega >\Omega _{c}$, the band
topology is equivalent to a double copy of Haldane's quantum anomalous Hall
model. Flat band structures are also found at $\Omega /t_{\parallel }=\frac{3%
}{4}$ whose localized eigenstates can be constructed as circulating
plaquette current states. The flat band structures may bring strong
correlation effects, such as Wigner crystal and ferromagnetism, when
interactions are turned on.

The effects of the spatial inhomogeneity to the $p$-orbital quantum
anomalous Hall states are also investigated. At each commensurate filling of
$\frac{1}{2},1,\frac{3}{2}$, and 2, the density profile exhibits insulating
plateaus, whose Hall conductances are quantized at integer values. In the
metallic regions between two adjacent plateaus, the anomalous Hall currents
are determined by the non-local response, which can be understood as the
combined effects of the gradients of the confining potential and particle
density.
We have also showed that the QAHE is robust 
at finite but low temperatures compared to band gaps.

We further point out that the generation of the quantum anomalous Hall
states from this ``orbital Zeeman'' effect is very general, not just for the
honeycomb lattice. The advantage of the $p$-orbital honeycomb lattice is
that an infinitesimal value of $\Omega$ is enough to generate the quantum
anomalous Hall states. For other generic lattice structures, beyond a
critical value of $\Omega$ which is comparable to the band width, the
orbital Zeeman effect generates inverted orbital bands of different 
orbital angular momenta. 
The further hybridization among them brings
non-trivial band topology, which is a similar mechanism to achieve
topological insulators in semi-conducting systems through spin-orbit
couplings.  A systematic study will be presented in a later publication.


\section*{Acknowledgement}

M. Z. acknowledges Prof. Shi-qun Li for the support and thanks Wei-Cheng Lee
for helpful discussions. H. H. H. and C. W. are supported by NSF under No.
DMR- 0804775 and AFOSR-YIP, M. Z. is supported by the NFRP-China
Grant(973Project) Nos.2006CB921404, and C.Z. is supported by ARO
(W911NF-09-1-0248) and DARPA-YFA (N66001-10-1-4025).

\bibliography{orbitalphysics}

\begin{thebibliography}{87}
\expandafter\ifx\csname natexlab\endcsname\relax\def\natexlab#1{#1}\fi
\expandafter\ifx\csname bibnamefont\endcsname\relax
  \def\bibnamefont#1{#1}\fi
\expandafter\ifx\csname bibfnamefont\endcsname\relax
  \def\bibfnamefont#1{#1}\fi
\expandafter\ifx\csname citenamefont\endcsname\relax
  \def\citenamefont#1{#1}\fi
\expandafter\ifx\csname url\endcsname\relax
  \def\url#1{\texttt{#1}}\fi
\expandafter\ifx\csname urlprefix\endcsname\relax\def\urlprefix{URL }\fi
\providecommand{\bibinfo}[2]{#2}
\providecommand{\eprint}[2][]{\url{#2}}

\bibitem[{\citenamefont{{Gemelke} et~al.}(2010)\citenamefont{{Gemelke},
  {Sarajlic}, and {Chu}}}]{gemelke2010}
\bibinfo{author}{\bibfnamefont{N.}~\bibnamefont{{Gemelke}}},
  \bibinfo{author}{\bibfnamefont{E.}~\bibnamefont{{Sarajlic}}},
  \bibnamefont{and} \bibinfo{author}{\bibfnamefont{S.}~\bibnamefont{{Chu}}},
  \emph{\bibinfo{title}{{Rotating Few-body Atomic Systems in the Fractional
  Quantum Hall Regime}}}, \bibinfo{howpublished}{arXiv:1007.2677}
  (\bibinfo{year}{2010}).

\bibitem[{\citenamefont{Gemelke}(2007)}]{gemelke2007}
\bibinfo{author}{\bibfnamefont{N.}~\bibnamefont{Gemelke}}, Ph.D. thesis
  (\bibinfo{year}{2007}).

\bibitem[{\citenamefont{Sarajic et~al.}(2009)\citenamefont{Sarajic, Gemelke,
  Chiow, Herrman, M\"uller, and Chu}}]{gemelke2009}
\bibinfo{author}{\bibfnamefont{E.}~\bibnamefont{Sarajic}},
  \bibinfo{author}{\bibfnamefont{N.}~\bibnamefont{Gemelke}},
  \bibinfo{author}{\bibfnamefont{S.-W.} \bibnamefont{Chiow}},
  \bibinfo{author}{\bibfnamefont{S.}~\bibnamefont{Herrman}},
  \bibinfo{author}{\bibfnamefont{H.}~\bibnamefont{M\"uller}}, \bibnamefont{and}
  \bibinfo{author}{\bibfnamefont{S.}~\bibnamefont{Chu}},
  \emph{\bibinfo{title}{Coherent control of ultracold matter: Fractional
  quantum hall physics and large-area atom interferometry}}
  (\bibinfo{year}{2009}).

\bibitem[{\citenamefont{Karplus and Luttinger}(1954)}]{karplus1954}
\bibinfo{author}{\bibfnamefont{R.}~\bibnamefont{Karplus}} \bibnamefont{and}
  \bibinfo{author}{\bibfnamefont{J.~M.} \bibnamefont{Luttinger}},
  \bibinfo{journal}{Phys. Rev.} \textbf{\bibinfo{volume}{95}},
  \bibinfo{pages}{1154} (\bibinfo{year}{1954}).

\bibitem[{\citenamefont{Smit}(1954)}]{smit1958}
\bibinfo{author}{\bibfnamefont{J.}~\bibnamefont{Smit}},
  \bibinfo{journal}{Physica} \textbf{\bibinfo{volume}{24}}, \bibinfo{pages}{39}
  (\bibinfo{year}{1954}).

\bibitem[{\citenamefont{Berger}(1970)}]{berger1970}
\bibinfo{author}{\bibfnamefont{L.}~\bibnamefont{Berger}},
  \bibinfo{journal}{Phys. Rev. B} \textbf{\bibinfo{volume}{2}},
  \bibinfo{pages}{4559} (\bibinfo{year}{1970}).

\bibitem[{\citenamefont{Jungwirth et~al.}(2002)\citenamefont{Jungwirth, Niu,
  and MacDonald}}]{jungwirth2002}
\bibinfo{author}{\bibfnamefont{T.}~\bibnamefont{Jungwirth}},
  \bibinfo{author}{\bibfnamefont{Q.}~\bibnamefont{Niu}}, \bibnamefont{and}
  \bibinfo{author}{\bibfnamefont{A.}~\bibnamefont{MacDonald}},
  \bibinfo{journal}{Phys. Rev. Lett.} \textbf{\bibinfo{volume}{88}},
  \bibinfo{pages}{207208} (\bibinfo{year}{2002}).

\bibitem[{\citenamefont{Nagaosa}(2006)}]{nagaosa2006}
\bibinfo{author}{\bibfnamefont{N.}~\bibnamefont{Nagaosa}}, \bibinfo{journal}{J.
  Phys. Soc. Jpn} \textbf{\bibinfo{volume}{75}}, \bibinfo{pages}{42001}
  (\bibinfo{year}{2006}).

\bibitem[{\citenamefont{Nagaosa et~al.}(2010)\citenamefont{Nagaosa, Sinova,
  Onoda, MacDonald, and Ong}}]{nagaosa2010}
\bibinfo{author}{\bibfnamefont{N.}~\bibnamefont{Nagaosa}},
  \bibinfo{author}{\bibfnamefont{J.}~\bibnamefont{Sinova}},
  \bibinfo{author}{\bibfnamefont{S.}~\bibnamefont{Onoda}},
  \bibinfo{author}{\bibfnamefont{A.~H.} \bibnamefont{MacDonald}},
  \bibnamefont{and} \bibinfo{author}{\bibfnamefont{N.~P.} \bibnamefont{Ong}},
  \bibinfo{journal}{Reviews of Modern Physics} \textbf{\bibinfo{volume}{82}},
  \bibinfo{pages}{1539} (\bibinfo{year}{2010}).

\bibitem[{\citenamefont{Xiao et~al.}(2009)\citenamefont{Xiao, Chang, and
  Niu}}]{niuqian2009}
\bibinfo{author}{\bibfnamefont{D.}~\bibnamefont{Xiao}},
  \bibinfo{author}{\bibfnamefont{M.-C.} \bibnamefont{Chang}}, \bibnamefont{and}
  \bibinfo{author}{\bibfnamefont{Q.}~\bibnamefont{Niu}},
  \bibinfo{journal}{arXiv:0907:2021}  (\bibinfo{year}{2009}).

\bibitem[{\citenamefont{{Qiao} et~al.}(2010)\citenamefont{{Qiao}, {Yang},
  {Feng}, {Tse}, {Ding}, {Yao}, {Wang}, and {Niu}}}]{niuqian2010}
\bibinfo{author}{\bibfnamefont{Z.}~\bibnamefont{{Qiao}}},
  \bibinfo{author}{\bibfnamefont{S.~A.} \bibnamefont{{Yang}}},
  \bibinfo{author}{\bibfnamefont{W.}~\bibnamefont{{Feng}}},
  \bibinfo{author}{\bibfnamefont{W.}~\bibnamefont{{Tse}}},
  \bibinfo{author}{\bibfnamefont{J.}~\bibnamefont{{Ding}}},
  \bibinfo{author}{\bibfnamefont{Y.}~\bibnamefont{{Yao}}},
  \bibinfo{author}{\bibfnamefont{J.}~\bibnamefont{{Wang}}}, \bibnamefont{and}
  \bibinfo{author}{\bibfnamefont{Q.}~\bibnamefont{{Niu}}},
  \emph{\bibinfo{title}{{Chern Number Creation in Graphene from Rashba and
  Exchange Effects}}}, \bibinfo{howpublished}{arXiv:1005.1672}
  (\bibinfo{year}{2010}).

\bibitem[{\citenamefont{Tomizawa and Kontani}(2009)}]{tomizawa2009}
\bibinfo{author}{\bibfnamefont{T.}~\bibnamefont{Tomizawa}} \bibnamefont{and}
  \bibinfo{author}{\bibfnamefont{H.}~\bibnamefont{Kontani}},
  \bibinfo{journal}{Phys. Rev. B} \textbf{\bibinfo{volume}{80}},
  \bibinfo{pages}{100401} (\bibinfo{year}{2009}).

\bibitem[{\citenamefont{Thouless et~al.}(1982)\citenamefont{Thouless, Kohmoto,
  Nightingale, and den Nijs}}]{thouless1982}
\bibinfo{author}{\bibfnamefont{D.~J.} \bibnamefont{Thouless}},
  \bibinfo{author}{\bibfnamefont{M.}~\bibnamefont{Kohmoto}},
  \bibinfo{author}{\bibfnamefont{M.~P.} \bibnamefont{Nightingale}},
  \bibnamefont{and} \bibinfo{author}{\bibfnamefont{M.}~\bibnamefont{den Nijs}},
  \bibinfo{journal}{Phys. Rev. Lett.} \textbf{\bibinfo{volume}{49}},
  \bibinfo{pages}{405} (\bibinfo{year}{1982}).

\bibitem[{\citenamefont{Kohmoto}(1985)}]{kohmoto1985}
\bibinfo{author}{\bibfnamefont{M.}~\bibnamefont{Kohmoto}},
  \bibinfo{journal}{Ann. Phys.} \textbf{\bibinfo{volume}{160}},
  \bibinfo{pages}{296} (\bibinfo{year}{1985}).

\bibitem[{\citenamefont{Jackiw}(1984)}]{jackiw1984}
\bibinfo{author}{\bibfnamefont{R.}~\bibnamefont{Jackiw}},
  \bibinfo{journal}{Phys. Rev. D} \textbf{\bibinfo{volume}{29}},
  \bibinfo{pages}{2375} (\bibinfo{year}{1984}).

\bibitem[{\citenamefont{Fradkin et~al.}(1986)\citenamefont{Fradkin, Dagotto,
  and Boyanovsky}}]{fradkin1986}
\bibinfo{author}{\bibfnamefont{E.}~\bibnamefont{Fradkin}},
  \bibinfo{author}{\bibfnamefont{E.}~\bibnamefont{Dagotto}}, \bibnamefont{and}
  \bibinfo{author}{\bibfnamefont{D.}~\bibnamefont{Boyanovsky}},
  \bibinfo{journal}{Phys. Rev. Lett.} \textbf{\bibinfo{volume}{57}},
  \bibinfo{pages}{2967} (\bibinfo{year}{1986}).

\bibitem[{\citenamefont{Haldane}(1988)}]{haldane1988}
\bibinfo{author}{\bibfnamefont{F.~D.~M.} \bibnamefont{Haldane}},
  \bibinfo{journal}{Phys. Rev. Lett.} \textbf{\bibinfo{volume}{61}},
  \bibinfo{pages}{2015} (\bibinfo{year}{1988}).

\bibitem[{\citenamefont{Dyakonov and Perel}(1971)}]{dyakonov1971}
\bibinfo{author}{\bibfnamefont{M.~I.} \bibnamefont{Dyakonov}} \bibnamefont{and}
  \bibinfo{author}{\bibfnamefont{V.~I.} \bibnamefont{Perel}},
  \bibinfo{journal}{Physics Letters A} \textbf{\bibinfo{volume}{35}},
  \bibinfo{pages}{459} (\bibinfo{year}{1971}).

\bibitem[{\citenamefont{Hirsch}(1999)}]{hirsch1989}
\bibinfo{author}{\bibfnamefont{J.~E.} \bibnamefont{Hirsch}},
  \bibinfo{journal}{Phys. Rev. Lett.} \textbf{\bibinfo{volume}{83}},
  \bibinfo{pages}{1834} (\bibinfo{year}{1999}).

\bibitem[{\citenamefont{Sinova et~al.}(2004)\citenamefont{Sinova, Culcer, Niu,
  Sinitsyn, Jungwirth, and MacDonald}}]{sinova2003}
\bibinfo{author}{\bibfnamefont{J.}~\bibnamefont{Sinova}},
  \bibinfo{author}{\bibfnamefont{D.}~\bibnamefont{Culcer}},
  \bibinfo{author}{\bibfnamefont{Q.}~\bibnamefont{Niu}},
  \bibinfo{author}{\bibfnamefont{N.~A.} \bibnamefont{Sinitsyn}},
  \bibinfo{author}{\bibfnamefont{T.}~\bibnamefont{Jungwirth}},
  \bibnamefont{and} \bibinfo{author}{\bibfnamefont{A.~H.}
  \bibnamefont{MacDonald}}, \bibinfo{journal}{Phys. Rev. Lett.}
  \textbf{\bibinfo{volume}{92}}, \bibinfo{pages}{126603}
  (\bibinfo{year}{2004}).

\bibitem[{\citenamefont{Murakami et~al.}(2003)\citenamefont{Murakami, Nagaosa,
  and Zhang}}]{murakami2003}
\bibinfo{author}{\bibfnamefont{S.}~\bibnamefont{Murakami}},
  \bibinfo{author}{\bibfnamefont{N.}~\bibnamefont{Nagaosa}}, \bibnamefont{and}
  \bibinfo{author}{\bibfnamefont{S.-C.} \bibnamefont{Zhang}},
  \bibinfo{journal}{Science} \textbf{\bibinfo{volume}{301}},
  \bibinfo{pages}{1348} (\bibinfo{year}{2003}).

\bibitem[{\citenamefont{Kato et~al.}(2004)\citenamefont{Kato, Myers, Gossard,
  and Awschalom}}]{kato2004}
\bibinfo{author}{\bibfnamefont{Y.~K.} \bibnamefont{Kato}},
  \bibinfo{author}{\bibfnamefont{R.~C.} \bibnamefont{Myers}},
  \bibinfo{author}{\bibfnamefont{A.~C.} \bibnamefont{Gossard}},
  \bibnamefont{and} \bibinfo{author}{\bibfnamefont{D.~D.}
  \bibnamefont{Awschalom}}, \bibinfo{journal}{Science}
  \textbf{\bibinfo{volume}{306}}, \bibinfo{pages}{1910} (\bibinfo{year}{2004}).

\bibitem[{\citenamefont{Wunderlich et~al.}(2005)\citenamefont{Wunderlich,
  Kaestner, Sinova, and Jungwirth}}]{wunderlich2005}
\bibinfo{author}{\bibfnamefont{J.}~\bibnamefont{Wunderlich}},
  \bibinfo{author}{\bibfnamefont{B.}~\bibnamefont{Kaestner}},
  \bibinfo{author}{\bibfnamefont{J.}~\bibnamefont{Sinova}}, \bibnamefont{and}
  \bibinfo{author}{\bibfnamefont{T.}~\bibnamefont{Jungwirth}},
  \bibinfo{journal}{Phys. Rev. Lett.} \textbf{\bibinfo{volume}{94}},
  \bibinfo{pages}{047204} (\bibinfo{year}{2005}).

\bibitem[{\citenamefont{Bernevig et~al.}(2006)\citenamefont{Bernevig, Hughes,
  and Zhang}}]{bernevig2006a}
\bibinfo{author}{\bibfnamefont{B.~A.} \bibnamefont{Bernevig}},
  \bibinfo{author}{\bibfnamefont{T.~L.} \bibnamefont{Hughes}},
  \bibnamefont{and} \bibinfo{author}{\bibfnamefont{S.-C.} \bibnamefont{Zhang}},
  \bibinfo{journal}{Science} \textbf{\bibinfo{volume}{314}},
  \bibinfo{pages}{1757} (\bibinfo{year}{2006}).

\bibitem[{\citenamefont{Qi et~al.}(2008)\citenamefont{Qi, Hughes, and
  Zhang}}]{qi2008a}
\bibinfo{author}{\bibfnamefont{X.-L.} \bibnamefont{Qi}},
  \bibinfo{author}{\bibfnamefont{T.~L.} \bibnamefont{Hughes}},
  \bibnamefont{and} \bibinfo{author}{\bibfnamefont{S.-C.} \bibnamefont{Zhang}},
  \bibinfo{journal}{Phys. Rev. B} \textbf{\bibinfo{volume}{78}},
  \bibinfo{pages}{195424} (\bibinfo{year}{2008}).

\bibitem[{\citenamefont{Kane and Mele}(2005)}]{kane2005}
\bibinfo{author}{\bibfnamefont{C.~L.} \bibnamefont{Kane}} \bibnamefont{and}
  \bibinfo{author}{\bibfnamefont{E.~J.} \bibnamefont{Mele}},
  \bibinfo{journal}{Phys. Rev. Lett.} \textbf{\bibinfo{volume}{95}},
  \bibinfo{pages}{146802} (\bibinfo{year}{2005}).

\bibitem[{\citenamefont{Sheng et~al.}(2006)\citenamefont{Sheng, Weng, Sheng,
  and Haldane}}]{sheng2006}
\bibinfo{author}{\bibfnamefont{D.~N.} \bibnamefont{Sheng}},
  \bibinfo{author}{\bibfnamefont{Z.~Y.} \bibnamefont{Weng}},
  \bibinfo{author}{\bibfnamefont{L.}~\bibnamefont{Sheng}}, \bibnamefont{and}
  \bibinfo{author}{\bibfnamefont{F.~D.~M.} \bibnamefont{Haldane}},
  \bibinfo{journal}{Phys. Rev. Lett.} \textbf{\bibinfo{volume}{97}},
  \bibinfo{pages}{036808} (\bibinfo{year}{2006}).

\bibitem[{\citenamefont{Moore and Balents}(2007)}]{moore2007}
\bibinfo{author}{\bibfnamefont{J.~E.} \bibnamefont{Moore}} \bibnamefont{and}
  \bibinfo{author}{\bibfnamefont{L.}~\bibnamefont{Balents}},
  \bibinfo{journal}{Phys. Rev. B} \textbf{\bibinfo{volume}{75}},
  \bibinfo{pages}{121306} (\bibinfo{year}{2007}).

\bibitem[{\citenamefont{Roy}(2009)}]{roy2009}
\bibinfo{author}{\bibfnamefont{R.}~\bibnamefont{Roy}}, \bibinfo{journal}{Phys.
  Rev. B} \textbf{\bibinfo{volume}{79}}, \bibinfo{pages}{195321}
  (\bibinfo{year}{2009}).

\bibitem[{\citenamefont{Fu et~al.}(2007)\citenamefont{Fu, Kane, and
  Mele}}]{fu2007}
\bibinfo{author}{\bibfnamefont{L.}~\bibnamefont{Fu}},
  \bibinfo{author}{\bibfnamefont{C.~L.} \bibnamefont{Kane}}, \bibnamefont{and}
  \bibinfo{author}{\bibfnamefont{E.~J.} \bibnamefont{Mele}},
  \bibinfo{journal}{Phys. Rev. Lett.} \textbf{\bibinfo{volume}{98}},
  \bibinfo{pages}{106803} (\bibinfo{year}{2007}).

\bibitem[{\citenamefont{Fu and Kane}(2007)}]{fu2007a}
\bibinfo{author}{\bibfnamefont{L.}~\bibnamefont{Fu}} \bibnamefont{and}
  \bibinfo{author}{\bibfnamefont{C.~L.} \bibnamefont{Kane}},
  \bibinfo{journal}{Phys. Rev. B} \textbf{\bibinfo{volume}{76}},
  \bibinfo{pages}{045302} (\bibinfo{year}{2007}).

\bibitem[{\citenamefont{Zhang et~al.}(2009{\natexlab{a}})\citenamefont{Zhang,
  Liu, Qi, Dai, Fang, and Zhang}}]{zhang2009}
\bibinfo{author}{\bibfnamefont{H.}~\bibnamefont{Zhang}},
  \bibinfo{author}{\bibfnamefont{C.-X.} \bibnamefont{Liu}},
  \bibinfo{author}{\bibfnamefont{X.-L.} \bibnamefont{Qi}},
  \bibinfo{author}{\bibfnamefont{X.}~\bibnamefont{Dai}},
  \bibinfo{author}{\bibfnamefont{Z.}~\bibnamefont{Fang}}, \bibnamefont{and}
  \bibinfo{author}{\bibfnamefont{S.-C.} \bibnamefont{Zhang}},
  \bibinfo{journal}{Nat Phys} \textbf{\bibinfo{volume}{5}},
  \bibinfo{pages}{438} (\bibinfo{year}{2009}{\natexlab{a}}).

\bibitem[{\citenamefont{Wu et~al.}(2006)\citenamefont{Wu, Bernevig, and
  Zhang}}]{wu2006}
\bibinfo{author}{\bibfnamefont{C.}~\bibnamefont{Wu}},
  \bibinfo{author}{\bibfnamefont{B.~A.} \bibnamefont{Bernevig}},
  \bibnamefont{and} \bibinfo{author}{\bibfnamefont{S.-C.} \bibnamefont{Zhang}},
  \bibinfo{journal}{Phys. Rev. Lett.} \textbf{\bibinfo{volume}{96}},
  \bibinfo{pages}{106401} (\bibinfo{year}{2006}).

\bibitem[{\citenamefont{Xu and Moore}(2006)}]{xu2006}
\bibinfo{author}{\bibfnamefont{C.}~\bibnamefont{Xu}} \bibnamefont{and}
  \bibinfo{author}{\bibfnamefont{J.~E.} \bibnamefont{Moore}},
  \bibinfo{journal}{Phys. Rev. B} \textbf{\bibinfo{volume}{73}},
  \bibinfo{pages}{064417} (\bibinfo{year}{2006}).

\bibitem[{\citenamefont{K\"onig et~al.}(2007)\citenamefont{K\"onig, Wiedmann,
  Br\"une, Roth, Buhmann, Molenkamp, Qi, and Zhang}}]{konig2007}
\bibinfo{author}{\bibfnamefont{M.}~\bibnamefont{K\"onig}},
  \bibinfo{author}{\bibfnamefont{S.}~\bibnamefont{Wiedmann}},
  \bibinfo{author}{\bibfnamefont{C.}~\bibnamefont{Br\"une}},
  \bibinfo{author}{\bibfnamefont{A.}~\bibnamefont{Roth}},
  \bibinfo{author}{\bibfnamefont{H.}~\bibnamefont{Buhmann}},
  \bibinfo{author}{\bibfnamefont{L.~W.} \bibnamefont{Molenkamp}},
  \bibinfo{author}{\bibfnamefont{X.~L.} \bibnamefont{Qi}}, \bibnamefont{and}
  \bibinfo{author}{\bibfnamefont{S.~C.} \bibnamefont{Zhang}},
  \bibinfo{journal}{Science} \textbf{\bibinfo{volume}{318}},
  \bibinfo{pages}{766} (\bibinfo{year}{2007}).

\bibitem[{\citenamefont{Hsieh et~al.}(2008)\citenamefont{Hsieh, Qian, Wray,
  Xia, Hor, Cava, and Hasan}}]{hsieh2008}
\bibinfo{author}{\bibfnamefont{D.}~\bibnamefont{Hsieh}},
  \bibinfo{author}{\bibfnamefont{D.}~\bibnamefont{Qian}},
  \bibinfo{author}{\bibfnamefont{L.}~\bibnamefont{Wray}},
  \bibinfo{author}{\bibfnamefont{Y.}~\bibnamefont{Xia}},
  \bibinfo{author}{\bibfnamefont{Y.~S.} \bibnamefont{Hor}},
  \bibinfo{author}{\bibfnamefont{R.~J.} \bibnamefont{Cava}}, \bibnamefont{and}
  \bibinfo{author}{\bibfnamefont{M.~Z.} \bibnamefont{Hasan}},
  \bibinfo{journal}{Nature} \textbf{\bibinfo{volume}{452}},
  \bibinfo{pages}{970} (\bibinfo{year}{2008}).

\bibitem[{\citenamefont{Hsieh et~al.}(2009)\citenamefont{Hsieh, Xia, Wray,
  Qian, Pal, Dil, Osterwalder, Meier, Bihlmayer, Kane et~al.}}]{hsieh2009}
\bibinfo{author}{\bibfnamefont{D.}~\bibnamefont{Hsieh}},
  \bibinfo{author}{\bibfnamefont{Y.}~\bibnamefont{Xia}},
  \bibinfo{author}{\bibfnamefont{L.}~\bibnamefont{Wray}},
  \bibinfo{author}{\bibfnamefont{D.}~\bibnamefont{Qian}},
  \bibinfo{author}{\bibfnamefont{A.}~\bibnamefont{Pal}},
  \bibinfo{author}{\bibfnamefont{J.~H.} \bibnamefont{Dil}},
  \bibinfo{author}{\bibfnamefont{J.}~\bibnamefont{Osterwalder}},
  \bibinfo{author}{\bibfnamefont{F.}~\bibnamefont{Meier}},
  \bibinfo{author}{\bibfnamefont{G.}~\bibnamefont{Bihlmayer}},
  \bibinfo{author}{\bibfnamefont{C.~L.} \bibnamefont{Kane}},
  \bibnamefont{et~al.}, \bibinfo{journal}{Science}
  \textbf{\bibinfo{volume}{323}}, \bibinfo{pages}{919} (\bibinfo{year}{2009}).

\bibitem[{\citenamefont{Xia et~al.}(2009)\citenamefont{Xia, Qian, Hsieh, Wray,
  Pal, Lin, Bansil, Grauer, Hor, Cava et~al.}}]{xia2009}
\bibinfo{author}{\bibfnamefont{Y.}~\bibnamefont{Xia}},
  \bibinfo{author}{\bibfnamefont{D.}~\bibnamefont{Qian}},
  \bibinfo{author}{\bibfnamefont{D.}~\bibnamefont{Hsieh}},
  \bibinfo{author}{\bibfnamefont{L.}~\bibnamefont{Wray}},
  \bibinfo{author}{\bibfnamefont{A.}~\bibnamefont{Pal}},
  \bibinfo{author}{\bibfnamefont{H.}~\bibnamefont{Lin}},
  \bibinfo{author}{\bibfnamefont{A.}~\bibnamefont{Bansil}},
  \bibinfo{author}{\bibfnamefont{D.}~\bibnamefont{Grauer}},
  \bibinfo{author}{\bibfnamefont{Y.~S.} \bibnamefont{Hor}},
  \bibinfo{author}{\bibfnamefont{R.~J.} \bibnamefont{Cava}},
  \bibnamefont{et~al.}, \bibinfo{journal}{Nat Phys}
  \textbf{\bibinfo{volume}{5}}, \bibinfo{pages}{398} (\bibinfo{year}{2009}).

\bibitem[{\citenamefont{Chen et~al.}(2009)\citenamefont{Chen, Analytis, Chu,
  Liu, Mo, Qi, Zhang, Lu, Dai, Fang et~al.}}]{chen2009}
\bibinfo{author}{\bibfnamefont{Y.~L.} \bibnamefont{Chen}},
  \bibinfo{author}{\bibfnamefont{J.~G.} \bibnamefont{Analytis}},
  \bibinfo{author}{\bibfnamefont{J.-H.} \bibnamefont{Chu}},
  \bibinfo{author}{\bibfnamefont{Z.~K.} \bibnamefont{Liu}},
  \bibinfo{author}{\bibfnamefont{S.-K.} \bibnamefont{Mo}},
  \bibinfo{author}{\bibfnamefont{X.~L.} \bibnamefont{Qi}},
  \bibinfo{author}{\bibfnamefont{H.~J.} \bibnamefont{Zhang}},
  \bibinfo{author}{\bibfnamefont{D.~H.} \bibnamefont{Lu}},
  \bibinfo{author}{\bibfnamefont{X.}~\bibnamefont{Dai}},
  \bibinfo{author}{\bibfnamefont{Z.}~\bibnamefont{Fang}}, \bibnamefont{et~al.},
  \bibinfo{journal}{Science} \textbf{\bibinfo{volume}{325}},
  \bibinfo{pages}{178} (\bibinfo{year}{2009}).

\bibitem[{\citenamefont{Roushan et~al.}(2009)\citenamefont{Roushan, Seo,
  Parker, Hor, Hsieh, Qian, Richardella, Hasan, Cava, and
  Yazdani}}]{roushan2009}
\bibinfo{author}{\bibfnamefont{P.}~\bibnamefont{Roushan}},
  \bibinfo{author}{\bibfnamefont{J.}~\bibnamefont{Seo}},
  \bibinfo{author}{\bibfnamefont{C.~V.} \bibnamefont{Parker}},
  \bibinfo{author}{\bibfnamefont{Y.~S.} \bibnamefont{Hor}},
  \bibinfo{author}{\bibfnamefont{D.}~\bibnamefont{Hsieh}},
  \bibinfo{author}{\bibfnamefont{D.}~\bibnamefont{Qian}},
  \bibinfo{author}{\bibfnamefont{A.}~\bibnamefont{Richardella}},
  \bibinfo{author}{\bibfnamefont{M.~Z.} \bibnamefont{Hasan}},
  \bibinfo{author}{\bibfnamefont{R.~J.} \bibnamefont{Cava}}, \bibnamefont{and}
  \bibinfo{author}{\bibfnamefont{A.}~\bibnamefont{Yazdani}},
  \bibinfo{journal}{Nature} \textbf{\bibinfo{volume}{460}},
  \bibinfo{pages}{1106} (\bibinfo{year}{2009}).

\bibitem[{\citenamefont{Alpichshev et~al.}(2010)\citenamefont{Alpichshev,
  Analytis, Chu, Fisher, Chen, Shen, Fang, and Kapitulnik}}]{alpichshev2010}
\bibinfo{author}{\bibfnamefont{Z.}~\bibnamefont{Alpichshev}},
  \bibinfo{author}{\bibfnamefont{J.~G.} \bibnamefont{Analytis}},
  \bibinfo{author}{\bibfnamefont{J.-H.} \bibnamefont{Chu}},
  \bibinfo{author}{\bibfnamefont{I.~R.} \bibnamefont{Fisher}},
  \bibinfo{author}{\bibfnamefont{Y.~L.} \bibnamefont{Chen}},
  \bibinfo{author}{\bibfnamefont{Z.~X.} \bibnamefont{Shen}},
  \bibinfo{author}{\bibfnamefont{A.}~\bibnamefont{Fang}}, \bibnamefont{and}
  \bibinfo{author}{\bibfnamefont{A.}~\bibnamefont{Kapitulnik}},
  \bibinfo{journal}{Phys. Rev. Lett.} \textbf{\bibinfo{volume}{104}},
  \bibinfo{pages}{016401} (\bibinfo{year}{2010}).

\bibitem[{\citenamefont{Zhang et~al.}(2009{\natexlab{b}})\citenamefont{Zhang,
  Cheng, Chen, Jia, Ma, He, Wang, Zhang, Dai, Fang
  et~al.}}]{zhang_chen_xue2009}
\bibinfo{author}{\bibfnamefont{T.}~\bibnamefont{Zhang}},
  \bibinfo{author}{\bibfnamefont{P.}~\bibnamefont{Cheng}},
  \bibinfo{author}{\bibfnamefont{X.}~\bibnamefont{Chen}},
  \bibinfo{author}{\bibfnamefont{J.-F.} \bibnamefont{Jia}},
  \bibinfo{author}{\bibfnamefont{X.}~\bibnamefont{Ma}},
  \bibinfo{author}{\bibfnamefont{K.}~\bibnamefont{He}},
  \bibinfo{author}{\bibfnamefont{L.}~\bibnamefont{Wang}},
  \bibinfo{author}{\bibfnamefont{H.}~\bibnamefont{Zhang}},
  \bibinfo{author}{\bibfnamefont{X.}~\bibnamefont{Dai}},
  \bibinfo{author}{\bibfnamefont{Z.}~\bibnamefont{Fang}}, \bibnamefont{et~al.},
  \bibinfo{journal}{Phys. Rev. Lett.} \textbf{\bibinfo{volume}{103}},
  \bibinfo{pages}{266803} (\bibinfo{year}{2009}{\natexlab{b}}).

\bibitem[{\citenamefont{Qi et~al.}(2006)\citenamefont{Qi, Wu, and
  Zhang}}]{qi2006a}
\bibinfo{author}{\bibfnamefont{X.-L.} \bibnamefont{Qi}},
  \bibinfo{author}{\bibfnamefont{Y.-S.} \bibnamefont{Wu}}, \bibnamefont{and}
  \bibinfo{author}{\bibfnamefont{S.-C.} \bibnamefont{Zhang}},
  \bibinfo{journal}{Phys. Rev. B} \textbf{\bibinfo{volume}{74}},
  \bibinfo{pages}{085308} (\bibinfo{year}{2006}).

\bibitem[{\citenamefont{Liu et~al.}(2008)\citenamefont{Liu, Qi, Dai, Fang, and
  Zhang}}]{liu2008}
\bibinfo{author}{\bibfnamefont{C.-X.} \bibnamefont{Liu}},
  \bibinfo{author}{\bibfnamefont{X.-L.} \bibnamefont{Qi}},
  \bibinfo{author}{\bibfnamefont{X.}~\bibnamefont{Dai}},
  \bibinfo{author}{\bibfnamefont{Z.}~\bibnamefont{Fang}}, \bibnamefont{and}
  \bibinfo{author}{\bibfnamefont{S.-C.} \bibnamefont{Zhang}},
  \bibinfo{journal}{Phys. Rev. Lett.} \textbf{\bibinfo{volume}{101}},
  \bibinfo{pages}{146802} (\bibinfo{year}{2008}).

\bibitem[{\citenamefont{Onoda and Nagaosa}(2003)}]{onoda2003a}
\bibinfo{author}{\bibfnamefont{M.}~\bibnamefont{Onoda}} \bibnamefont{and}
  \bibinfo{author}{\bibfnamefont{N.}~\bibnamefont{Nagaosa}},
  \bibinfo{journal}{Phys. Rev. Lett.} \textbf{\bibinfo{volume}{90}},
  \bibinfo{pages}{206601} (\bibinfo{year}{2003}).

\bibitem[{\citenamefont{Yu et~al.}(2010)\citenamefont{Yu, Zhang, Zhang, Zhang,
  Dai, and Fang}}]{yu2010}
\bibinfo{author}{\bibfnamefont{R.}~\bibnamefont{Yu}},
  \bibinfo{author}{\bibfnamefont{W.}~\bibnamefont{Zhang}},
  \bibinfo{author}{\bibfnamefont{H.-J.} \bibnamefont{Zhang}},
  \bibinfo{author}{\bibfnamefont{S.-C.} \bibnamefont{Zhang}},
  \bibinfo{author}{\bibfnamefont{X.}~\bibnamefont{Dai}}, \bibnamefont{and}
  \bibinfo{author}{\bibfnamefont{Z.}~\bibnamefont{Fang}},
  \bibinfo{journal}{Science} \textbf{\bibinfo{volume}{329}},
  \bibinfo{pages}{61} (\bibinfo{year}{2010}).

\bibitem[{\citenamefont{Ho and Yip}(2000)}]{ho2000}
\bibinfo{author}{\bibfnamefont{T.~L.} \bibnamefont{Ho}} \bibnamefont{and}
  \bibinfo{author}{\bibfnamefont{S.~K.} \bibnamefont{Yip}},
  \bibinfo{journal}{Phys. Rev. Lett.} \textbf{\bibinfo{volume}{84}},
  \bibinfo{pages}{4031} (\bibinfo{year}{2000}).

\bibitem[{\citenamefont{Scarola and Sarma}(2007)}]{scarola2007}
\bibinfo{author}{\bibfnamefont{V.~W.} \bibnamefont{Scarola}} \bibnamefont{and}
  \bibinfo{author}{\bibfnamefont{S.~D.} \bibnamefont{Sarma}},
  \bibinfo{journal}{Phys. Rev. Lett.} \textbf{\bibinfo{volume}{98}},
  \bibinfo{pages}{210403} (\bibinfo{year}{2007}).

\bibitem[{\citenamefont{Umucalilar
  et~al.}(2008{\natexlab{a}})\citenamefont{Umucalilar, Zhai, and
  Oktel}}]{umucalilar2008}
\bibinfo{author}{\bibfnamefont{R.~O.} \bibnamefont{Umucalilar}},
  \bibinfo{author}{\bibfnamefont{H.}~\bibnamefont{Zhai}}, \bibnamefont{and}
  \bibinfo{author}{\bibfnamefont{M.~O.} \bibnamefont{Oktel}},
  \bibinfo{journal}{Phys. Rev. Lett.} \textbf{\bibinfo{volume}{100}},
  \bibinfo{pages}{070402} (\bibinfo{year}{2008}{\natexlab{a}}).

\bibitem[{\citenamefont{Zhu et~al.}(2006)\citenamefont{Zhu, Fu, Wu, Zhang, and
  Duan}}]{zhu2006}
\bibinfo{author}{\bibfnamefont{S.-L.} \bibnamefont{Zhu}},
  \bibinfo{author}{\bibfnamefont{H.}~\bibnamefont{Fu}},
  \bibinfo{author}{\bibfnamefont{C.~J.} \bibnamefont{Wu}},
  \bibinfo{author}{\bibfnamefont{S.~C.} \bibnamefont{Zhang}}, \bibnamefont{and}
  \bibinfo{author}{\bibfnamefont{L.~M.} \bibnamefont{Duan}},
  \bibinfo{journal}{Phys. Rev. Lett.} \textbf{\bibinfo{volume}{97}},
  \bibinfo{pages}{240401} (\bibinfo{year}{2006}).

\bibitem[{\citenamefont{Zhang}(2010)}]{zhangcw2010}
\bibinfo{author}{\bibfnamefont{C.}~\bibnamefont{Zhang}},
  \bibinfo{journal}{arXiv:1004.4231}  (\bibinfo{year}{2010}).

\bibitem[{\citenamefont{Shao et~al.}(2008)\citenamefont{Shao, Zhu, Sheng, Xing,
  and Wang}}]{shao2008}
\bibinfo{author}{\bibfnamefont{L.~B.} \bibnamefont{Shao}},
  \bibinfo{author}{\bibfnamefont{S.-L.} \bibnamefont{Zhu}},
  \bibinfo{author}{\bibfnamefont{L.}~\bibnamefont{Sheng}},
  \bibinfo{author}{\bibfnamefont{D.~Y.} \bibnamefont{Xing}}, \bibnamefont{and}
  \bibinfo{author}{\bibfnamefont{Z.~D.} \bibnamefont{Wang}},
  \bibinfo{journal}{Phys. Rev. Lett.} \textbf{\bibinfo{volume}{101}},
  \bibinfo{pages}{246810} (\bibinfo{year}{2008}).

\bibitem[{\citenamefont{Stanescu et~al.}(2010)\citenamefont{Stanescu, Galitski,
  and Das~Sarma}}]{stanescu2010}
\bibinfo{author}{\bibfnamefont{T.~D.} \bibnamefont{Stanescu}},
  \bibinfo{author}{\bibfnamefont{V.}~\bibnamefont{Galitski}}, \bibnamefont{and}
  \bibinfo{author}{\bibfnamefont{S.}~\bibnamefont{Das~Sarma}},
  \bibinfo{journal}{Phys. Rev. A} \textbf{\bibinfo{volume}{82}},
  \bibinfo{pages}{013608} (\bibinfo{year}{2010}).

\bibitem[{\citenamefont{Liu et~al.}(2010)\citenamefont{Liu, Liu, Wu, and
  Sinova}}]{liu_liu2010}
\bibinfo{author}{\bibfnamefont{X.-J.} \bibnamefont{Liu}},
  \bibinfo{author}{\bibfnamefont{X.}~\bibnamefont{Liu}},
  \bibinfo{author}{\bibfnamefont{C.}~\bibnamefont{Wu}}, \bibnamefont{and}
  \bibinfo{author}{\bibfnamefont{J.}~\bibnamefont{Sinova}},
  \bibinfo{journal}{Phys. Rev. A} \textbf{\bibinfo{volume}{81}},
  \bibinfo{pages}{033622} (\bibinfo{year}{2010}).

\bibitem[{\citenamefont{Goldman et~al.}(2010)\citenamefont{Goldman, Satija,
  Nikolic, Bermudez, Martin-Delgado, Lewenstein, and Spielman}}]{goldman2010}
\bibinfo{author}{\bibfnamefont{N.}~\bibnamefont{Goldman}},
  \bibinfo{author}{\bibfnamefont{I.}~\bibnamefont{Satija}},
  \bibinfo{author}{\bibfnamefont{P.}~\bibnamefont{Nikolic}},
  \bibinfo{author}{\bibfnamefont{A.}~\bibnamefont{Bermudez}},
  \bibinfo{author}{\bibfnamefont{M.~A.} \bibnamefont{Martin-Delgado}},
  \bibinfo{author}{\bibfnamefont{M.}~\bibnamefont{Lewenstein}},
  \bibnamefont{and} \bibinfo{author}{\bibfnamefont{I.~B.}
  \bibnamefont{Spielman}}, \bibinfo{journal}{Phys. Rev. Lett.}
  \textbf{\bibinfo{volume}{105}}, \bibinfo{pages}{255302}
  (\bibinfo{year}{2010}).

\bibitem[{\citenamefont{Wu}(2008{\natexlab{a}})}]{wu2008}
\bibinfo{author}{\bibfnamefont{C.}~\bibnamefont{Wu}}, \bibinfo{journal}{Phys.
  Rev. Lett.} \textbf{\bibinfo{volume}{101}}, \bibinfo{pages}{186807}
  (\bibinfo{year}{2008}{\natexlab{a}}).

\bibitem[{\citenamefont{Grynberg et~al.}(1993)\citenamefont{Grynberg, Lounis,
  Verkerk, Courtois, and Salomon}}]{grynberg1993}
\bibinfo{author}{\bibfnamefont{G.}~\bibnamefont{Grynberg}},
  \bibinfo{author}{\bibfnamefont{B.}~\bibnamefont{Lounis}},
  \bibinfo{author}{\bibfnamefont{P.}~\bibnamefont{Verkerk}},
  \bibinfo{author}{\bibfnamefont{J.~Y.} \bibnamefont{Courtois}},
  \bibnamefont{and} \bibinfo{author}{\bibfnamefont{C.}~\bibnamefont{Salomon}},
  \bibinfo{journal}{Phys. Rev. Lett.} \textbf{\bibinfo{volume}{70}},
  \bibinfo{pages}{2249} (\bibinfo{year}{1993}).

\bibitem[{\citenamefont{Soltan-Panahi et~al.}(2010)\citenamefont{Soltan-Panahi,
  Struck, Hauke, Bick, Plenkers, Meineke, Becker, Windpassinger, Lewenstein,
  and Sengstock}}]{sengstock2010}
\bibinfo{author}{\bibfnamefont{P.}~\bibnamefont{Soltan-Panahi}},
  \bibinfo{author}{\bibfnamefont{J.}~\bibnamefont{Struck}},
  \bibinfo{author}{\bibfnamefont{P.}~\bibnamefont{Hauke}},
  \bibinfo{author}{\bibfnamefont{A.}~\bibnamefont{Bick}},
  \bibinfo{author}{\bibfnamefont{W.}~\bibnamefont{Plenkers}},
  \bibinfo{author}{\bibfnamefont{G.}~\bibnamefont{Meineke}},
  \bibinfo{author}{\bibfnamefont{C.}~\bibnamefont{Becker}},
  \bibinfo{author}{\bibfnamefont{P.}~\bibnamefont{Windpassinger}},
  \bibinfo{author}{\bibfnamefont{M.}~\bibnamefont{Lewenstein}},
  \bibnamefont{and}
  \bibinfo{author}{\bibfnamefont{K.}~\bibnamefont{Sengstock}},
  \bibinfo{journal}{arXiv:1005.1276}  (\bibinfo{year}{2010}).

\bibitem[{\citenamefont{Liu and Wu}(2006)}]{liu2006}
\bibinfo{author}{\bibfnamefont{W.~V.} \bibnamefont{Liu}} \bibnamefont{and}
  \bibinfo{author}{\bibfnamefont{C.}~\bibnamefont{Wu}}, \bibinfo{journal}{Phys.
  Rev. A} \textbf{\bibinfo{volume}{74}}, \bibinfo{pages}{013607}
  (\bibinfo{year}{2006}).

\bibitem[{\citenamefont{Stojanovi\'c et~al.}(2008)\citenamefont{Stojanovi\'c,
  Wu, Liu, and Das~Sarma}}]{stojanovic2008}
\bibinfo{author}{\bibfnamefont{V.~M.} \bibnamefont{Stojanovi\'c}},
  \bibinfo{author}{\bibfnamefont{C.}~\bibnamefont{Wu}},
  \bibinfo{author}{\bibfnamefont{W.~V.} \bibnamefont{Liu}}, \bibnamefont{and}
  \bibinfo{author}{\bibfnamefont{S.}~\bibnamefont{Das~Sarma}},
  \bibinfo{journal}{Phys. Rev. Lett.} \textbf{\bibinfo{volume}{101}},
  \bibinfo{pages}{125301} (\bibinfo{year}{2008}).

\bibitem[{\citenamefont{Wu}(2009)}]{wu2009}
\bibinfo{author}{\bibfnamefont{C.}~\bibnamefont{Wu}}, \bibinfo{journal}{Modern
  Physics Letters B} \textbf{\bibinfo{volume}{23}}, \bibinfo{pages}{1}
  (\bibinfo{year}{2009}).

\bibitem[{\citenamefont{Isacsson and Girvin}(2005)}]{isacsson2005}
\bibinfo{author}{\bibfnamefont{A.}~\bibnamefont{Isacsson}} \bibnamefont{and}
  \bibinfo{author}{\bibfnamefont{S.~M.} \bibnamefont{Girvin}},
  \bibinfo{journal}{Phys. Rev. A} \textbf{\bibinfo{volume}{72}},
  \bibinfo{pages}{053604} (\bibinfo{year}{2005}).

\bibitem[{\citenamefont{Kuklov}(2006)}]{kuklov2006}
\bibinfo{author}{\bibfnamefont{A.~B.} \bibnamefont{Kuklov}},
  \bibinfo{journal}{Phys. Rev. Lett.} \textbf{\bibinfo{volume}{97}},
  \bibinfo{pages}{110405} (\bibinfo{year}{2006}).

\bibitem[{\citenamefont{M\"uller et~al.}(2007)\citenamefont{M\"uller,
  F\"olling, Widera, and Bloch}}]{mueller2007}
\bibinfo{author}{\bibfnamefont{T.}~\bibnamefont{M\"uller}},
  \bibinfo{author}{\bibfnamefont{S.}~\bibnamefont{F\"olling}},
  \bibinfo{author}{\bibfnamefont{A.}~\bibnamefont{Widera}}, \bibnamefont{and}
  \bibinfo{author}{\bibfnamefont{I.}~\bibnamefont{Bloch}},
  \bibinfo{journal}{Phys. Rev. Lett.} \textbf{\bibinfo{volume}{99}},
  \bibinfo{pages}{200405} (\bibinfo{year}{2007}).

\bibitem[{\citenamefont{Wirth et~al.}(2010)\citenamefont{Wirth, \"Olschl\"ager,
  and Hemmerich}}]{wirth2010}
\bibinfo{author}{\bibfnamefont{G.}~\bibnamefont{Wirth}},
  \bibinfo{author}{\bibfnamefont{M.}~\bibnamefont{\"Olschl\"ager}},
  \bibnamefont{and}
  \bibinfo{author}{\bibfnamefont{A.}~\bibnamefont{Hemmerich}},
  \bibinfo{journal}{arXiv:1006.0509}  (\bibinfo{year}{2010}).

\bibitem[{\citenamefont{Wu et~al.}(2007)\citenamefont{Wu, Bergman, Balents, and
  Das~Sarma}}]{wu2007}
\bibinfo{author}{\bibfnamefont{C.}~\bibnamefont{Wu}},
  \bibinfo{author}{\bibfnamefont{D.}~\bibnamefont{Bergman}},
  \bibinfo{author}{\bibfnamefont{L.}~\bibnamefont{Balents}}, \bibnamefont{and}
  \bibinfo{author}{\bibfnamefont{S.}~\bibnamefont{Das~Sarma}},
  \bibinfo{journal}{Phys. Rev. Lett.} \textbf{\bibinfo{volume}{99}},
  \bibinfo{pages}{070401} (\bibinfo{year}{2007}).

\bibitem[{\citenamefont{Wu}(2008{\natexlab{b}})}]{wu2008a}
\bibinfo{author}{\bibfnamefont{C.}~\bibnamefont{Wu}}, \bibinfo{journal}{Phys.
  Rev. Lett.} \textbf{\bibinfo{volume}{100}}, \bibinfo{pages}{200406}
  (\bibinfo{year}{2008}{\natexlab{b}}).

\bibitem[{\citenamefont{Wu and Das~Sarma}(2008)}]{wu2008b}
\bibinfo{author}{\bibfnamefont{C.}~\bibnamefont{Wu}} \bibnamefont{and}
  \bibinfo{author}{\bibfnamefont{S.}~\bibnamefont{Das~Sarma}},
  \bibinfo{journal}{Phys. Rev. B} \textbf{\bibinfo{volume}{77}},
  \bibinfo{pages}{235107} (\bibinfo{year}{2008}).

\bibitem[{\citenamefont{Zhang et~al.}(2010)\citenamefont{Zhang, Hung, and
  Wu}}]{zhang_hung_wu2010}
\bibinfo{author}{\bibfnamefont{S.}~\bibnamefont{Zhang}},
  \bibinfo{author}{\bibfnamefont{H.-h.} \bibnamefont{Hung}}, \bibnamefont{and}
  \bibinfo{author}{\bibfnamefont{C.}~\bibnamefont{Wu}}, \bibinfo{journal}{Phys.
  Rev. A} \textbf{\bibinfo{volume}{82}}, \bibinfo{pages}{053618}
  (\bibinfo{year}{2010}).

\bibitem[{\citenamefont{Lee et~al.}(2010)\citenamefont{Lee, Wu, and
  Das~Sarma}}]{lee2009}
\bibinfo{author}{\bibfnamefont{W.-C.} \bibnamefont{Lee}},
  \bibinfo{author}{\bibfnamefont{C.}~\bibnamefont{Wu}}, \bibnamefont{and}
  \bibinfo{author}{\bibfnamefont{S.}~\bibnamefont{Das~Sarma}},
  \bibinfo{journal}{Phys. Rev. A} \textbf{\bibinfo{volume}{82}},
  \bibinfo{pages}{053611} (\bibinfo{year}{2010}).

\bibitem[{\citenamefont{Hung et~al.}(2009)\citenamefont{Hung, Lee, and
  Wu}}]{hung2009}
\bibinfo{author}{\bibfnamefont{H.-h.} \bibnamefont{Hung}},
  \bibinfo{author}{\bibfnamefont{W.-C.} \bibnamefont{Lee}}, \bibnamefont{and}
  \bibinfo{author}{\bibfnamefont{C.}~\bibnamefont{Wu}},
  \bibinfo{journal}{arXiv:0910.0507}  (\bibinfo{year}{2009}).

\bibitem[{\citenamefont{Zhu et~al.}(2007)\citenamefont{Zhu, Wang, and
  Duan}}]{zhu2007}
\bibinfo{author}{\bibfnamefont{S.-L.} \bibnamefont{Zhu}},
  \bibinfo{author}{\bibfnamefont{B.}~\bibnamefont{Wang}}, \bibnamefont{and}
  \bibinfo{author}{\bibfnamefont{L.~M.} \bibnamefont{Duan}},
  \bibinfo{journal}{Phys. Rev. Lett.} \textbf{\bibinfo{volume}{98}},
  \bibinfo{pages}{260402} (\bibinfo{year}{2007}).

\bibitem[{\citenamefont{Lee et~al.}(2009)\citenamefont{Lee, Gr\'{e}maud, Han,
  Englert, and Miniatura}}]{kllee2009}
\bibinfo{author}{\bibfnamefont{K.~L.} \bibnamefont{Lee}},
  \bibinfo{author}{\bibfnamefont{B.}~\bibnamefont{Gr\'{e}maud}},
  \bibinfo{author}{\bibfnamefont{R.}~\bibnamefont{Han}},
  \bibinfo{author}{\bibfnamefont{B.-G.} \bibnamefont{Englert}},
  \bibnamefont{and}
  \bibinfo{author}{\bibfnamefont{C.}~\bibnamefont{Miniatura}},
  \bibinfo{journal}{Phys. Rev. A} \textbf{\bibinfo{volume}{80}},
  \bibinfo{pages}{043411} (\bibinfo{year}{2009}).

\bibitem[{\citenamefont{{Sun} et~al.}(2010)\citenamefont{{Sun}, {Gu},
  {Katsura}, and {Das Sarma}}}]{sun2010}
\bibinfo{author}{\bibfnamefont{K.}~\bibnamefont{{Sun}}},
  \bibinfo{author}{\bibfnamefont{Z.}~\bibnamefont{{Gu}}},
  \bibinfo{author}{\bibfnamefont{H.}~\bibnamefont{{Katsura}}},
  \bibnamefont{and} \bibinfo{author}{\bibfnamefont{S.}~\bibnamefont{{Das
  Sarma}}}, \bibinfo{journal}{ArXiv e-prints}  (\bibinfo{year}{2010}),
  \eprint{1012.5864}.

\bibitem[{\citenamefont{{Neupert} et~al.}(2010)\citenamefont{{Neupert},
  {Santos}, {Chamon}, and {Mudry}}}]{neupert2010}
\bibinfo{author}{\bibfnamefont{T.}~\bibnamefont{{Neupert}}},
  \bibinfo{author}{\bibfnamefont{L.}~\bibnamefont{{Santos}}},
  \bibinfo{author}{\bibfnamefont{C.}~\bibnamefont{{Chamon}}}, \bibnamefont{and}
  \bibinfo{author}{\bibfnamefont{C.}~\bibnamefont{{Mudry}}},
  \bibinfo{journal}{ArXiv e-prints}  (\bibinfo{year}{2010}),
  \eprint{1012.4723}.

\bibitem[{\citenamefont{{Tang} et~al.}(2010)\citenamefont{{Tang}, {Mei}, and
  {Wen}}}]{tang2010}
\bibinfo{author}{\bibfnamefont{E.}~\bibnamefont{{Tang}}},
  \bibinfo{author}{\bibfnamefont{J.}~\bibnamefont{{Mei}}}, \bibnamefont{and}
  \bibinfo{author}{\bibfnamefont{X.}~\bibnamefont{{Wen}}},
  \bibinfo{journal}{ArXiv e-prints}  (\bibinfo{year}{2010}),
  \eprint{1012.2930}.

\bibitem[{\citenamefont{Leggett}(2001)}]{leggett2001}
\bibinfo{author}{\bibfnamefont{A.~J.} \bibnamefont{Leggett}},
  \bibinfo{journal}{Rev. Mod. Phys.} \textbf{\bibinfo{volume}{73}},
  \bibinfo{pages}{307} (\bibinfo{year}{2001}).

\bibitem[{\citenamefont{Bak}(1982)}]{bak1982}
\bibinfo{author}{\bibfnamefont{P.}~\bibnamefont{Bak}}, \bibinfo{journal}{Rep.
  Prog. Phys.} \textbf{\bibinfo{volume}{45}}, \bibinfo{pages}{587}
  (\bibinfo{year}{1982}).

\bibitem[{\citenamefont{Bak}(1986)}]{bak1986}
\bibinfo{author}{\bibfnamefont{P.}~\bibnamefont{Bak}}, \bibinfo{journal}{Phys.
  Today} \textbf{\bibinfo{volume}{39}}, \bibinfo{pages}{38}
  (\bibinfo{year}{1986}).

\bibitem[{\citenamefont{J\"ordens et~al.}(2008)\citenamefont{J\"ordens,
  Strohmaier, G\"unter, Moritz, and Esslinger}}]{Esslinger2008}
\bibinfo{author}{\bibfnamefont{R.}~\bibnamefont{J\"ordens}},
  \bibinfo{author}{\bibfnamefont{N.}~\bibnamefont{Strohmaier}},
  \bibinfo{author}{\bibfnamefont{K.}~\bibnamefont{G\"unter}},
  \bibinfo{author}{\bibfnamefont{H.}~\bibnamefont{Moritz}}, \bibnamefont{and}
  \bibinfo{author}{\bibfnamefont{T.}~\bibnamefont{Esslinger}},
  \bibinfo{journal}{Nature} \textbf{\bibinfo{volume}{455}},
  \bibinfo{pages}{204} (\bibinfo{year}{2008}).

\bibitem[{\citenamefont{Sherson et~al.}(2010)\citenamefont{Sherson, Weitenberg,
  Endres, Cheneau, Bloch, and Kuhr}}]{Kuhr2010}
\bibinfo{author}{\bibfnamefont{J.~F.} \bibnamefont{Sherson}},
  \bibinfo{author}{\bibfnamefont{C.}~\bibnamefont{Weitenberg}},
  \bibinfo{author}{\bibfnamefont{M.}~\bibnamefont{Endres}},
  \bibinfo{author}{\bibfnamefont{M.}~\bibnamefont{Cheneau}},
  \bibinfo{author}{\bibfnamefont{I.}~\bibnamefont{Bloch}}, \bibnamefont{and}
  \bibinfo{author}{\bibfnamefont{S.}~\bibnamefont{Kuhr}},
  \bibinfo{journal}{Nature} \textbf{\bibinfo{volume}{467}}, \bibinfo{pages}{68}
  (\bibinfo{year}{2010}).

\bibitem[{\citenamefont{Schneider et~al.}(2008)\citenamefont{Schneider,
  Hackermuller, Will, Best, Bloch, Costi, Helmes, Rasch, and
  Rosch}}]{Bloch2008}
\bibinfo{author}{\bibfnamefont{U.}~\bibnamefont{Schneider}},
  \bibinfo{author}{\bibfnamefont{L.}~\bibnamefont{Hackermuller}},
  \bibinfo{author}{\bibfnamefont{S.}~\bibnamefont{Will}},
  \bibinfo{author}{\bibfnamefont{T.}~\bibnamefont{Best}},
  \bibinfo{author}{\bibfnamefont{I.}~\bibnamefont{Bloch}},
  \bibinfo{author}{\bibfnamefont{T.~A.} \bibnamefont{Costi}},
  \bibinfo{author}{\bibfnamefont{R.~W.} \bibnamefont{Helmes}},
  \bibinfo{author}{\bibfnamefont{D.}~\bibnamefont{Rasch}}, \bibnamefont{and}
  \bibinfo{author}{\bibfnamefont{A.}~\bibnamefont{Rosch}},
  \bibinfo{journal}{Science} \textbf{\bibinfo{volume}{322}},
  \bibinfo{pages}{1520} (\bibinfo{year}{2008}).

\bibitem[{\citenamefont{Bakr et~al.}(2009)\citenamefont{Bakr, Gillen, Peng,
  Foelling, and Greiner}}]{Greiner2009}
\bibinfo{author}{\bibfnamefont{W.~S.} \bibnamefont{Bakr}},
  \bibinfo{author}{\bibfnamefont{J.~I.} \bibnamefont{Gillen}},
  \bibinfo{author}{\bibfnamefont{A.}~\bibnamefont{Peng}},
  \bibinfo{author}{\bibfnamefont{S.}~\bibnamefont{Foelling}}, \bibnamefont{and}
  \bibinfo{author}{\bibfnamefont{M.}~\bibnamefont{Greiner}},
  \bibinfo{journal}{Nature} \textbf{\bibinfo{volume}{462}}, \bibinfo{pages}{74}
  (\bibinfo{year}{2009}).

\bibitem[{\citenamefont{Umucalilar
  et~al.}(2008{\natexlab{b}})\citenamefont{Umucalilar, Zhai, and
  Oktel}}]{Zhai2008}
\bibinfo{author}{\bibfnamefont{R.~O.} \bibnamefont{Umucalilar}},
  \bibinfo{author}{\bibfnamefont{H.}~\bibnamefont{Zhai}}, \bibnamefont{and}
  \bibinfo{author}{\bibfnamefont{M.~O.} \bibnamefont{Oktel}},
  \bibinfo{journal}{Phys. rev. lett.} \textbf{\bibinfo{volume}{100}},
  \bibinfo{pages}{070402} (\bibinfo{year}{2008}{\natexlab{b}}).

\bibitem[{\citenamefont{Haljan et~al.}(2001)\citenamefont{Haljan, Coddington,
  Engels, and Cornell}}]{Haljan}
\bibinfo{author}{\bibfnamefont{P.~C.} \bibnamefont{Haljan}},
  \bibinfo{author}{\bibfnamefont{I.}~\bibnamefont{Coddington}},
  \bibinfo{author}{\bibfnamefont{P.}~\bibnamefont{Engels}}, \bibnamefont{and}
  \bibinfo{author}{\bibfnamefont{E.~A.} \bibnamefont{Cornell}},
  \bibinfo{journal}{Phys. Rev. lett.} \textbf{\bibinfo{volume}{87}},
  \bibinfo{pages}{210403} (\bibinfo{year}{2001}).

\bibitem[{\citenamefont{Lin et~al.}(2009)\citenamefont{Lin, Compton, Garcia,
  Porto, and Spielman}}]{Lin2009}
\bibinfo{author}{\bibfnamefont{Y.-J.} \bibnamefont{Lin}},
  \bibinfo{author}{\bibfnamefont{R.~L.} \bibnamefont{Compton}},
  \bibinfo{author}{\bibfnamefont{K.~J.} \bibnamefont{Garcia}},
  \bibinfo{author}{\bibfnamefont{J.~V.} \bibnamefont{Porto}}, \bibnamefont{and}
  \bibinfo{author}{\bibfnamefont{I.~B.} \bibnamefont{Spielman}},
  \bibinfo{journal}{Nature} \textbf{\bibinfo{volume}{462}},
  \bibinfo{pages}{628} (\bibinfo{year}{2009}).

\bibitem[{\citenamefont{Streda}(1982)}]{Streda}
\bibinfo{author}{\bibfnamefont{P.}~\bibnamefont{Streda}}, \bibinfo{journal}{J.
  Phys. C} \textbf{\bibinfo{volume}{15}}, \bibinfo{pages}{L717}
  (\bibinfo{year}{1982}).

\end{thebibliography}

\end{document}